# Multiple scattering theory for heterogeneous elastic continua with strong property fluctuation: theoretical fundamentals and applications

Huijing He[†]
Department of Mechanical and Materials Engineering
University of Nebraska-Lincoln, Lincoln, Nebraska 68588, USA
[†] he.hui.jing@hotmail.com

## ABSTRACT

Scattering of elastic waves in heterogeneous media has become one of the most important problems in the field of wave propagation due to its broad applications in seismology, natural resource exploration, ultrasonic nondestructive evaluation and biomedical ultrasound. Nevertheless, it is one of the most challenging problems because of the complicated medium inhomogeneity and the complexity of the elastodynamic equations. A widely accepted model for the propagation and scattering of elastic waves, which properly incorporates the multiple scattering phenomenon and the statistical information of the inhomogeneities is still missing. In this work, the author developed a multiple scattering model for heterogeneous elastic continua with strong property fluctuation and obtained the exact solution to the dispersion equation under the first-order smoothing approximation. The model establishes an accurate quantitative relation between the microstructural properties and the coherent wave propagation parameters and can be used for characterization or inversion of microstructures. Starting from the elastodynamic differential equations, a system of integral equation for the Green functions of the heterogeneous medium was developed by using Green's functions of a homogeneous reference medium. After properly eliminating the singularity of the Green tensor and introducing a new set of renormalized field variables, the original integral equation is reformulated into a system of renormalized integral equations. Dyson's equation and its first-order smoothing approximation, describing the ensemble averaged response of the heterogeneous system, are then derived with the aid of Feynman's diagram technique. The dispersion equations for the longitudinal and transverse coherent waves are then obtained by applying Fourier transform to the Dyson equation. The exact solution to the dispersion equations are obtained numerically. To validate the new model, the results for weak-property-fluctuation materials are compared to the predictions given by an improved weak-fluctuation multiple scattering theory. It is shown that the new model is capable of giving a more robust and accurate prediction of the dispersion behavior of weak-property-fluctuation materials. Numerical results further show that the new model is still able to provide accurate results for strong-property-fluctuation materials while the weak-fluctuation model is completely failed. As applications of the new model, dispersion and attenuation curves for coherent waves in the Earth's lithosphere, the porous and two-phase alloys, and human cortical bone are calculated. Detailed analysis shows the model can capture the major dispersion and attenuation characteristics, such as the longitudinal and transverse wave Q-factors and their ratios, existence of two propagation modes, anomalous negative dispersion, nonlinear attenuation-frequency relation, and even the disappearance of coherent waves. Additionally, it helps gain new insights into a series of longstanding problems, such as the dominant mechanism of seismic attenuation and the existence of the Mohorovičić discontinuity. This work provides a general and accurate theoretical framework for quantitative characterization of microstructures in a broad spectrum of heterogeneous materials and it is anticipated to have vital applications in seismology, ultrasonic nondestructive evaluation and biomedical ultrasound.

## I. INTRODUCTION

Scattering of elastic waves is the central topic in the theory of elastodynamics and its applications cover a broad spectrum of physical and engineering disciplines, ranging from seismology, natural resource exploration, ultrasonic nondestructive evaluation to biomedical ultrasound. Multiple scattering is a universal phenomenon which occurs in any materials that exhibit spatial fluctuation of elastic moduli or density. The planet Earth is a typical heterogeneous media. Geophysical studies show the rocks constituting the lithosphere, typically with a characteristic size from several hundred meters to tens of kilometers, exhibit strong fluctuations in both the elastic stiffness and the density. A seismic wave gets scattered by large amounts of such inhomogeneities during its travel from the source, normally tens of kilometers beneath the Earth's surface, to the observation station located up to thousands of kilometers away from the epicenter. As a consequence, the seismic signal experiences multiple scattering and appears as a wavetrain consists of several dispersive and attenuating direct arrivals followed by a long duration of coda waves. Multiple scattering theory is a necessary tool for the explanation of measured signals and extracting statistical information of the heterogeneous lithosphere. A series of classics in seismology [1-4] are devoted to seismic wave scattering and attenuation and numerous research papers have also been published [5-33]. Geophysicists also make use of man-made earthquakes and borehole explosions to explore oil and gas reservoir [34-36]. Porous rocks saturated by hydrocarbon-enriched liquid mixtures are the major constituent of oil and gas reservoir, of which the mass density and elastic modulus are significantly different from that of tight rocks. By analyzing the unique dispersion and attenuation of the artificial seismic waves, geophysicists can identify the reservoirs, infer its mineral constituents and evaluate reserves of the resources. Ultrasonic waves with a central frequency of several megahertz are extensively used in industry for nondestructive characterization of microstructures in heterogeneous materials, such as high-temperature alloys, composite materials or ceramics [37-68]. The typical size of the microstructure in these materials, such as pores, grains, crystallites and microtextured regions, lies between a few micrometers to several millimeters, thus the wavelengths are comparable to the characteristic dimension of the microstructures. In this frequency regime, the wave-scatterer interaction is relatively strong and the multiple scattering theory is extremely important for the analysis of measured signals. In analogy to its industrial applications, ultrasound is also used in medicine to diagnose and monitor certain diseases, like cancer and osteoporosis [69-77]. Bone quantitative ultrasound is a typical example of such applications. Cortical bone is a typical two-phase elastic material composed of a porous solid frame saturated by marrow. Clinical studies reveal that the porosity and average pore size are the key microstructural parameters that control the risk of bone fracture. Consequently, there is an increasing interest in quantitatively characterizing the microstructures in cortical bone [70]. X-ray microtomography (XMT) and nuclear magnetic resonance (NMR) analysis show the diameters of the majority of pores lie between 20 μm and 300 μm, while the volume fraction of pores varies between 5% and 30% [78]. These two factors combine to determine that the propagation of ultrasound in cortical bone is dominated by multiple scattering.

    From the above discussion, we see an accurate and quantitative description of multiple scattering is of vital importance to the success of ultrasound applications. Nevertheless, theoretical analysis of multiple scattering in heterogeneous media is exceedingly challenging. Complexities of the scattering problem arise from two aspects. First, the mass density and/or elastic stiffness vary randomly with spatial coordinates, thus the governing equations are stochastic differential equations, conventional mathematical physics approaches developed for deterministic equations become invalid for these materials. Second, there exist large amounts of interfaces among different phases, where the boundary conditions, i.e., the continuity of displacement and traction must be enforced. As a consequence, obtaining an accurate analytical solution to the complete set of boundary value problems quickly becomes intractable. Solving for the numerical solution to this problem is equally difficult. The material properties (density and/or elastic moduli) exhibit a sharp jump across the boundaries, and accurate description of the discontinuity is impossible for certain numerical method, like the finite-difference time domain approach. In addition, large quantities of irregular boundaries exist in the medium, for which an accurate depiction in numerical models is also challenging. Moreover, large amount of meshes or elements makes the numerical simulation of the wave scattering extremely time-consuming, especially for the case of strong-property fluctuation materials and high frequency signals. Finally, numerical dispersion and artificial excitation or consumption frequently contaminate the numerical algorithms and result in distorted signals. Therefore, significant efforts have been directed to developing approximate analytical models for wave propagation in heterogeneous media.

    During the last century, pioneering scientists have made tremendous efforts in studying the multiple scattering of waves and proposed numerous theoretical models to explain the rich phenomena. Based on the fundamental assumptions and methodologies, theoretical models can be classified into two categories, one is called the macroscopic phenomenological models, the other is known as the scattering models. Phenomenological models are also known as homogenized effective medium models, which focus on searching for different homogenization schemes to obtain the effective constitutive properties. The most representative example of phenomenological models is developed by Biot in his pioneering research on fluid-saturated porous materials [79-81]. In Biot's theory, two sets of field variables (displacement, strain and stress) are introduced to describe the motion of the solid phase and the fluid phase, respectively. The equations of motion of a representative volume element for the two-phase material were derived from the Lagrange equations of a specific kinetic energy density, in which a set of mass coefficients were introduced to account for the coupling interactions between the two phases. After taking divergence and curl of the governing equations, Biot obtained two independent equations for the dilatational waves and one for the transverse waves, by which he further predicted the existence of two longitudinal propagation modes, known as the Biot waves of the first and the second type, and a transverse mode. In 1980, Plona [82] conducted a series of ultrasonic immersion tests on a slab of porous elastic material. After analyzing the transmitted signals with different incident angles, he identified

a slow longitudinal mode in addition to the fast longitudinal wave and the transverse wave. He concluded that these waves correspond to those predicted by Biot. Since then, Biot's model becomes the standard method for dealing with wave propagation in fluid-saturated porous elastic materials and is widely used in geomechanics, hydrogeology, petroleum exploration and more recently in biomechanics [70, 83-88]. However, it is generally acknowledged that Biot's model have a number of drawbacks. One shortcoming is that the fluid viscosity is regarded as the sole mechanism for attenuation, while the portion contributed by scattering is completely neglected. It is reported that Biot's model always underestimate the attenuation [86-88]. Another noteworthy defect is that a number of phenomenological parameters, such as the frequency correction factor of the viscosity coefficient and the structural factor are involved, which either have obscured physical meaning or the accurate values are difficult to measure experimentally. Furthermore, Biot's model cannot predict the anomalous negative dispersion as observed in cortical and trabecular bone [70, 73, 89], and fluid-saturated sediment sand [90]. Additionally, numerical results show it cannot predict the velocity and attenuation accurately and simultaneously [70]. In order to remedy these deficiencies, researchers have proposed different versions of modified Biot's model. Dvorkin and coworkers incorporate the effects of local squirt flow into Biot's model and developed the so-called BISQ model [19-21]. The BISQ model achieves certain degree of success in regards to improving the estimation of seismic attenuation. Muller and Gurevich [26-27] considered the effects of both the wave-induced flow and scattering on the seismic attenuation and derived a statistical smoothing approximation theory based on the Biot equations. Johnson and coworkers introduced the concepts of dynamic permeability and dynamic tortuosity to describe the fluid-saturated pores in the solid frame [91]. It has been applied in analysis of ultrasound dispersion and attenuation of trabecular bones and achieved intermediate degree of success [70]. Another type of macroscopic phenomenological method is known as the dynamic self-consistent theory. In analogous to the static self-consistent theory [92-95], in which the stress polarization as introduced, Willis and Sabino [59-64] adopted the concept of the momentum polarization for dynamic problems and derived a system of coupled integral equations for the ensemble-averaged displacement and strain. A set of self-consistent conditions are derived to fix the properties of a homogeneous reference medium. Formulas for waves propagating in general two-phase composite materials are then specialized for the cases of aligned spheroidal inclusions and randomly oriented spheroids. Numerical results for velocity and attenuation shows that their model can successfully capture the negative dispersion, which is a common feature of two-phase materials. An obvious drawback of the model is that at high frequencies, the attenuation predicted by the model approaches zero, which is physically impractical. In light of Willis and Sabino's work, Zhuck and Lakhtakia [96-97] incorporated the second-order statistics of the random medium and developed a system of renormalized integral equations for the estimation of the effective constitutive properties. In the proposed model, they first considered the singularity of the elasodynamic Green's tensor, and thus their results are applicable for elastic materials with strong property fluctuation. As pointed out in their work, a major drawback of the homogenization approach is that it is only valid for heterogeneous materials in which the characteristic dimension of the heterogeneities is less than the wavelength of the excitation signals, i.e., at relatively low frequencies. Through the above discussion, it is seen different versions of effective medium models have several shortcomings in common. The most noteworthy is that they are valid either at relatively low frequencies or in a relatively narrow frequency band, none of them can predict the effective properties in the whole frequency range. This puts server limits on practical applications, for instance, analysis of high frequency signals. Another disadvantage is the introduction of phenomenological parameters which require to be measured experimentally or need to be known in advance. These limitations pose challenges for practical applications and consequently, they are rarely used as inversion models for microstructure characterization.

Contrary to phenomenological models, whose attention is concentrated on searching for an equivalent homogeneous medium, scattering models are focused on investigating the interaction of wave fields with randomly distributed small-scale scatterers. Parameters of practical importance, such as backscattering coefficients, scattering cross section, diffusivity and mean free path, in addition to the velocity and attenuation, are extracted through the analysis of the scattering process. At present, a vast variety of scattering theories and approximations have been proposed. All these scattering theories can be classified into several categories following different classification schemes. For instance, according to the frequency range in which they are applicable, scattering theories can be classified into low frequency, stochastic and high frequency theories. Based on the fractional fluctuation of the material properties of the constituent phases, scattering theories are also classified into weak scattering theories [2, 16, 48, 98] and strong scattering theories [99-109]. In view of stochastic medium model, scattering models can also be classified into discrete scatterer models [110-119] and continuum models [2, 16, 48, 98]. In order to obtain a comprehensive understanding of the background of this area, here we give a brief overview of the most popular models and approximations, and compare their advantages and disadvantages. The Born approximation is probably the most famous approximation used in the scattering community. When the fluctuations of the materials properties are weak, and the wavelength is large compared to the typical dimension of the heterogeneities, the field in the inclusions can be approximated by the unperturbed field. This approximation is called the Born approximation. The Born approximation is an ideal tool for dealing with weak scattering problems, and it has been used in nearly all involved disciplines, including seismology [1-3], ultrasonic NDE [37-41] and electromagnetic remote sensing [120-122]. Explicit expressions for measurable quantities like the scattering section and the backscattering coefficient of polycrystalline materials are derived in [37-41]. Born approximation fails in the high frequency regime and for strong fluctuation materials. Complementary to the Born approximation, the Rytov approximation, the geometric approximation and the parabolic approximation [3, 123-124] are developed to analyze scattering behavior in the high frequency regime, although they are still limited to weak scattering problems. When the concentration of the inclusions is low, or in other words, the solid is a dilute solution of the dispersed inclusions, a straightforward approximation is to omit the interactions among different inclusions, and assume that each scatterer interacts with the incident wave independently, this approximation is called the independent scattering approximation (ISA) [18, 33, 41]. ISA is applicable to inclusions with weak property perturbation or predict the dispersion and attenuation accurately when the volumetric concentration of the point-scatterers is very small.

Multiple scattering theories account for the sophisticated scattered wavefield faithfully and aim at giving an accurate and consistent description of the propagation behaviors. Rossum and Nieuwenhuizen [125] and Barabanenkov et. al. [126] gave two excellent reviews for the historical development and the current state of general multiple scattering theories. The first multiple scattering theory for elastic waves was developed by Karal and Keller [98]. This model relies on the small-perturbation expansion of differential operators and their inverse operators. After successive iteration of the series expansion of the wave operator, the differential equations are transferred into a system of integral equations, from which the Christoffel equation for the coherent plane waves is then derived. Based on Keller's approximation, they obtained analytical formulas for the velocity and attenuation which are valid in the whole frequency range. Stanke and Kino [46-47] developed a unified model for polycrystalline materials in parallel to the Karal-Keller approach. This model has been applied in nondestructive characterization of microstructures in polycrystalline alloys. Exact dispersion and attenuation of alloys with cubic crystallites are obtained in the whole frequency range. Weaver [48] developed the first multiple scattering theory based on Dyson's equation and the first-order-smoothing-approximation (FOSA) [127-128]. The Dyson equation of the ensemble-averaged Green's function for polycrystalline medium was obtained by introducing a multiple scattering series expansion of the stochastic differential equations. By invoking the First-Order-Smoothing approximation (FOSA), the explicit expressions of the mass operators were derived for longitudinal and transverse coherent waves. Closed forms of longitudinal and transverse Green's functions and dispersion equations were then obtained based on the FOSA Dyson's equation. Because of the extreme complexity of the resulting expressions, the Born approximation was adopted in the last step and the closed-form analytical expressions for longitudinal and transverse wave attenuation were obtained. As pointed out in [48], the final expressions are only valid for weak fluctuation materials and low frequency waves. Since then, Weaver's model has become one of the most widely used model in the nondestructive evaluation community. Weaver's original work considered untextured polycrystals with spherical cubic grains only. Turner [57] extended Weaver's model to incorporate the effects of textures. He studied equal-axed cubic polycrystal with the crystallographic axis aligned along a preferred direction. Green's functions for transversely isotropic reference medium are incorporated to account for the macroscopically transverse isotropy of the textured polycrystals. Turner and Anugonda further considered two-phase materials with microscopically isotropic components [58]. It is noted that both [57] and [58] focus on the attenuation under the Born approximation, while the exact solution to the exact dispersion equation is still missing. Moreover, it is seen that only one root for attenuation is found for all propagation directions. Calvet and Margerin calculated both the velocity and the Q-factors of polycrystalline materials with cubic and hexagonal crystallites using Weaver's model, with an ultimate goal to analyze the grain anisotropy of the Earth's uppermost inner core [13]. All the previous calculations in the framework of Weaver's model adopt the Born approximation and valid in the low frequency regime only. In order to obtain the propagation characteristics at high frequencies, Calvet and Margerin [12] proposed a spectral function approach, in which the imaginary part of the ensemble averaged Green's function derived in Weaver's model is identified as the spectral function. By introducing a simplified spectral function of the coherent waves and invoking the method of least square fitting, the dispersion and attenuation of the coherent wave are reconstructed from the exact spectral function. They first discovered that at high frequencies there exist two sets of longitudinal and transverse modes, although at the end of the paper they questioned the discovery by providing a subtle example for which there should be one propagation mode but their calculations gave two. Calvet and Margerin further extended Weaver's model to incorporate the effects of grain shape [14]. An interesting phenomenon is that the attenuation and dispersion exhibit obvious anisotropy as a result of the geometric anisotropy. A noteworthy feature of the spectral function approach is that it is only an approximation to the exact solution. Solving for the exact solution to the dispersion equation derived from the FOSA Dyson's equation is still very challenging. Recently the author conducted comprehensive study on Weaver's model and the spectral function approach and obtained the exact solution to the dispersion equations [129]. By comparing the accurate solution with the results obtained using the spectral function approach, it is recovered that the spectral function method is valid at low and high frequencies only, and in the stochastic transition region it gives incorrect results. Most importantly, the performance of Weaver's model for two-phase materials is quite unstable, in certain cases it gives physically impractical results. Both Weaver's model and Stanke-Kino's model use the Voigt-average material as the homogeneous reference medium. Turner pointed out the Voigt-averaged velocities always overestimate the velocities that measured in experiments [49]. To remedy this discrepancy, he introduced a self-consistent scheme to calculate the properties of the homogeneous reference medium, and incorporated these material properties into Weaver's model [50]. Through comparison of the results obtained by different homogenization schemes, such as the Voigt, Reuss, Hill and self-consistent techniques, he concluded that the self-consistent method significantly improves the predictions of the dispersion and attenuation. It is worth mentioning that all these scattering theories are valid only for weak scattering media because the small-perturbation expansion is adopted for the description of property perturbation. A number of multiple scattering theories are also developed based on the discrete-scatterer model. One such theory is known as the generalized coherent potential approximation (GCPA) [110-112]. Based on the observation that waves propagating in the homogeneous effective medium should have no scattering at low frequencies and have very little scattering in the high frequency range, Sheng [110] proposed that the dispersion equation can be obtained by enforcing the mass operator vanish at low frequency and assuming its minimum in the high frequency range. Further approximations are introduced in order to make the resulting equations solvable, such as the T-matrix approach. The dispersion curve for liquid suspensions with monodispersed methylmethacrylate spheres are obtained and it shows that in the intermediate to high frequency range, there exist two propagation modes, which was observed experimentally by [113]. Sheng and coworkers developed the theory for liquid suspensions and electromagnetic waves only while the counterpart theory for elastic materials is still missing. Foldy [114] and Twersky [115-116] developed another multiple scattering theory based on the discrete scatterer model. In their model, the complex wavenumber is expressed explicitly using the far-field scattering amplitude of a single inclusion. The results are applicable when the volume concentration of the scatterers is low. Quasi-crystal approximation [60-61, 117-118] is another discrete scattering model applicable for heterogeneous media in which inclusions are located on a regular lattice. The range of validation of different versions of scattering theories are summarized in Tab. 1.

Table 1. Ranges of validation of common scattering models and approximations.

| Scattering Theories | $\lambda \ll d$ | $\lambda \approx d$ | $\lambda \gg d$ | WPFM | SPFM | LVF | HVF | $V$ | $\alpha$ |
|---|---|---|---|---|---|---|---|---|---|
| Biot's model | × | √ | √ | | solid-fluid | | | √ | × |
| Weaver's model | √ | √ | √ | √[a] | × | √ | √ | √ | √ |
| KA, S-K model | √ | √ | √ | √[b] | × | √ | √ | √ | √ |
| Foldy-Twersky model | × | × | √ | √ | √ | √ | × | √ | √ |
| SFA | √ | × | √ | √ | × | √ | √ | √ | × |
| BA | × | √ | √ | √ | × | | | | |
| RA, PA | √ | × | × | √ | × | | | | |
| ISA, SSA | × | × | √ | √ | √ | √ | × | × | √ |
| FSA, GA | √ | × | × | √ | × | | | | |
| GCPA | × | × | √ | √ | √ | √ | × | √ | × |
| QCA | × | × | √ | | √ | | | √ | √ |

Acronyms: KA=Keller's Approximation, S-K model= Stanke-Kino's model, SFA=Spectral Function Approach, BA=Born Approximation, RA=Rytov Approximation, PA=Parabolic Approximation, ISA=Independent Scattering Approximation, SSA=Single Scattering Approximation, FSA=Forward scattering approximation, GA=Geometric Approximation, GCPA=Generalized Coherent Potential Approximation, QCA= Quasi-crystal approximation, LVC=low volume concentration, HVC=high volume concentration, WPFM=Weak-Property-Fluctuation-Medium, SPFM=Strong-Property-Fluctuation-Medium

[a] Weaver's model gives reasonable predictions for the velocity and attenuation of polycrystalline alloys in the whole frequency range [129], but it gives unstable predictions for two-phase materials, as shown in this work.

[b] Stanke-Kino's model has been shown to be equivalent to Weaver's model. For polycrystalline alloys, it gives nearly the same results as that given by Weaver's model [50]. However, numerical results for two-phase materials calculated using this model are unavailable.

From Tab. 1 it is seen that the adoption of different types of approximation puts severe limitations on the scope of applications of the existing models. Although each model achieves a certain degree of success, a general multiple scattering theory for heterogeneous materials with strong property fluctuations and valid in the whole frequency range is still missing. As a consequence, the exact dispersion and attenuation behavior of coherent waves have not been obtained yet. Moreover, there are also a number of fundamental problems, such as the dominant mechanism of seismic wave attenuation [9], the existence of the Mohorovičić discontinuity [130-132], applicability of the Kramers-Kronig relation to multiple scattered elastic waves [65-66, 133-134], and anomalous negative dispersion of cortical bone [70, 73] are full of controversy in the scientific community. Intrigued by the limitations of the existing models, in the present work the author aims at developing a most universal theoretical framework that applicable to a large variety of heterogeneous elastic materials, regardless of weak or strong property fluctuation, with low or high volumetric concentration of inclusions, excited by low or high frequency signals, while giving up all the conventional approximations. Based on the exact solutions to the new model, this work will also provide a series of completely new explanations to the longstanding problems. In addition, it will be demonstrated that the new model is a real prediction model that can be used for microstructure inversion and seismic/ultrasonic data interpretation.

The contents of subsequent sections are arranged as follows: Section II presents the rigorous development of the new model, including Green's function of the heterogeneous medium and its integral representation, singularity of the Green tensor, derivation of the renormalized Dyson's equation using Feynman's diagram and the first-order-smoothing-approximation, solution of the Dyson equation and the dispersion equations. In Section III, the exact solutions for the dispersion and attenuation of longitudinal and transverse waves will be discussed. The advantages of the new model will be demonstrated through comparison of the numerical results with that calculated from a multiple scattering theory for weak-property-fluctuation materials. Section IV presents a series of practical applications of the new model in seismology, ultrasonic NDE and bone quantitative ultrasound. To show the power of the model, we calculate the exact dispersion and Q-factor curves of seismic waves propagating in the Earth's lithosphere, the dispersion and attenuation of ultrasonic waves in porous aluminum and Cu-Al alloy, and the dispersion and attenuation curves of ultrasound in human cortical bone. Through comprehensive parametric study, we show that the new model can give a concise and consistent explanation for a broad spectrum of observed scattering phenomena, such as the Q-factors of seismic waves, negative dispersion in bone, spectral splitting and occurrence of two modes. In section V, we give a brief discussion on how to perform successful numerical and experimental verifications of the model. Future work on extension of the current model to incorporate more complicated microstructures are also presented. Finally, the unique features of the new model are summarized and the major conclusions are highlighted in Section VI.

## II. THEORETICAL FUNDAMENTALS

In this section, we present rigorously the theoretical development of the new multiple scattering theory for strong-property-fluctuation elastic continua. It needs to mention that mathematics plays a central role in developing the new multiple scattering model. A series of advanced mathematical techniques, including tensor analysis, system of integral equations, singularity techniques, Feynman's diagram, statistical theory of heterogeneous medium, Fourier transform, theory of complex variables and advanced numerical techniques provide us a precious opportunity to develop an accurate and universal theory while introducing as less approximations as possible.

From the theory of elastodynamics we know, the local wave motion in an inhomogeneous media is governed by a set of partial differential equations (PDEs), known as the Navier-Lamé equations [135]. Boundary conditions are specified by the continuity of stress and displacement across the phase boundaries. Thus, the multiple scattering problem is essentially a sophisticated PDE boundary value problem (BVP). From a mathematical point of view, the integral equation (IE) description and the PDE BVP description comprise two equivalent and complementary representations for the same problem. While the PDE BVP description is frequently utilized to solve scattering problems of a single or a finite number of scatterers with regular geometry [135-136], the IE description has proven to be more convenient for dealing with scattering problems in inhomogeneous media with multiple irregular inclusions or even randomly distributed scatterers [48, 98, 137-138]. Transforming the multiple scattering problem from the PDE BVP description into the IE description is the first key step for the development of the new theory. Green's function plays a vital role in this transformation procedure. As a fundamental solution to the elastodynamic equation, Green's function contains the complete information of the whole system and can be used to analyze more complicated source distributions or time varying excitations. When transferred into the frequency-wavenumber domain, Green's function can also be used to extract dispersion behaviors of plane wave components propagating along different directions. Due to its extreme significance for the theoretical development of this work, we first perform comprehensive study on Green's functions of heterogeneous media.

### A. Green's function and its integral representation

Before embarking on the analysis of wave scattering in a general heterogeneous medium, we first consider the most fundamental problem: scattering of elastic waves by a single inclusion in an infinite homogeneous elastic medium.

### 1. Green's function of a homogeneous medium with a single inclusion

The time-harmonic wave propagation in a heterogeneous medium is governed by the classic elastodynamics equation [135], given by:

$$[c_{ijkl}(\mathbf{x})u_{k,l}]_{,j} + \rho(\mathbf{x})\omega^2 u_i = 0, \tag{1}$$

where $\omega$ is the circular frequency, $u_i$ denotes the displacement components, $\rho(\mathbf{x})$ and $c_{ijkl}(\mathbf{x})$ are the mass density and elastic stiffness. The elastic stiffness has the following symmetries:

$$c_{ijkl}(\mathbf{x}) = c_{jikl}(\mathbf{x}) = c_{ijlk}(\mathbf{x}) = c_{klij}(\mathbf{x}). \tag{2}$$

Throughout this work the Cartesian tensor notation is used. A bold-faced letter represents a vector, tensor or matrix, and italic letters with subscript indices represent tensor components or matrix elements. A comma followed by a coordinate index means taking partial derivative with respect to the corresponding spatial coordinate. The Einstein summation convention, i.e., a repeated index implies summation over that index from 1 to 3, is assumed in this work.

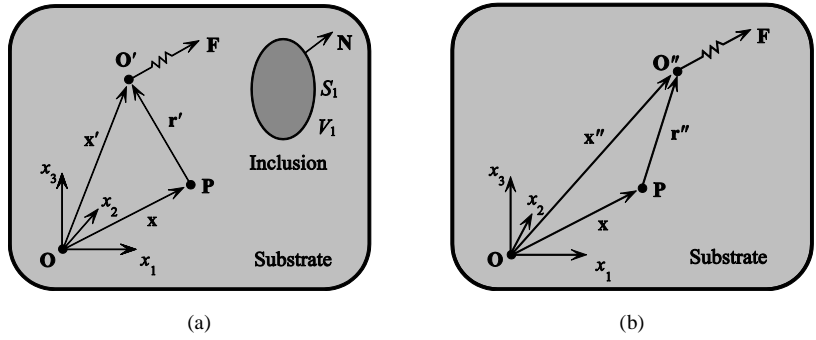

FIG. 1. Definition of the single-inclusion Green's function (a) and the homogeneous Green's function (b).

Consider an infinite homogeneous elastic medium in which an inclusion with different density and elastic stiffness is embedded, as shown in Fig. 1(a). The quantities pertaining to the homogeneous medium and that to the inclusion are discriminated by an attached index "(0)" or "(1)", respectively.

Green's function is defined as the resulting field excited by a time-harmonic unit concentrated force $\mathbf{F}$ applied at a generic point $\mathbf{O'}$ along $\mathbf{e}_{a'}$, where $\mathbf{e}_{a'}$ represents the coordinate basis of a source-region coordinated system with its origin located at $\mathbf{O'}$. The coordinate basis of the coordinate system $\mathbf{O'}x'y'z'$ by no means needs to be the same as that of the coordinate system $\mathbf{O}xyz$, so at this point we assume they are different from each other. The portions of the field in the substrate and in the inclusion are governed by two different sets of equations, which are given by:

$$c_{ijkl}^{(0)} G_{ka',lj} + \rho^{(0)}\omega^2 G_{ia'} + F a_{ia'}\delta(\mathbf{x}-\mathbf{x}') = 0, \quad \text{in } V_0, \tag{3}$$

$$c_{ijkl}^{(1)} G_{ka',lj} + \rho^{(1)}\omega^2 G_{ia'} + F a_{ia'}\delta(\mathbf{x}-\mathbf{x}') = 0, \quad \text{in } V_1, \tag{4}$$

where $V_0$ and $V_1$ represent the volumes occupied by the substrate and the inclusion, respectively, and $a_{ia'}$ denotes the directional cosine of $\mathbf{e}_{a'}$, i.e., $a_{ia'} = \cos(\mathbf{e}_{a'}, \mathbf{e}_i)$.

Suppose the inclusion and the substrate are perfectly bounded, so the stress and displacement are continuous at the boundary of the inclusion, thus we have

$$c_{ijkl}^{(0)} G_{ka',l}(\mathbf{x}\to S^+)N_j = c_{ijkl}^{(1)} G_{ka',l}(\mathbf{x}\to S^-)N_j, \quad G_{ka'}(\mathbf{x}\to S^+) = G_{ka'}(\mathbf{x}\to S^-), \quad \text{on } S. \tag{5}$$

In order to obtain the integral representation of the perturbed Green's function, we now consider the corresponding homogeneous Green's function, which is defined as the field induced in the same medium but in the absence of the inclusion by a unit concentrated force $\mathbf{F}$ applied at another point $\mathbf{O''}$ along $\mathbf{e}_{a''}$, as shown in Fig. 1(b). $\mathbf{e}_{a''}$ is the coordinate basis of a new coordinate system $\mathbf{O''}x''y''z''$ which is used to describe the source distribution in the homogeneous medium. This function satisfies:

$$c_{ijkl}^{(0)} G_{ka'',lj}^{(0)} + F a_{ia''}\delta(\mathbf{x}-\mathbf{x}'') + \rho^{(0)}\omega^2 G_{ia''}^{(0)} = 0. \tag{6}$$

Multiplying Eq. (6) and Eq. (3) by $G_{ia'}$ and $G_{ia''}^{(0)}$, respectively, and then subtracting the resulting equations, we obtain:

$$c_{ijkl}^{(0)}G_{k\alpha'',lj}^{(0)}G_{i\alpha'} - c_{ijkl}^{(0)}G_{k\alpha',lj}G_{i\alpha''}^{(0)} + Fa_{i\alpha''}G_{i\alpha'}\delta(\mathbf{x}-\mathbf{x}'') - Fa_{i\alpha'}G_{i\alpha''}^{(0)}\delta(\mathbf{x}-\mathbf{x}') = 0. \tag{7}$$

Multiplying Eq. (6) and Eq. (4) by $G_{i\alpha'}$ and $G_{i\alpha''}^{(0)}$, respectively, and then subtracting the resulting equations, we obtain:

$$c_{ijkl}^{(0)}G_{k\alpha'',lj}^{(0)}G_{i\alpha'} - c_{ijkl}^{(1)}G_{k\alpha',lj}G_{i\alpha''}^{(0)} + \rho^{(0)}\omega^2 G_{i\alpha''}^{(0)}G_{i\alpha'} - \rho^{(1)}\omega^2 G_{i\alpha'}G_{i\alpha''}^{(0)} + Fa_{i\alpha''}G_{i\alpha'}\delta(\mathbf{x}-\mathbf{x}'') - Fa_{i\alpha'}G_{i\alpha''}^{(0)}\delta(\mathbf{x}-\mathbf{x}') = 0. \tag{8}$$

Appling product rule of partial derivatives to Eqs. (7) and (8), we get:

$$c_{ijkl}^{(0)}[G_{k\alpha'',l}^{(0)}G_{i\alpha'}]_{,j} - c_{ijkl}^{(0)}G_{k\alpha'',l}^{(0)}G_{i\alpha',j} - [c_{ijkl}^{(0)}G_{k\alpha',l}G_{i\alpha''}^{(0)}]_{,j} + c_{ijkl}^{(0)}G_{k\alpha',l}G_{i\alpha'',j}^{(0)} + Fa_{i\alpha''}G_{i\alpha'}\delta(\mathbf{x}-\mathbf{x}'') - Fa_{i\alpha'}G_{i\alpha''}^{(0)}\delta(\mathbf{x}-\mathbf{x}') = 0, \tag{9}$$

$$c_{ijkl}^{(0)}[G_{k\alpha'',l}^{(0)}G_{i\alpha'}]_{,j} - c_{ijkl}^{(0)}G_{k\alpha'',l}^{(0)}G_{i\alpha',j} - [c_{ijkl}^{(1)}G_{k\alpha',l}G_{i\alpha''}^{(0)}]_{,j} + c_{ijkl}^{(1)}G_{k\alpha',l}G_{i\alpha'',j}^{(0)}$$
$$+ \rho^{(0)}\omega^2 G_{i\alpha''}^{(0)}G_{i\alpha'} - \rho^{(1)}\omega^2 G_{i\alpha'}G_{i\alpha''}^{(0)} + Fa_{i\alpha''}G_{i\alpha'}\delta(\mathbf{x}-\mathbf{x}'') - Fa_{i\alpha'}G_{i\alpha''}^{(0)}\delta(\mathbf{x}-\mathbf{x}') = 0. \tag{10}$$

Considering the symmetry of the elastic stiffness, Eqs. (9) and (10) can be simplified as:

$$c_{ijkl}^{(0)}[G_{k\alpha'',l}^{(0)}G_{i\alpha'}]_{,j} - [c_{ijkl}^{(0)}G_{k\alpha',l}G_{i\alpha''}^{(0)}]_{,j} + Fa_{i\alpha''}G_{i\alpha'}\delta(\mathbf{x}-\mathbf{x}'') - Fa_{i\alpha'}G_{i\alpha''}^{(0)}\delta(\mathbf{x}-\mathbf{x}') = 0, \tag{11}$$

$$c_{ijkl}^{(0)}[G_{k\alpha'',l}^{(0)}G_{i\alpha'}]_{,j} - [c_{ijkl}^{(1)}G_{k\alpha',l}G_{i\alpha''}^{(0)}]_{,j} + (c_{ijkl}^{(1)} - c_{ijkl}^{(0)})G_{k\alpha',l}G_{i\alpha'',j}^{(0)} - (\rho^{(1)} - \rho^{(0)})\omega^2 G_{i\alpha'}G_{i\alpha''}^{(0)} + Fa_{i\alpha''}G_{i\alpha'}\delta(\mathbf{x}-\mathbf{x}'') - Fa_{i\alpha'}G_{i\alpha''}^{(0)}\delta(\mathbf{x}-\mathbf{x}') = 0. \tag{12}$$

Integrating Eq. (11) over the domain $V_0$, and then applying Gauss's divergence theorem, we obtain:

$$\iiint_{V_0} Fa_{i\alpha''}\delta(\mathbf{x}-\mathbf{x}')G_{i\alpha'}dV - \iiint_{V_0} Fa_{i\alpha'}\delta(\mathbf{x}-\mathbf{x}')G_{i\alpha''}dV$$
$$+ \oiint_{S^\infty}[c_{ijkl}^{(0)}G_{k\alpha'',l}^{(0)}G_{i\alpha'} - c_{ijkl}^{(0)}G_{k\alpha',l}G_{i\alpha''}^{(0)}]N_j dS + \oiint_{S_1}[c_{ijkl}^{(0)}G_{k\alpha'',l}^{(0)}G_{i\alpha'} - c_{ijkl}^{(0)}G_{k\alpha',l}G_{i\alpha''}^{(0)}](-N_j)dS = 0. \tag{13}$$

A negative sign is introduced in front of $N_j$ in the integral over $S_1$ because the out normal of volume $V_0$ corresponds to the negative of **N**, which is shown in Fig. 1. $S^\infty$ represents an imaginary surface that lies infinitely far from the source and the inclusion.

Integrating Eq. (12) over the volume $V_1$ and applying Gauss's divergence theorem, we obtain:

$$\iiint_{V_1} Fa_{i\alpha''}\delta(\mathbf{x}-\mathbf{x}')G_{i\alpha'}dV - \iiint_{V_1} Fa_{i\alpha'}G_{i\alpha''}^{(0)}\delta(\mathbf{x}-\mathbf{x}')dV + \oiint_{S_1}[c_{ijkl}^{(0)}G_{k\alpha'',l}^{(0)}G_{i\alpha'} - c_{ijkl}^{(1)}G_{k\alpha',l}G_{i\alpha''}^{(0)}]N_j dS$$
$$+ \iiint_{V_1}[(c_{ijkl}^{(1)} - c_{ijkl}^{(0)})G_{k\alpha',l}G_{i\alpha'',j}^{(0)} - (\rho^{(1)} - \rho^{(0)})\omega^2 G_{i\alpha'}G_{i\alpha''}^{(0)}]dV = 0. \tag{14}$$

Adding Eqs. (13) and (14) and then invoking the boundary conditions Eq. (5), we have:

$$\iiint_{V_0+V_1} Fa_{i\alpha''}\delta(\mathbf{x}-\mathbf{x}'')G_{i\alpha'}dV - \iiint_{V_0+V_1} Fa_{i\alpha'}\delta(\mathbf{x}-\mathbf{x}')G_{i\alpha''}^{(0)}dV + \oiint_{S^\infty}[c_{ijkl}^{(0)}G_{k\alpha'',l}^{(0)}G_{i\alpha'} - c_{ijkl}^{(0)}G_{k\alpha',l}G_{i\alpha''}^{(0)}]N_j dS$$
$$+ \iiint_{V_1}[(c_{ijkl}^{(1)} - c_{ijkl}^{(0)})G_{k\alpha',l}G_{i\alpha'',j}^{(0)} - (\rho^{(1)} - \rho^{(0)})\omega^2 G_{i\alpha'}G_{i\alpha''}^{(0)}]dV = 0. \tag{15}$$

Throughout this work we assume that all the materials have a small damping, so both the homogeneous and the inhomogeneous Green's functions decay exponentially as the distance between the source point and the field point increases. Consequently, these functions tend to zero at infinity and the integral over $S^\infty$ vanishes. Furthermore, we assume that the coordinate basis of the coordinate systems $\mathbf{O}'x_1'x_2'x_3'$ and $\mathbf{O}''x_1''x_2''x_3''$ coincide with that of the coordinate $Ox_1x_2x_3$, thus $a_{i\alpha'} = \delta_{i\alpha'}$ and $a_{i\alpha''} = \delta_{i\alpha''}$, Eq. (15) can be simplified as:

$$FG_{\alpha''\alpha'} - FG_{\alpha'\alpha''}^{(0)} + \iiint_{V_1}[(c_{ijkl}^{(1)} - c_{ijkl}^{(0)})G_{k\alpha',l}G_{i\alpha'',j}^{(0)} - (\rho^{(1)} - \rho^{(0)})\omega^2 G_{i\alpha'}G_{i\alpha''}^{(0)}]dV = 0. \tag{16}$$

Finally, we obtain the integral representation of the inhomogeneous Green function expressed in terms of the reference homogeneous Green's function,

$$G_{\alpha''\alpha'} = G_{\alpha'\alpha''}^{(0)} + \frac{1}{F}\iiint_{V_1}[-(c_{ijkl}^{(1)} - c_{ijkl}^{(0)})G_{k\alpha',l}G_{i\alpha'',j}^{(0)} + (\rho^{(1)} - \rho^{(0)})\omega^2 G_{i\alpha'}G_{i\alpha''}^{(0)}]dV. \tag{17}$$

To further simplify the expression of Eq. (17), we decompose the density and elastic stiffness of the inclusion into two parts, one corresponds to the ensemble averaged value, for this case it equals to the properties of the substrate, and the other corresponds to the property fluctuations,

$$c_{ijkl}^{(1)} = c_{ijkl}^{(0)} + \delta c_{ijkl}^{(1)}, \quad \rho^{(1)} = \rho^{(0)} + \delta\rho^{(1)}. \tag{18}$$

Substitution of (18) into (17) yields:

$$G_{\alpha''\alpha'}(\mathbf{x}'',\mathbf{x}') = G_{\alpha'\alpha''}^{(0)}(\mathbf{x}',\mathbf{x}'') + \frac{1}{F}\iiint_{V_1}[-\delta c_{ijkl}^{(1)}G_{k\alpha',l}(\mathbf{x},\mathbf{x}')G_{i\alpha'',j}^{(0)}(\mathbf{x},\mathbf{x}'') + \delta\rho^{(1)}\omega^2 G_{i\alpha'}(\mathbf{x},\mathbf{x}')G_{i\alpha''}^{(0)}(\mathbf{x},\mathbf{x}'')]dV, \tag{19}$$

where Green's function of the homogeneous medium is given by:

$$G_{i\alpha'}^0(\mathbf{x},\mathbf{x}',\omega) = \frac{F}{4\pi}\frac{\delta_{i\alpha'}}{\mu}\frac{e^{ik_T|\mathbf{x}-\mathbf{x}'|}}{|\mathbf{x}-\mathbf{x}'|} + \frac{F}{4\pi\rho\omega^2}\frac{\partial^2}{\partial x_i \partial x_a'}\left[\frac{e^{ik_L|\mathbf{x}-\mathbf{x}'|}}{|\mathbf{x}-\mathbf{x}'|} - \frac{e^{ik_T|\mathbf{x}-\mathbf{x}'|}}{|\mathbf{x}-\mathbf{x}'|}\right]. \tag{20}$$

For a detailed derivation of this expression, readers are referred to the author's monograph [129]. It is seen from Eq. (20) that the homogeneous Green's function has the following symmetry:

$$G^0_{i\alpha'}(\mathbf{x}, \mathbf{x}') = G^0_{\alpha'i}(\mathbf{x}, \mathbf{x}') = G^0_{\alpha'i}(\mathbf{x}', \mathbf{x}) = G^0_{i\alpha'}(\mathbf{x}', \mathbf{x}). \tag{21}$$

Applying these relations to Eq. (19), we obtain:

$$G_{\alpha''\alpha'}(\mathbf{x}'', \mathbf{x}') = G^{(0)}_{\alpha''\alpha'}(\mathbf{x}'', \mathbf{x}') + \frac{1}{F}\iiint_{V_1}[-\delta c^{(1)}_{ijkl} G_{k\alpha',l}(\mathbf{x}, \mathbf{x}') G^{(0)}_{\alpha''i,j}(\mathbf{x}'', \mathbf{x}) + \delta\rho^{(1)}\omega^2 G_{i\alpha'}(\mathbf{x}, \mathbf{x}') G^{(0)}_{\alpha''i}(\mathbf{x}'', \mathbf{x}')]dV. \tag{22}$$

Eq. (22) is the major result in this section. It establishes the relationship between the perturbed Green's function and Green's function of the homogeneous medium. From a mathematical point of view, Eq. (22) is a system of Fredholm equations of the second kind, which can be solved numerically by introducing certain discretization schemes. At this point it is meaningful to give a physical explanation of Eq. (22). It tells us the perturbed Green's function can be expressed as a summation of the unperturbed Green's function and a perturbation term, and the perturbation term is given by a spatial convolution of the perturbed and unperturbed Green's function weighted by the property fluctuation. When the fluctuation of the material properties is weak, the perturbation of the inhomogeneous Green's function is small compared to the homogeneous one. Based on this observation, M. Born suggested to replace the perturbed Green's function appeared in the kernel by the unperturbed Green's function, and thus an explicit expression for the perturbed function is obtained. This approximation is known as the Born approximation. Here we do not use this approximation and proceed in an accurate manner.

### *2. Green's function for a heterogeneous medium with multiple inclusions*

Now we consider a medium with multiple scatterers. A schematic diagram is shown in Fig. 2.

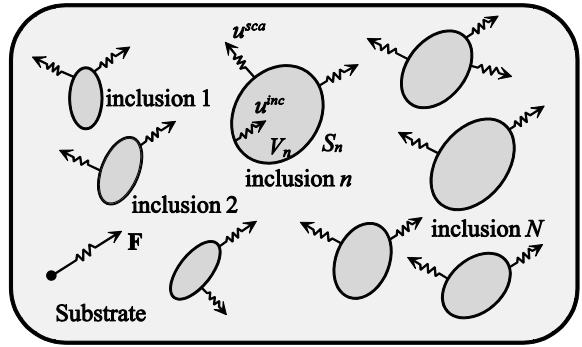

FIG. 2. Schematic diagram of an inhomogeneous medium with multiple inclusions.

Obviously, the above derivation can be straightforwardly extended to inhomogeneous medium with multiple inclusions, and the resulting equation is:

$$G_{\beta''\alpha'}(\mathbf{x}'', \mathbf{x}') = G^0_{\beta''\alpha'}(\mathbf{x}'', \mathbf{x}') + \frac{1}{F}\sum_{n=1}^{N}\int_{V_n}[\delta\rho^{(n)}\omega^2 G_{i\alpha'}(\mathbf{x}, \mathbf{x}') G^0_{\beta''i}(\mathbf{x}'', \mathbf{x}) - \delta c^{(n)}_{ijkl} G_{k\alpha',l}(\mathbf{x}, \mathbf{x}') G^0_{\beta''i,j}(\mathbf{x}'', \mathbf{x})]dV, \tag{23}$$

where $\delta\rho^{(n)}$ and $\delta c^{(n)}_{ijkl}$ represent the material property fluctuation of the $n$-th inclusion.

One needs to keep in mind that all the boundary conditions are naturally incorporated into this equation. From Eq. (23) it is seen that the perturbed Green's function is expressed as the summation of the unperturbed Green's function and a perturbation term which includes contribution of all the inclusions. Meanwhile, the internal field of each inclusion is dependent on all other inclusions and the unperturbed Green's function. At this point we need to point out that by neglecting the interactions among different scatterers, we obtain the independent scattering approximation (ISA) of the perturbed field [38-41]. ISA is frequently used in combination with the Born approximation and yields an explicit expression for $G_{\beta''\alpha'}$. If the property fluctuation is weak or the volume concentration of the scatterers is sufficiently low, each scatterer approximately interacts with the incident wave independently and ISA can give reasonable predictions. However, if these conditions are not satisfied, i.e., material exhibits strong property fluctuation or contains notable volume fraction of inclusions, the interactions among all the scatterers cannot be ignored and the multiple scattering theory is necessary to fully capture the propagation behavior.

For a general heterogeneous material in which the inclusions are densely distributed in the whole space, the material property fluctuations are most conveniently represented by functions of spatial coordinates, and Eq. (23) is rewritten as:

$$G_{\beta''\alpha'}(\mathbf{x}'',\mathbf{x}') = G^0_{\beta''\alpha'}(\mathbf{x}'',\mathbf{x}') + \frac{1}{F}\int_V [\delta\rho(\mathbf{x})\omega^2 G_{i\alpha'}(\mathbf{x},\mathbf{x}')G^0_{\beta''i}(\mathbf{x}'',\mathbf{x}) - \delta c_{ijkl}(\mathbf{x})G_{k\alpha',l}(\mathbf{x},\mathbf{x}')G^0_{\beta''i,j}(\mathbf{x}'',\mathbf{x})]dV. \tag{24}$$

Here comes a peculiar problem. Since the inclusions are densely distributed, and may occupy a finite percentage of the total volume of the whole medium, which material should be chosen as the reference medium? The answer to this question will be clear after we find out the singularity of the Green tensor and introduce the renormalization scheme in Section II.C. Contrary to our intuition, the proper choice of the reference material is neither either phase of the constituent materials nor the volumetric average of the component phases.

### *3. Integral representation of the perturbed field*

Equation (24) is the governing equation for the displacement Green's function. To complete the description of the elastic field, we also need the strain Green's function. The strain tensor of a general elastic medium is defined by:

$$\varepsilon_{\alpha''\beta''} = \frac{1}{2}(u_{\alpha'',\beta''} + u_{\beta'',\alpha''}). \tag{25}$$

Consequently, the Green function for strain is given by:

$$\frac{1}{2}[G_{\beta''\alpha',\alpha''}(\mathbf{x}'',\mathbf{x}') + G_{\alpha''\alpha',\beta''}(\mathbf{x}'',\mathbf{x}')] = \frac{1}{2}[G^0_{\beta''\alpha',\alpha''}(\mathbf{x}'',\mathbf{x}') + G^0_{\alpha''\alpha',\beta''}(\mathbf{x}'',\mathbf{x}')]$$

$$+ \frac{1}{F}\int_V \left[ \delta\rho(\mathbf{x})\omega^2 \frac{1}{2}[G^0_{\beta''i,\alpha''}(\mathbf{x}'',\mathbf{x}) + G^0_{\alpha''i,\beta''}(\mathbf{x}'',\mathbf{x})]G_{i\alpha'}(\mathbf{x},\mathbf{x}') - \frac{1}{2}[G^0_{\beta''i,j\alpha''}(\mathbf{x}'',\mathbf{x}) + G^0_{\alpha''i,j\beta''}(\mathbf{x}'',\mathbf{x})]\delta c_{ijkl}(\mathbf{x})G_{k\alpha',l}(\mathbf{x},\mathbf{x}') \right]dV. \tag{26}$$

Considering the elastic stiffness variation tensor $\delta c_{ijkl}(\mathbf{x})$ has the same symmetry as $c_{ijkl}(\mathbf{x})$, we can write (24) and (26) as

$$G_{\beta''\alpha'}(\mathbf{x}'',\mathbf{x}') = G^0_{\beta''\alpha'}(\mathbf{x}'',\mathbf{x}') + \frac{1}{F}\int_V \left\{ \delta\rho(\mathbf{x})\omega^2 G^0_{\beta''i}(\mathbf{x}'',\mathbf{x})G_{i\alpha'}(\mathbf{x},\mathbf{x}') \right.$$

$$\left. - \frac{1}{2}[G^0_{\beta''i,j}(\mathbf{x}'',\mathbf{x}) + G^0_{\beta''j,i}(\mathbf{x}'',\mathbf{x})]\delta c_{ijkl}(\mathbf{x})\frac{1}{2}[G_{k\alpha',l}(\mathbf{x},\mathbf{x}') + G_{l\alpha',k}(\mathbf{x},\mathbf{x}')] \right\}dV, \tag{27}$$

$$\frac{1}{2}[G_{\beta''\alpha',\alpha''}(\mathbf{x}'',\mathbf{x}') + G_{\alpha''\alpha',\beta''}(\mathbf{x}'',\mathbf{x}')]$$

$$= \frac{1}{2}[G^0_{\beta''\alpha',\alpha''}(\mathbf{x}'',\mathbf{x}') + G^0_{\alpha''\alpha',\beta''}(\mathbf{x}'',\mathbf{x}')] + \frac{1}{F}\int_V \left\{ \delta\rho(\mathbf{x})\omega^2 \frac{1}{2}[G^0_{\beta''i,\alpha''}(\mathbf{x}'',\mathbf{x}) + G^0_{\alpha''i,\beta''}(\mathbf{x}'',\mathbf{x})]G_{i\alpha'}(\mathbf{x},\mathbf{x}') \right. \tag{28}$$

$$\left. - \frac{1}{4}[G^0_{\beta''i,j\alpha''}(\mathbf{x}'',\mathbf{x}) + G^0_{\beta''j,i\alpha''}(\mathbf{x}'',\mathbf{x}) + G^0_{\alpha''i,j\beta''}(\mathbf{x}'',\mathbf{x}) + G^0_{\alpha''j,i\beta''}(\mathbf{x}'',\mathbf{x})]\delta c_{ijkl}(\mathbf{x})\frac{1}{2}[G_{k\alpha',l}(\mathbf{x},\mathbf{x}') + G_{l\alpha',k}(\mathbf{x},\mathbf{x}')] \right\}dV.$$

Equations (27) and (28) can be expressed in a matrix form:

$$\begin{bmatrix} G_{\beta''\alpha'}(\mathbf{x}'',\mathbf{x}') \\ \varepsilon_{\alpha''\beta''\alpha'}(\mathbf{x}'',\mathbf{x}') \end{bmatrix} = \begin{bmatrix} G^0_{\beta''\alpha'}(\mathbf{x}'',\mathbf{x}') \\ \varepsilon^0_{\alpha''\beta''\alpha'}(\mathbf{x}'',\mathbf{x}') \end{bmatrix} + \frac{1}{F}\int_V \begin{bmatrix} G^0_{\beta''i}(\mathbf{x}'',\mathbf{x}) & -\varepsilon^0_{\beta''ij}(\mathbf{x}'',\mathbf{x}) \\ \varepsilon^0_{\alpha''\beta''i}(\mathbf{x}'',\mathbf{x}) & -E^0_{\alpha''\beta''ij}(\mathbf{x}'',\mathbf{x}) \end{bmatrix} \begin{bmatrix} \delta\rho(\mathbf{x})\omega^2\delta_{ij} & 0 \\ 0 & \delta c_{ijkl}(\mathbf{x}) \end{bmatrix} \begin{bmatrix} G_{j\alpha'}(\mathbf{x},\mathbf{x}') \\ \varepsilon_{kl\alpha'}(\mathbf{x},\mathbf{x}') \end{bmatrix} dV, \tag{29}$$

where

$$\varepsilon^0_{\beta''ij}(\mathbf{x}'',\mathbf{x}) = \frac{1}{2}[G^0_{\beta''i,j}(\mathbf{x}'',\mathbf{x}) + G^0_{\beta''j,i}(\mathbf{x}'',\mathbf{x})], \quad \varepsilon^0_{\alpha''\beta''i}(\mathbf{x}'',\mathbf{x}) = \frac{1}{2}[G^0_{\beta''i,\alpha''}(\mathbf{x}'',\mathbf{x}) + G^0_{\alpha''i,\beta''}(\mathbf{x}'',\mathbf{x})], \tag{30}$$

$$E^0_{\alpha''\beta''ij}(\mathbf{x}'',\mathbf{x}) = \frac{1}{4}[G^0_{\beta''i,j\alpha''}(\mathbf{x}'',\mathbf{x}) + G^0_{\beta''j,i\alpha''}(\mathbf{x}'',\mathbf{x}) + G^0_{\alpha''i,j\beta''}(\mathbf{x}'',\mathbf{x}) + G^0_{\alpha''j,i\beta''}(\mathbf{x}'',\mathbf{x})], \tag{31}$$

$$\varepsilon_{\alpha''\beta''\alpha'}(\mathbf{x}'',\mathbf{x}') = \frac{1}{2}[G_{\beta''\alpha',\alpha''}(\mathbf{x}'',\mathbf{x}') + G_{\alpha''\alpha',\beta''}(\mathbf{x}'',\mathbf{x}')], \quad \varepsilon_{kl\alpha'}(\mathbf{x},\mathbf{x}') = \frac{1}{2}[G_{k\alpha',l}(\mathbf{x},\mathbf{x}') + G_{l\alpha',k}(\mathbf{x},\mathbf{x}')]. \tag{32}$$

Note the homogeneous Green's function $G^0_{\beta''i}(\mathbf{x}'',\mathbf{x})$ given by Eq. (20) can be rewritten as $G^0_{\beta''i''}(\mathbf{x}'',\mathbf{x})$, where:

$$G^0_{\beta''i''}(\mathbf{x}'',\mathbf{x},\omega) = \frac{F}{4\pi}\frac{\delta_{\beta i}}{\mu}\frac{e^{ik_T|\mathbf{x}''-\mathbf{x}|}}{|\mathbf{x}''-\mathbf{x}|} - \frac{F}{4\pi\rho\omega^2}\frac{\partial^2}{\partial x''_i \partial x''_\beta}\left[\frac{e^{ik_L|\mathbf{x}''-\mathbf{x}|}}{|\mathbf{x}''-\mathbf{x}|} - \frac{e^{ik_T|\mathbf{x}''-\mathbf{x}|}}{|\mathbf{x}''-\mathbf{x}|}\right]. \tag{33}$$

From Eq. (33) one can easily find:

$$\frac{\partial G^0_{\beta''i''}}{\partial x_j} = -\frac{\partial G^0_{\beta''i''}}{\partial x''_j}. \tag{34}$$

Consequently, we have the following identities:

$$\varepsilon^0_{\beta''ij}(\mathbf{x}'',\mathbf{x}) = -\varepsilon^0_{\beta''i'j'}(\mathbf{x}'',\mathbf{x}), \quad \varepsilon^0_{\alpha''\beta''i}(\mathbf{x}'',\mathbf{x}) = \varepsilon^0_{\alpha''\beta''i'}(\mathbf{x}'',\mathbf{x}), \quad E^0_{\alpha''\beta''ij}(\mathbf{x}'',\mathbf{x}) = -E^0_{\alpha''\beta''i'j'}(\mathbf{x}'',\mathbf{x}), \tag{35}$$

where

$$\varepsilon^0_{\beta''i'j'}(\mathbf{x}'',\mathbf{x}) = \frac{1}{2}[G^0_{\beta''i'',j''}(\mathbf{x}'',\mathbf{x}) + G^0_{\beta''j'',i''}(\mathbf{x}'',\mathbf{x})], \quad \varepsilon^0_{\alpha''\beta''i''}(\mathbf{x}'',\mathbf{x}) = \frac{1}{2}[G^0_{\beta''i'',\alpha''}(\mathbf{x}'',\mathbf{x}) + G^0_{\alpha''i'',\beta''}(\mathbf{x}'',\mathbf{x})], \tag{36}$$

$$E^0_{\alpha''\beta''i'j'}(\mathbf{x}'',\mathbf{x}) = \frac{1}{4}[G^0_{\beta''i'',j''\alpha''}(\mathbf{x}'',\mathbf{x}) + G^0_{\beta''j'',i''\alpha''}(\mathbf{x}'',\mathbf{x}) + G^0_{\alpha''i'',j''\beta''}(\mathbf{x}'',\mathbf{x}) + G^0_{\alpha''j'',i''\beta''}(\mathbf{x}'',\mathbf{x})]. \tag{37}$$

Substituting F=1 newton into Eq. (29) and considering the relations given by Eq. (35), we can simplify Eq. (29) as:

$$\begin{bmatrix} G_{\beta'a'}(\mathbf{x}'',\mathbf{x}') \\ \varepsilon_{\alpha''\beta'a'}(\mathbf{x}'',\mathbf{x}') \end{bmatrix} = \begin{bmatrix} G^0_{\beta'a'}(\mathbf{x}'',\mathbf{x}') \\ \varepsilon^0_{\alpha''\beta'a'}(\mathbf{x}'',\mathbf{x}') \end{bmatrix} + \int_V \begin{bmatrix} G^0_{\beta''i''}(\mathbf{x}'',\mathbf{x}) & \varepsilon^0_{\beta''i'j'}(\mathbf{x}'',\mathbf{x}) \\ \varepsilon^0_{\alpha''\beta''i''}(\mathbf{x}'',\mathbf{x}) & E^0_{\alpha''\beta''i'j'}(\mathbf{x}'',\mathbf{x}) \end{bmatrix} \begin{bmatrix} \delta\rho(\mathbf{x})\omega^2\delta_{ij} & 0 \\ 0 & \delta c_{ijkl}(\mathbf{x}) \end{bmatrix} \begin{bmatrix} G_{ja'}(\mathbf{x},\mathbf{x}') \\ \varepsilon_{kla'}(\mathbf{x},\mathbf{x}') \end{bmatrix} dV. \tag{38}$$

For the convenience of subsequent discussion, we rewrite Eq. (38) in a more compact form:

$$\boldsymbol{\Psi}(\mathbf{x}'' - \mathbf{x}') = \boldsymbol{\Psi}^0(\mathbf{x}'' - \mathbf{x}') + \iiint_{V(\mathbf{x})} \boldsymbol{\Gamma}(\mathbf{x}'' - \mathbf{x}) \boldsymbol{\Pi}(\mathbf{x}) \boldsymbol{\Psi}(\mathbf{x} - \mathbf{x}') d^3\mathbf{x}, \tag{39}$$

where

$$\boldsymbol{\Psi}(\mathbf{x}'' - \mathbf{x}') = \begin{bmatrix} G_{11} & G_{12} & G_{13} \\ G_{21} & G_{22} & G_{23} \\ G_{31} & G_{32} & G_{33} \\ G_{11,1} & G_{12,1} & G_{13,1} \\ G_{21,2} & G_{22,2} & G_{23,2} \\ G_{31,3} & G_{32,3} & G_{33,3} \\ G_{21,3}+G_{31,2} & G_{22,3}+G_{32,2} & G_{23,3}+G_{33,2} \\ G_{11,3}+G_{31,1} & G_{12,3}+G_{32,1} & G_{13,3}+G_{33,1} \\ G_{21,1}+G_{11,2} & G_{22,1}+G_{12,2} & G_{23,1}+G_{13,2} \end{bmatrix}, \quad \boldsymbol{\Psi}^0(\mathbf{x}'' - \mathbf{x}') = \begin{bmatrix} G^0_{11} & G^0_{12} & G^0_{13} \\ G^0_{21} & G^0_{22} & G^0_{23} \\ G^0_{31} & G^0_{32} & G^0_{33} \\ G^0_{11,1} & G^0_{12,1} & G^0_{13,1} \\ G^0_{21,2} & G^0_{22,2} & G^0_{23,2} \\ G^0_{31,3} & G^0_{32,3} & G^0_{33,3} \\ G^0_{21,3}+G^0_{31,2} & G^0_{22,3}+G^0_{32,2} & G^0_{23,3}+G^0_{33,2} \\ G^0_{11,3}+G^0_{31,1} & G^0_{12,3}+G^0_{32,1} & G^0_{13,3}+G^0_{33,1} \\ G^0_{21,1}+G^0_{11,2} & G^0_{22,1}+G^0_{12,2} & G^0_{23,1}+G^0_{13,2} \end{bmatrix}, \tag{40}$$

$$\boldsymbol{\Pi}(\mathbf{x}) = \begin{bmatrix} \delta\rho\omega^2 & 0 & 0 & 0 & 0 & 0 & 0 & 0 & 0 \\ 0 & \delta\rho\omega^2 & 0 & 0 & 0 & 0 & 0 & 0 & 0 \\ 0 & 0 & \delta\rho\omega^2 & 0 & 0 & 0 & 0 & 0 & 0 \\ 0 & 0 & 0 & \delta c_{11} & \delta c_{12} & \delta c_{13} & \delta c_{14} & \delta c_{15} & \delta c_{16} \\ 0 & 0 & 0 & \delta c_{12} & \delta c_{22} & \delta c_{23} & \delta c_{24} & \delta c_{25} & \delta c_{26} \\ 0 & 0 & 0 & \delta c_{13} & \delta c_{23} & \delta c_{33} & \delta c_{34} & \delta c_{35} & \delta c_{36} \\ 0 & 0 & 0 & \delta c_{14} & \delta c_{24} & \delta c_{34} & \delta c_{44} & \delta c_{45} & \delta c_{46} \\ 0 & 0 & 0 & \delta c_{15} & \delta c_{25} & \delta c_{35} & \delta c_{45} & \delta c_{55} & \delta c_{56} \\ 0 & 0 & 0 & \delta c_{16} & \delta c_{26} & \delta c_{36} & \delta c_{46} & \delta c_{56} & \delta c_{66} \end{bmatrix}, \tag{41}$$

$$\boldsymbol{\Gamma}(\mathbf{x}'' - \mathbf{x}) = \begin{bmatrix} \mathbf{A} & \mathbf{B} \\ \mathbf{B}^T & \mathbf{D} \end{bmatrix}, \tag{42}$$

and

$$\mathbf{A} = \begin{bmatrix} G^0_{11} & G^0_{12} & G^0_{13} \\ G^0_{21} & G^0_{22} & G^0_{23} \\ G^0_{31} & G^0_{32} & G^0_{33} \end{bmatrix}, \quad \mathbf{B} = \begin{bmatrix} G^0_{11,1} & G^0_{12,2} & G^0_{13,3} & G^0_{12,3}+G^0_{13,2} & G^0_{11,3}+G^0_{13,1} & G^0_{12,1}+G^0_{11,2} \\ G^0_{21,1} & G^0_{22,2} & G^0_{23,3} & G^0_{22,3}+G^0_{23,2} & G^0_{21,3}+G^0_{23,1} & G^0_{22,1}+G^0_{21,2} \\ G^0_{31,1} & G^0_{32,2} & G^0_{33,3} & G^0_{32,3}+G^0_{33,2} & G^0_{31,3}+G^0_{33,1} & G^0_{32,1}+G^0_{31,2} \end{bmatrix}, \tag{43}$$

$$\mathbf{D} = \begin{bmatrix} G^0_{11,11} & G^0_{12,21} & G^0_{13,31} & G^0_{12,31}+G^0_{13,21} & G^0_{11,31}+G^0_{13,11} & G^0_{12,11}+G^0_{11,21} \\ G^0_{21,12} & G^0_{22,22} & G^0_{23,32} & G^0_{22,32}+G^0_{23,22} & G^0_{21,32}+G^0_{23,12} & G^0_{22,12}+G^0_{21,22} \\ G^0_{31,13} & G^0_{32,23} & G^0_{33,33} & G^0_{32,33}+G^0_{33,23} & G^0_{31,33}+G^0_{33,13} & G^0_{32,13}+G^0_{31,23} \\ G^0_{21,13}+G^0_{31,12} & G^0_{22,23}+G^0_{32,22} & G^0_{23,33}+G^0_{33,32} & G^0_{22,33}+G^0_{33,22}+2G^0_{23,23} & G^0_{21,33}+G^0_{23,13}+G^0_{31,32}+G^0_{33,12} & G^0_{21,23}+G^0_{22,13}+G^0_{31,22}+G^0_{32,12} \\ G^0_{11,31}+G^0_{31,11} & G^0_{12,32}+G^0_{32,12} & G^0_{13,33}+G^0_{33,13} & G^0_{12,33}+G^0_{32,31}+G^0_{13,23}+G^0_{33,21} & G^0_{11,33}+G^0_{33,11}+2G^0_{13,13} & G^0_{12,13}+G^0_{11,23}+G^0_{32,11}+G^0_{31,21} \\ G^0_{11,12}+G^0_{21,11} & G^0_{22,21}+G^0_{12,22} & G^0_{23,31}+G^0_{13,32} & G^0_{12,32}+G^0_{22,31}+G^0_{13,22}+G^0_{23,21} & G^0_{21,31}+G^0_{11,32}+G^0_{23,11}+G^0_{13,12} & G^0_{11,22}+G^0_{22,11}+2G^0_{12,12} \end{bmatrix}.$$

(44)

One can easily verify that $\boldsymbol{\Gamma}$ is a symmetric matrix, i.e., $\boldsymbol{\Gamma}^T = \boldsymbol{\Gamma}$.

From Eq. (39) it is seen that the following integral is involved:

$$\iiint_{V(\mathbf{x})} \mathbf{\Gamma}(\mathbf{x}'' - \mathbf{x}) \mathbf{\Pi}(\mathbf{x}) \mathbf{\Psi}(\mathbf{x} - \mathbf{x}') d^3\mathbf{x} ,$$

where $\mathbf{\Gamma}(\mathbf{x}'' - \mathbf{x})$ is composed of homogeneous Green's function and its spatial derivatives up to second order. Meanwhile, we see that the domain of the integration is the whole space of the heterogeneous medium. A question naturally arises here: Since Green's functions and their derivatives are not well defined at $\mathbf{x}'' = \mathbf{x}$, how can we calculate these integrals? As one can see, Green's functions and their derivatives become infinite when $\mathbf{x}$ approaches $\mathbf{x}''$, this property is called the singularity of Green's functions. Obviously, Green's function and its derivatives of different orders tend to infinity with different speed as $\mathbf{x}$ approaches $\mathbf{x}''$, this means that they have different degrees of singularity. To properly define and calculate these integrals, we need to introduce the concept of shape-dependent principal value, which will be detailed in the next section.

### B. Singularity of Green's tensor and the renormalized integral equation

The singularity of electromagnetic (EM) Green's function was first discovered by Bladel [139]. He found that the field distribution in the source region cannot be calculated using the conventional dyadic Green's function, instead, a term proportional to Dirac-delta function must be subtracted in order to obtain the correct result. Finkel'berg [140] incorporated the singularity of EM Green's functions in the development of an effective medium theory for dielectric mixtures. Kong and Tsang [99-109] further calculated the singularities of EM Green's function of heterogeneous dielectrics with different spatial correlation functions, including the Gaussian, exponential and Von-Karman correlation functions, and derived effective dielectric constants for inhomogeneous media with spherical and ellipsoidal inclusions. They also derived the explicit expressions of backscattering cross section and applied them in satellite remote sensing. Lakhtakia and coworkers extended Kong and Tsang's theory to EM anisotropic media, and developed effective medium theory correspondingly [141-142]. Analogous to the development of EM effective medium theory, Zhunk analyzed the singularity of Green's function of the acoustic waves [139-140] and the elastodynamic Green's tensor [96] and developed an effective medium theory for heterogeneous acoustic and elastic materials. It has been pointed out that any multiple scattering model without considering the singularity of the Green tensor is only applicable to materials with small property fluctuation. In this work, the singularity of the elastodynamic Green tensor is introduced into the multiple scattering theory of elastic waves for the first time. Because of its vital importance to the success of the new model, we carry out comprehensive study on the singularity of the elastodynamic Green's tensor and propose three different but equivalent methods for the calculation of the singularity tensor. Although the final results are the same, each method provides a unique view of physical meaning of the singularity. In order to avoid interrupting the main content, the detailed calculation is presented in Appendix A. Here we use the final results directly.

From the calculation in Appendix A, we know the singularities of $G^0_{\beta''i'}(\mathbf{x}'', \mathbf{x})$ and $\varepsilon^0_{\beta''i'j'}(\mathbf{x}'', \mathbf{x})$ are relatively weak and do not cause problems in the calculation of the integrals, but $E^0_{\alpha'\beta''i'j'}(\mathbf{x}'', \mathbf{x})$ has the δ-singularity which needs to be carefully dealt with. By introducing the concept of shape-dependent principal value, the correct definition of the matrix $\mathbf{D}$ is given by:

$$\mathbf{D} := P.S.\mathbf{D} - \tilde{\mathbf{S}}\delta(\mathbf{x}'' - \mathbf{x}) , \tag{45}$$

where $P.S.\mathbf{D}$ is the shape-dependent principal value of the Green tensor, $\tilde{\mathbf{S}}$ is the singularity of the Green tensor, which is also dependent on the shape of the inclusions. For instance, the singularities of heterogeneous media with spherical, spheroidal, ellipsoidal and cylindrical inclusions all have distinct expressions and possess different symmetries. In the current work, we only consider random medium with spherical inclusions, for which the macroscopic properties are isotropic. Correspondingly, $P.S.\mathbf{D}$ and $\tilde{\mathbf{S}}$ have the following form:

$$P.S.\mathbf{D} = \begin{bmatrix} P.S.E^0_{1111} & P.S.E^0_{1221} & P.S.E^0_{1331} & 2E^0_{1231} & 2E^0_{1131} & 2E^0_{1211} \\ P.S.E^0_{2112} & P.S.E^0_{2222} & P.S.E^0_{2332} & 2E^0_{2232} & 2E^0_{2132} & 2E^0_{2212} \\ P.S.E^0_{3113} & P.S.E^0_{3223} & P.S.E^0_{3333} & 2E^0_{3233} & 2E^0_{3133} & 2E^0_{3213} \\ 2E^0_{2113} & 2E^0_{2223} & 2E^0_{2333} & P.S.4E^0_{2233} & 4E^0_{2133} & 4E^0_{2123} \\ 2E^0_{1131} & 2E^0_{1232} & 2E^0_{1333} & 4E^0_{1233} & P.S.4E^0_{1133} & 4E^0_{1213} \\ 2E^0_{1112} & 2E^0_{2221} & 2E^0_{2331} & 4E^0_{1232} & 4E^0_{2131} & P.S.4E^0_{1122} \end{bmatrix}, \tag{46}$$

$$\tilde{\mathbf{S}} = \begin{bmatrix} S_{1111} & S_{1221} & S_{1221} & 0 & 0 & 0 \\ S_{1221} & S_{1111} & S_{1221} & 0 & 0 & 0 \\ S_{1221} & S_{1221} & S_{1111} & 0 & 0 & 0 \\ 0 & 0 & 0 & 4S_{2233} & 0 & 0 \\ 0 & 0 & 0 & 0 & 4S_{2233} & 0 \\ 0 & 0 & 0 & 0 & 0 & 4S_{2233} \end{bmatrix}, \tag{47}$$

where

$$S_{1111} = \frac{2\lambda + 7\mu}{15\mu(\lambda + 2\mu)}, \quad S_{1221} = -\frac{\lambda + \mu}{15\mu(\lambda + 2\mu)}, \quad S_{2233} = \frac{3\lambda + 8\mu}{30\mu(\lambda + 2\mu)}. \tag{48}$$

From Eq. (48) we can see that $S_{1111} = S_{1221} + 2S_{2233}$, thus $S_{ijkl}$ is an isotropic tensor. This conclusion is consistent with the assumption that the heterogeneous medium is macroscopically isotropic.

In analogous to the isotropic elastic stiffness tensor, we can rewrite $S_{ijkl}$ as

$$S_{\alpha ij\beta} = S_1 \delta_{\alpha i} \delta_{j\beta} + S_2 (\delta_{\alpha j} \delta_{i\beta} + \delta_{\alpha\beta} \delta_{ij}), \tag{49}$$

where

$$S_1 = -\frac{\lambda + \mu}{15(\lambda + 2\mu)\mu}, \quad S_2 = \frac{3\lambda + 8\mu}{30(\lambda + 2\mu)\mu}. \tag{50}$$

Similarly, we can give the correct definition of the matrix $\mathbf{\Gamma}(\mathbf{x}'' - \mathbf{x})$ by introducing its shape-dependent principal value $P.S.\mathbf{\Gamma}(\mathbf{x}'' - \mathbf{x})$, i.e.,

$$\mathbf{\Gamma}(\mathbf{x}'' - \mathbf{x}) := P.S.\mathbf{\Gamma}(\mathbf{x}'' - \mathbf{x}) - \mathbf{S}\delta(\mathbf{x}'' - \mathbf{x}), \tag{51}$$

where

$$P.S.\mathbf{\Gamma}(\mathbf{x}'' - \mathbf{x}) = \begin{bmatrix} \mathbf{A} & \mathbf{B} \\ \mathbf{B}^T & P.S.\mathbf{D} \end{bmatrix}, \quad \mathbf{S} = \begin{bmatrix} \mathbf{0} & \mathbf{0} \\ \mathbf{0} & \tilde{\mathbf{S}} \end{bmatrix}. \tag{52}$$

Substitution of (51) into (39) yields:

$$\mathbf{\Psi}(\mathbf{x}'' - \mathbf{x}') = \mathbf{\Psi}^0(\mathbf{x}'' - \mathbf{x}') + \iiint_{V(\mathbf{x})} [P.S.\mathbf{\Gamma}(\mathbf{x}'' - \mathbf{x}) - \mathbf{S}\delta(\mathbf{x}'' - \mathbf{x})]\mathbf{\Pi}(\mathbf{x})\mathbf{\Psi}(\mathbf{x} - \mathbf{x}')d^3\mathbf{x}. \tag{53}$$

Invoking the definition of the Dirac-$\delta$ function, we get

$$\mathbf{\Psi}(\mathbf{x}'' - \mathbf{x}') = \mathbf{\Psi}^0(\mathbf{x}'' - \mathbf{x}') + \iiint_{V(\mathbf{x})} P.S.\mathbf{\Gamma}(\mathbf{x}'' - \mathbf{x})\mathbf{\Pi}(\mathbf{x})\mathbf{\Psi}(\mathbf{x} - \mathbf{x}')d^3\mathbf{x} - \mathbf{S}\mathbf{\Pi}(\mathbf{x}'')\mathbf{\Psi}(\mathbf{x}'' - \mathbf{x}'). \tag{54}$$

Introducing a new quantity, called the renormalized field variable,

$$\mathbf{\Phi}(\mathbf{x}'' - \mathbf{x}') = \mathbf{\Psi}(\mathbf{x}'' - \mathbf{x}') + \mathbf{S}\mathbf{\Pi}(\mathbf{x}'')\mathbf{\Psi}(\mathbf{x}'' - \mathbf{x}'), \tag{55}$$

the integral equation (54) is rewritten as:

$$\mathbf{\Phi}(\mathbf{x}'' - \mathbf{x}') = \mathbf{\Psi}^0(\mathbf{x}'' - \mathbf{x}') + \iiint_{V(\mathbf{x})} P.S.\mathbf{\Gamma}(\mathbf{x}'' - \mathbf{x})\mathbf{\Xi}(\mathbf{x})\mathbf{\Phi}(\mathbf{x} - \mathbf{x}')d^3\mathbf{x}, \tag{56}$$

where

$$\mathbf{\Xi}(\mathbf{x}) = \mathbf{\Pi}(\mathbf{x})[\mathbf{I} + \mathbf{S}\mathbf{\Pi}(\mathbf{x})]^{-1}. \tag{57}$$

In this work, we consider heterogeneous media whose component phases are all isotropic materials, for which the matrix $\mathbf{\Xi}(\mathbf{x})$ and its elements are given by:

$$\mathbf{\Xi}(\mathbf{x}) = \begin{bmatrix} \delta\rho\omega^2 & 0 & 0 & 0 & 0 & 0 & 0 & 0 & 0 \\ 0 & \delta\rho\omega^2 & 0 & 0 & 0 & 0 & 0 & 0 & 0 \\ 0 & 0 & \delta\rho\omega^2 & 0 & 0 & 0 & 0 & 0 & 0 \\ 0 & 0 & 0 & \Xi_{11} & \Xi_{12} & \Xi_{12} & 0 & 0 & 0 \\ 0 & 0 & 0 & \Xi_{12} & \Xi_{11} & \Xi_{12} & 0 & 0 & 0 \\ 0 & 0 & 0 & \Xi_{12} & \Xi_{12} & \Xi_{11} & 0 & 0 & 0 \\ 0 & 0 & 0 & 0 & 0 & 0 & \Xi_{44} & 0 & 0 \\ 0 & 0 & 0 & 0 & 0 & 0 & 0 & \Xi_{44} & 0 \\ 0 & 0 & 0 & 0 & 0 & 0 & 0 & 0 & \Xi_{44} \end{bmatrix}, \tag{58}$$

$$\Xi_{11} = \frac{\delta c_{12} + 2\delta c_{44} + 4(S_1 + S_2)\delta c_{44}(3\delta c_{12} + 2\delta c_{44})}{(1 + 4S_2\delta c_{44})[1 + (3S_1 + 2S_2)(3\delta c_{12} + 2\delta c_{44})]},$$

$$\Xi_{12} = \frac{\delta c_{12}(1 + 4S_2\delta c_{44}) - 2\delta c_{44}[(3S_1 + 2S_2)\delta c_{12} + 2S_1\delta c_{44}]}{(1 + 4S_2\delta c_{44})[1 + (3S_1 + 2S_2)(3\delta c_{12} + 2\delta c_{44})]}, \quad (59)$$

$$\Xi_{44} = \frac{\delta c_{44}}{1 + 4S_2\delta c_{44}}.$$

One can easily verify the relation: $\Xi_{11} = \Xi_{12} + 2\Xi_{44}$. Eq. (56) is known as the renormalized integral equation [141-142]. In contrast to the development of effective medium models, in the current work we strive to develop a multiple scattering theory, which should be able to describe the coherent wave propagation in the whole frequency range. To do so, we first perform multiple scattering series expansion of Eq. (56), and then derive the Dyson equation with the help of Feynman's diagram and the first-order smoothing approximation.

## C. Feynman's diagram, Dyson's equation and the First-Order-Smoothing-Approximation

Eq. (56) is an integral equation for the unknown quantity $\mathbf{\Phi}(\mathbf{x}'' - \mathbf{x}')$. We can see that this quantity appears on both sides of the equation and this naturally forms an iteration scheme. By performing iteration successively, i.e., substituting the righthand side of Eq. (56) into the renormalized quantity in the integrand, we get an infinite series Eq. (60). In Eq. (60), the renormalized field is expressed using the unperturbed field variables only. In this work, Eq. (60) is called the multiple scattering series expansion of the perturbed field.

$$\begin{aligned}
\mathbf{\Phi}(\mathbf{x}'' - \mathbf{x}') &= \mathbf{\Psi}^0(\mathbf{x}'' - \mathbf{x}') + \iiint_{V(\mathbf{x})} P.S.\mathbf{\Gamma}(\mathbf{x}'' - \mathbf{x})\mathbf{\Xi}(\mathbf{x})\mathbf{\Psi}^0(\mathbf{x} - \mathbf{x}')d^3\mathbf{x} + \\
&\quad \iiint_{V(\mathbf{x}_1)}\iiint_{V(\mathbf{x}_2)} P.S.\mathbf{\Gamma}(\mathbf{x}'' - \mathbf{x}_1)\mathbf{\Xi}(\mathbf{x}_1)P.S.\mathbf{\Gamma}(\mathbf{x}_1 - \mathbf{x}_2)\mathbf{\Xi}(\mathbf{x}_2)\mathbf{\Psi}^0(\mathbf{x}_2 - \mathbf{x}')d^3\mathbf{x}_2 d^3\mathbf{x}_1 + \\
&\quad \iiint_{V(\mathbf{x}_1)}\iiint_{V(\mathbf{x}_2)}\iiint_{V(\mathbf{x}_3)} P.S.\mathbf{\Gamma}(\mathbf{x}'' - \mathbf{x}_1)\mathbf{\Xi}(\mathbf{x}_1)P.S.\mathbf{\Gamma}(\mathbf{x}_1 - \mathbf{x}_2)\mathbf{\Xi}(\mathbf{x}_2)P.S.\mathbf{\Gamma}(\mathbf{x}_2 - \mathbf{x}_3)\mathbf{\Xi}(\mathbf{x}_3)\mathbf{\Psi}^0(\mathbf{x}_3 - \mathbf{x}')d^3\mathbf{x}_3 d^3\mathbf{x}_2 d^3\mathbf{x}_1 + \cdots \\
&\quad \iiint_{V(\mathbf{x}_1)}\iiint_{V(\mathbf{x}_2)}\cdots\iiint_{V(\mathbf{x}_n)} P.S.\mathbf{\Gamma}(\mathbf{x}'' - \mathbf{x}_1)\mathbf{\Xi}(\mathbf{x}_1)P.S.\mathbf{\Gamma}(\mathbf{x}_1 - \mathbf{x}_2)\mathbf{\Xi}(\mathbf{x}_2)\cdots P.S.\mathbf{\Gamma}(\mathbf{x}_{n-1} - \mathbf{x}_n)\mathbf{\Xi}(\mathbf{x}_n)\mathbf{\Psi}^0(\mathbf{x}_n - \mathbf{x}')d^3\mathbf{x}_n d^3\mathbf{x}_{n-1}\cdots d^3\mathbf{x}_2 d^3\mathbf{x}_1 + \cdots
\end{aligned} \quad (60)$$

The coherent field, also called the ensemble-averaged field, is defined by averaging the perturbed field over the whole space of realizations of the random medium. Taking ensemble average of Eq. (60) gives:

$$\begin{aligned}
\langle\mathbf{\Phi}(\mathbf{x}'' - \mathbf{x}')\rangle &= \mathbf{\Psi}^0(\mathbf{x}'' - \mathbf{x}') + \iiint_{V(\mathbf{x})} \langle P.S.\mathbf{\Gamma}(\mathbf{x}'' - \mathbf{x})\mathbf{\Xi}(\mathbf{x})\mathbf{\Psi}^0(\mathbf{x} - \mathbf{x}')\rangle d^3\mathbf{x} + \\
&\quad \iiint_{V(\mathbf{x}_1)}\iiint_{V(\mathbf{x}_2)} \langle P.S.\mathbf{\Gamma}(\mathbf{x}'' - \mathbf{x}_1)\mathbf{\Xi}(\mathbf{x}_1)P.S.\mathbf{\Gamma}(\mathbf{x}_1 - \mathbf{x}_2)\mathbf{\Xi}(\mathbf{x}_2)\mathbf{\Psi}^0(\mathbf{x}_2 - \mathbf{x}')\rangle d^3\mathbf{x}_2 d^3\mathbf{x}_1 + \\
&\quad \iiint_{V(\mathbf{x}_1)}\iiint_{V(\mathbf{x}_2)}\iiint_{V(\mathbf{x}_3)} \langle P.S.\mathbf{\Gamma}(\mathbf{x}'' - \mathbf{x}_1)\mathbf{\Xi}(\mathbf{x}_1)P.S.\mathbf{\Gamma}(\mathbf{x}_1 - \mathbf{x}_2)\mathbf{\Xi}(\mathbf{x}_2)P.S.\mathbf{\Gamma}(\mathbf{x}_2 - \mathbf{x}_3)\mathbf{\Xi}(\mathbf{x}_3)\mathbf{\Psi}^0(\mathbf{x}_3 - \mathbf{x}')\rangle d^3\mathbf{x}_3 d^3\mathbf{x}_2 d^3\mathbf{x}_1 + \cdots \\
&\quad \iiint_{V(\mathbf{x}_1)}\iiint_{V(\mathbf{x}_2)}\cdots\iiint_{V(\mathbf{x}_n)} \langle P.S.\mathbf{\Gamma}(\mathbf{x}'' - \mathbf{x}_1)\mathbf{\Xi}(\mathbf{x}_1)P.S.\mathbf{\Gamma}(\mathbf{x}_1 - \mathbf{x}_2)\mathbf{\Xi}(\mathbf{x}_2)\cdots P.S.\mathbf{\Gamma}(\mathbf{x}_{n-1} - \mathbf{x}_n)\mathbf{\Xi}(\mathbf{x}_n)\mathbf{\Psi}^0(\mathbf{x}_n - \mathbf{x}')\rangle d^3\mathbf{x}_n d^3\mathbf{x}_{n-1}\cdots d^3\mathbf{x}_2 d^3\mathbf{x}_1 + \cdots
\end{aligned} \quad (61)$$

where the angular bracket means the ensemble average of the enclosed quantity. According to the original definition, we need to calculate the random field variables in the whole configuration space, which includes an infinite number of realizations of the random medium. According to the ergodic hypothesis, the ensemble average of a quantity is equal to the volumetric average of the perturbed quantity, thus the angular bracket can also be understood as taking volumetric average of the enclosed quantity. Since $P.S.\mathbf{\Gamma}(\mathbf{x}'' - \mathbf{y})$ is a deterministic quantity, we can take it out from the bracket and get:

$$\begin{aligned}
\langle\Phi_{ai}(\mathbf{x}'' - \mathbf{x}')\rangle &= \Psi_{ai}^0(\mathbf{x}'' - \mathbf{x}') + \iiint_{V(\mathbf{x})} P.S.\Gamma_{ab}(\mathbf{x}'' - \mathbf{x})\langle\Xi_{bc}(\mathbf{x})\rangle\Psi_{ci}^0(\mathbf{x} - \mathbf{x}')d^3\mathbf{x} + \\
&\quad \iiint_{V(\mathbf{x}_1)}\iiint_{V(\mathbf{x}_2)} P.S.\Gamma_{ab}(\mathbf{x}'' - \mathbf{x}_1)P.S.\Gamma_{cd}(\mathbf{x}_1 - \mathbf{x}_2)\langle\Xi_{bc}(\mathbf{x}_1)\Xi_{de}(\mathbf{x}_2)\rangle\Psi_{ei}^0(\mathbf{x}_2 - \mathbf{x}')d^3\mathbf{x}_2 d^3\mathbf{x}_1 + \\
&\quad \iiint_{V(\mathbf{x}_1)}\iiint_{V(\mathbf{x}_2)}\iiint_{V(\mathbf{x}_3)} P.S.\Gamma_{ab}(\mathbf{x}'' - \mathbf{x}_1)P.S.\Gamma_{cd}(\mathbf{x}_1 - \mathbf{x}_2)P.S.\Gamma_{ef}(\mathbf{x}_2 - \mathbf{x}_3)\langle\Xi_{bc}(\mathbf{x}_1)\Xi_{de}(\mathbf{x}_2)\Xi_{fg}(\mathbf{x}_3)\rangle\Psi_{gi}^0(\mathbf{x}_3 - \mathbf{x}')d^3\mathbf{x}_3 d^3\mathbf{x}_2 d^3\mathbf{x}_1 + \cdots \\
&\quad \iiint_{V(\mathbf{x}_1)}\iiint_{V(\mathbf{x}_2)}\cdots\iiint_{V(\mathbf{x}_n)} P.S.\Gamma_{ab}(\mathbf{x}'' - \mathbf{x}_1)P.S.\Gamma_{cd}(\mathbf{x}_1 - \mathbf{x}_2)\cdots P.S.\Gamma_{fg}(\mathbf{x}_{n-1} - \mathbf{x}_n)\cdot \\
&\quad \langle\Xi_{bc}(\mathbf{x}_1)\Xi_{de}(\mathbf{x}_2)\cdots\Xi_{gh}(\mathbf{x}_n)\rangle\Psi_{hi}^0(\mathbf{x}_n - \mathbf{x}')d^3\mathbf{x}_n d^3\mathbf{x}_{n-1}\cdots d^3\mathbf{x}_2 d^3\mathbf{x}_1 + \cdots
\end{aligned} \quad (62)$$

One need to be informed that in Eq. (62), the subscripts expressed using Latin letters are matrix indices instead of tensor indices. In Eq. (62), an infinite number of multi-point correlation functions are involved:

$$\langle\Xi_{ab}(\mathbf{x})\rangle, \quad \langle\Xi_{ab}(\mathbf{x}_1)\Xi_{cd}(\mathbf{x}_2)\rangle, \quad \langle\Xi_{ab}(\mathbf{x}_1)\Xi_{cd}(\mathbf{x}_2)\Xi_{ef}(\mathbf{x}_3)\rangle, \quad \langle\Xi_{ab}(\mathbf{x}_1)\Xi_{cd}(\mathbf{x}_2)\Xi_{ef}(\mathbf{x}_3)\Xi_{gh}(\mathbf{x}_4)\rangle, \quad \cdots \quad (63)$$

Multi-point correlation functions, also called higher-order moments, provide key information about the statistical characteristic of heterogeneous media, and they have found broad applications in different subjects, such as characterization of galaxy distribution in cosmology [145], describing random EM media for satellite remote sensing [101, 109, 122], analyzing small-scale heterogeneities in the Earth's lithosphere [3, 29, 146], and characterizing microstructures in metal polycrystals [46-48]. For a detailed discussion on how to calculate higher-order moments readers are referred to [47, 107, 147-148]. No matter what type of statistics the inhomogeneous medium complies with, one can always choose a reference homogeneous medium such that $\langle \Xi_{ab}(\mathbf{x}) \rangle = 0$, so we only need to consider the second and higher order moments. Generally speaking, multi-point moments have no simple relations with lower order moments. However, if the random medium following the Gaussian statistics, i.e., the two-point correlation function can be expressed as a Gaussian function, all the moments of odd order vanish [123, 127]:

$$\langle \Xi_{ab}(\mathbf{x}_1)\Xi_{cd}(\mathbf{x}_2)\Xi_{ef}(\mathbf{x}_3) \rangle = 0, \quad \cdots, \quad \langle \Xi_{ab}(\mathbf{x}_1)\Xi_{cd}(\mathbf{x}_2)\Xi_{ef}(\mathbf{x}_3)\cdots\Xi_{gh}(\mathbf{x}_{2n-1}) \rangle = 0, \tag{64}$$

and all the moments of even order can be expressed as products of the second order moments $\langle \Xi_{ab}(\mathbf{x}_1)\Xi_{cd}(\mathbf{x}_2) \rangle$, i.e.:

$$\langle \Xi_{ab}(\mathbf{x}_1)\Xi_{cd}(\mathbf{x}_2)\Xi_{ef}(\mathbf{x}_3)\Xi_{gh}(\mathbf{x}_4) \rangle$$
$$= \langle \Xi_{ab}(\mathbf{x}_1)\Xi_{cd}(\mathbf{x}_2) \rangle\langle \Xi_{ef}(\mathbf{x}_3)\Xi_{gh}(\mathbf{x}_4) \rangle + \langle \Xi_{ab}(\mathbf{x}_1)\Xi_{ef}(\mathbf{x}_3) \rangle\langle \Xi_{cd}(\mathbf{x}_2)\Xi_{gh}(\mathbf{x}_4) \rangle + \langle \Xi_{ab}(\mathbf{x}_1)\Xi_{gh}(\mathbf{x}_4) \rangle\langle \Xi_{cd}(\mathbf{x}_2)\Xi_{ef}(\mathbf{x}_3) \rangle, \tag{65}$$

$$\langle \Xi_{ab}(\mathbf{x}_1)\Xi_{cd}(\mathbf{x}_2)\cdots\Xi_{ef}(\mathbf{x}_{2n-1})\Xi_{gh}(\mathbf{x}_{2n}) \rangle = \langle \Xi_{ab}(\mathbf{x}_1)\Xi_{cd}(\mathbf{x}_2) \rangle \cdots \langle \Xi_{ef}(\mathbf{x}_{2n-1})\Xi_{gh}(\mathbf{x}_{2n}) \rangle + all\ permutations. \tag{66}$$

To proceed further, it is convenient to introduce Feynman's diagram technique, which was first developed by Richard Feynman to investigate interactions of elementary particles with quantum many body systems. Rytov adopted this method to study multiple scattering of classic scalar waves [123]. In this work, we extend Rytov's implementation of Feynman's diagram technique to investigate the multiple scattering of elastic waves. The symbolic representations of all the involved quantities are shown in Tab. 2.

Table 2. Symbolic representation of field variables used in Feynman's diagram.

| Reference field: $\Psi^0(\mathbf{x}''-\mathbf{x}')$ | Propagator: $P.S.\Gamma(\mathbf{x}''-\mathbf{x}')$ | Scattering point: $\Pi(\mathbf{x})$ | Ensemble averaged field: $\langle\Phi(\mathbf{x}''-\mathbf{x}')\rangle$ |
|---|---|---|---|
| 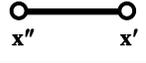 | 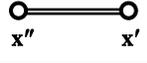 | 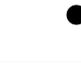 | 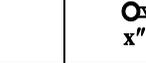 |

In addition to Tab. 2, we also adopt the convention that two scattering points lie in the same inclusion, i.e., located in the same bracket, are linked by a dashed line.

Considering the relations given by Eqs. (64)-(66), the diagrammatic representation of Eq. (62) is:

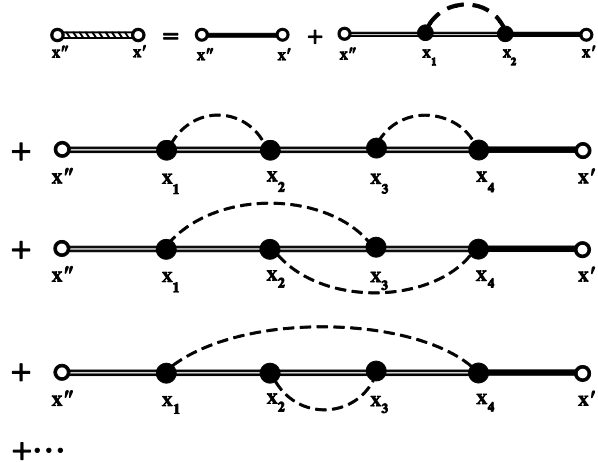

$+\cdots$

FIG. 3. Feynman's diagram representation of the multiple scattering series.

To proceed further, we have to introduce the concepts of weakly- and strongly-connected diagrams. If a diagram can be divided into two subdiagrams, each of which has two or more scattering points, without breaking any dashed lines, then the diagram is called a weakly-connected diagram. Diagrams do not possess this property are called strongly-connected diagrams. For instance, the third term on the righthand side of the equation is a weakly-connected diagram, the second, fourth and fifth terms are strongly-connected diagrams. For simplicity, we introduce an additional symbol to represent the sum of all the strongly-connected diagrams:

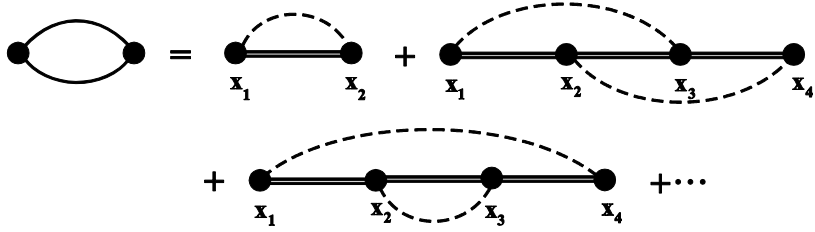

FIG. 4. Summation of strongly connected diagrams.

It can be verified that all the weakly connected diagrams can be expressed as the product of two or more strongly connected diagrams. Now the ensemble averaged Green's function can be expressed solely by strongly connected diagrams:

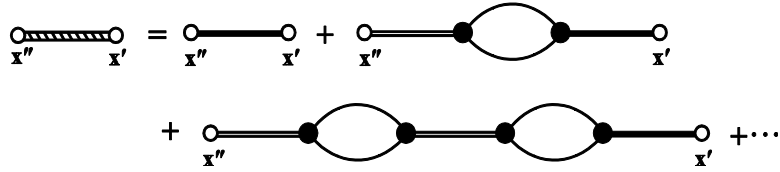

FIG. 5. Representation of the ensemble averaged Green's function using strongly connected diagrams.

It is easy to verify that this infinite series is the solution to the following equation:

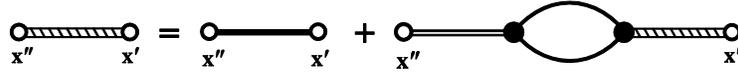

FIG. 6. Diagrammatic representation of Dyson's equation.

Equation shown in Fig. 6 is the famous Dyson's equation, which has been studied extensively in numerous fields wherever multiple scattering is considered, such as scattering of electromagnetic waves [101, 122, 141-142], scattering of photons and other subatomic particles [149-152]. So far, no approximation is introduced and consequently, the Dyson equation is accurate in the sense that multiple scattering events of all orders are included. Nevertheless, this equation is extremely difficult to solve because an infinite number of multipoint correlation functions are involved and their calculation is exceedingly tedious. If only the first term of the strongly connected diagram is retained, we obtain a simplified Dyson's equation:

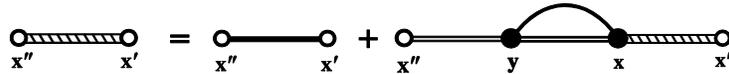

FIG. 7. Diagram representation of Dyson's equation under FOSA.

This approximation is called the first-order smoothing approximation [123, 127], also known as the Bourret approximation [153] and bi-local approximation [122]. Collin [154] compared different selective summation techniques and concluded that the FOSA has the fastest convergence rate. We can write down the explicit equation corresponding to the diagram shown in Fig. 7:

$$\langle \Phi(\mathbf{x}''-\mathbf{x}') \rangle = \Psi^0(\mathbf{x}''-\mathbf{x}') + \langle \iiint_{V(\mathbf{y})} P.S.\Gamma(\mathbf{x}''-\mathbf{y})\Xi(\mathbf{y}) \iiint_{V(\mathbf{x})} P.S.\Gamma(\mathbf{y}-\mathbf{x})\Xi(\mathbf{x}) \rangle \langle \Phi(\mathbf{x}-\mathbf{x}') \rangle d^3\mathbf{x}. \tag{67}$$

Eq. (67) is the so-called Dyson's equation under FOSA, in subsequent discussion, we call it Dyson's equation for simplicity. A primary study on the accuracy of this approximation was carried out in [156] in the context of the homogenized continuum model. It is found the difference between the third-order and second-order (corresponding to the FOSA here) estimates of the homogenized constitutive parameters are very small for all values of characteristic size, even when the constitutive contrast between the component materials is as large as 30. Similarly, it shows that higher order (fourth-order, fifth-order …) approximation does not add significantly to the estimation given by the second-order approximation. A rigorous analysis of the accuracy and error estimation in the multiple scattering scenario is out of the scope of this work. It is noted that different from the homogenization theories, no constraint on the frequency is introduced here, so the conclusions drawn above are assumed to hold in the whole frequency range. This conjecture will be examined and verified by numerical results for a variety of materials, see Sections III and IV. Another point we need to stress is that Eq. (67) is valid for random media with any types of two-point correlation functions, such as the exponential and the Von-Karman correlation functions. This is because all the third- and higher-order moments are completely neglected.

Considering $P.S.\Gamma_{ab}(\mathbf{x}''-\mathbf{y})$ and $P.S.\Gamma_{cd}(\mathbf{y}-\mathbf{x})$ are deterministic quantities, Eq. (67) can be simplified as:

$$\langle\Phi_{af}(\mathbf{x}''-\mathbf{x}')\rangle = \Psi_{af}^0(\mathbf{x}''-\mathbf{x}') + \iiint_{V(\mathbf{y})} P.S.\Gamma_{ab}(\mathbf{x}''-\mathbf{y})\iiint_{V(\mathbf{x})} P.S.\Gamma_{cd}(\mathbf{y}-\mathbf{x})\langle\Xi_{bc}(\mathbf{y})\Xi_{de}(\mathbf{x})\rangle\langle\Phi_{ef}(\mathbf{x}-\mathbf{x}')\rangle d^3\mathbf{x}. \tag{68}$$

For a statically homogeneous medium, the two-point correlation function is given by [47-48, 155]:

$$\langle\Xi_{ab}(\mathbf{y})\Xi_{cd}(\mathbf{x})\rangle = \langle\Xi_{ab}\Xi_{cd}\rangle P(\mathbf{y}-\mathbf{x}), \tag{69}$$

where $P(\mathbf{y}-\mathbf{x})$ is called the spatial autocorrelation function (SAF). Heterogeneous materials with different microstructural distributions are described by different SAFs. For example, random media with blurred interfaces are most conveniently described by Gaussian-type SAF, inclusions or scatterers with sharp boundaries follow exponential-type SAF, and Von Karman-type are introduced to describe self-similar random media like turbulent media [3]. In this work, we only consider random media with exponential correlation functions since it describes the correlation features of most random media discussed in this work to an accurate or acceptable level. If the inclusions in the medium are of spherical geometry, then the random media is statistically isotropic, and the exponential correlation function is a function of the distance between the two points [3, 47-48]:

$$P(\mathbf{y}-\mathbf{x}) = e^{-\frac{|\mathbf{y}-\mathbf{x}|}{a}}, \tag{70}$$

where $a$ is the correlation length.

The Fourier transform of SAF represents the power spectrum of the medium fluctuation, which is given by:

$$\tilde{P}(\mathbf{k}) = \frac{8\pi a^3}{(1+k^2 a^2)^2}. \tag{71}$$

As pointed out before, we can always choose the properties of the reference medium such that the first-order moment vanishes. This choice also ensures the fastest convergence rate of the multiple-scattering series, thus we take:

$$\langle\Xi_{ab}(\mathbf{x})\rangle = 0. \tag{72}$$

For a two-phase heterogeneous medium, Eq. (72) gives three independent relations [12]:

$\langle\delta\rho\rangle = 0$:
$$\rho = f_1\rho_1 + f_2\rho_2, \tag{73}$$

$\langle\Xi_{12}\rangle = 0$:
$$\frac{\delta c_{12}^{(1)}(1+4S_2\delta c_{44}^{(1)}) - 2\delta c_{44}^{(1)}[(3S_1+2S_2)\delta c_{12}^{(1)}+2S_1\delta c_{44}^{(1)}]}{(1+4S_2\delta c_{44}^{(1)})[1+(3S_1+2S_2)(3\delta c_{12}^{(1)}+2\delta c_{44}^{(1)})]}f_1 + \frac{\delta c_{12}^{(2)}(1+4S_2\delta c_{44}^{(2)}) - 2\delta c_{44}^{(2)}[(3S_1+2S_2)\delta c_{12}^{(2)}+2S_1\delta c_{44}^{(2)}]}{(1+4S_2\delta c_{44}^{(2)})[1+(3S_1+2S_2)(3\delta c_{12}^{(2)}+2\delta c_{44}^{(2)})]}f_2 = 0, \tag{74}$$

$\langle\Xi_{44}\rangle = 0$:
$$\frac{\delta c_{44}^{(1)}}{1+4S_2\delta c_{44}^{(1)}}f_1 + \frac{\delta c_{44}^{(2)}}{1+4S_2\delta c_{44}^{(2)}}f_2 = 0, \tag{75}$$

where

$$\delta c_{12}^{(1)} = \lambda_1 - \lambda, \quad \delta c_{44}^{(1)} = \mu_1 - \mu, \quad \delta c_{12}^{(2)} = \lambda_2 - \lambda, \quad \delta c_{44}^{(2)} = \mu_2 - \mu. \tag{76}$$

As can be seen, the mass density of the reference medium is still defined by the volumetric average of the component materials, but the elastic stiffness is obtained by solving a system of complicated nonlinear algebraic equations (74) and (75), which is largely different from the Voigt-averaged values. In Sections III and IV we will see that the Voigt-average estimation always overestimates the quasi-static limit of the velocities of the coherent waves. This conclusion is in agreement with that given by other works [49-50].

### D. Solution of Dyson's equation and the dispersion equations

Equation (68) is a system of integral equations of convolution type, which is most conveniently solved by the Fourier transformation technique. In this work, we use the following spatial Fourier transform pair:

$$\tilde{\Psi}(\mathbf{k}) = \iiint_{V(\mathbf{x})}\Psi(\mathbf{x})e^{-i\mathbf{k}\cdot\mathbf{x}}d^3\mathbf{x}, \quad \Psi(\mathbf{x}) = \frac{1}{8\pi^3}\iiint_{V(\mathbf{k})}\tilde{\Psi}(\mathbf{k})e^{i\mathbf{k}\cdot\mathbf{x}}d^3\mathbf{k}. \tag{77}$$

Applying Fourier transform to the Dyson equation (68), and considering (69), we get:

$$\langle\tilde{\Phi}_{af}(\mathbf{k})\rangle = \tilde{\Psi}_{af}^0(\mathbf{k}) + P.S.\tilde{\Gamma}_{ab}(\mathbf{k})\langle\Xi_{bc}\Xi_{de}\rangle\left[\frac{1}{8\pi^3}\iiint_{V(\mathbf{s})}P.S.\tilde{\Gamma}_{cd}(\mathbf{s})\tilde{P}(\mathbf{k}-\mathbf{s})d^3\mathbf{s}\right]\langle\tilde{\Phi}_{ef}(\mathbf{k})\rangle. \tag{78}$$

Multiplying both sides of Eq. (78) by $\left[P.S.\tilde{\Gamma}_{ga}(\mathbf{k})\right]^{-1}$, we get:

$$\left[P.S.\tilde{\Gamma}_{ga}(\mathbf{k})\right]^{-1}\langle\tilde{\Phi}_{af}(\mathbf{k})\rangle=\left[P.S.\tilde{\Gamma}_{ga}(\mathbf{k})\right]^{-1}\tilde{\Psi}_{af}^{0}(\mathbf{k})+\langle\Xi_{gc}\Xi_{de}\rangle\left[\frac{1}{8\pi^{3}}\iiint_{V(\mathbf{s})}P.S.\tilde{\Gamma}_{cd}(\mathbf{s})\tilde{P}(\mathbf{k}-\mathbf{s})d^{3}\mathbf{s}\right]\langle\tilde{\Phi}_{ef}(\mathbf{k})\rangle. \quad (79)$$

Rearranging this equation, we have:

$$\left\{\left[P.S.\tilde{\Gamma}_{ge}(\mathbf{k})\right]^{-1}-\langle\Xi_{gc}\Xi_{de}\rangle\left[\frac{1}{8\pi^{3}}\iiint_{V(\mathbf{s})}P.S.\tilde{\Gamma}_{cd}(\mathbf{s})\tilde{P}(\mathbf{k}-\mathbf{s})d^{3}\mathbf{s}\right]\right\}\langle\tilde{\Phi}_{ef}(\mathbf{k})\rangle=\left[P.S.\tilde{\Gamma}_{ga}(\mathbf{k})\right]^{-1}\tilde{\Psi}_{af}^{0}(\mathbf{k}). \quad (80)$$

The ensemble averaged response in the frequency-wavenumber domain can be solved as:

$$\langle\tilde{\Phi}_{hf}(\mathbf{k})\rangle=\left\{\left[P.S.\tilde{\Gamma}_{hg}(\mathbf{k})\right]^{-1}-\langle\Xi_{hc}\Xi_{dg}\rangle\left[\frac{1}{8\pi^{3}}\iiint_{V(\mathbf{s})}P.S.\tilde{\Gamma}_{cd}(\mathbf{s})\tilde{P}(\mathbf{k}-\mathbf{s})d^{3}\mathbf{s}\right]\right\}^{-1}\left[P.S.\tilde{\Gamma}_{ga}(\mathbf{k})\right]^{-1}\tilde{\Psi}_{af}^{0}(\mathbf{k}). \quad (81)$$

Simultaneously we obtain the dispersion equation:

$$\det\left\{\left[P.S.\tilde{\Gamma}_{ge}(\mathbf{k})\right]^{-1}-\langle\Xi_{gc}\Xi_{de}\rangle\left[\frac{1}{8\pi^{3}}\iiint_{V(\mathbf{s})}P.S.\tilde{\Gamma}_{cd}(\mathbf{s})\tilde{P}(\mathbf{k}-\mathbf{s})d^{3}\mathbf{s}\right]\right\}=0. \quad (82)$$

The solution in the spatial-frequency domain is given by:

$$\langle\Phi_{hf}(\mathbf{x},\mathbf{x}',\omega)\rangle=\frac{1}{8\pi^{3}}\iiint_{V(\mathbf{k})}\left\{\left[P.S.\tilde{\Gamma}_{hg}(\mathbf{k})\right]^{-1}-\langle\Xi_{hc}\Xi_{dg}\rangle\left[\frac{1}{8\pi^{3}}\iiint_{V(\mathbf{s})}P.S.\tilde{\Gamma}_{cd}(\mathbf{s})\tilde{P}(\mathbf{k}-\mathbf{s})d^{3}\mathbf{s}\right]\right\}^{-1}\left[P.S.\tilde{\Gamma}_{ga}(\mathbf{k})\right]^{-1}\tilde{\Psi}_{af}^{0}(\mathbf{k})e^{i\mathbf{k}\cdot(\mathbf{x}-\mathbf{x}')}d^{3}\mathbf{k}. \quad (83)$$

If the source is a time varying signal $\mathbf{F}(t)$, and it is correlated to its spectrum through the following temporal Fourier transform pair:

$$\tilde{F}(\omega)=\int_{-\infty}^{+\infty}F(t)e^{i\omega t}dt\;,\quad F(t)=\frac{1}{2\pi}\int_{-\infty}^{+\infty}\tilde{F}(\omega)e^{-i\omega t}d\omega. \quad (84)$$

Thus, the complete wavefield in spacetime induced by a general time-varying point source is given by:

$$\langle\Phi_{hf}(\mathbf{x},\mathbf{x}',t)\rangle=$$
$$\frac{1}{2\pi}\int_{-\infty}^{+\infty}\tilde{F}(\omega)\left(\frac{1}{8\pi^{3}}\iiint_{V(\mathbf{k})}\left\{\left[P.S.\tilde{\Gamma}_{hg}(\mathbf{k})\right]^{-1}-\langle\Xi_{hc}\Xi_{dg}\rangle\left[\frac{1}{8\pi^{3}}\iiint_{V(\mathbf{s})}P.S.\tilde{\Gamma}_{cd}(\mathbf{s})\tilde{P}(\mathbf{k}-\mathbf{s})d^{3}\mathbf{s}\right]\right\}^{-1}\left[P.S.\tilde{\Gamma}_{ga}(\mathbf{k})\right]^{-1}\tilde{\Psi}_{af}^{0}(\mathbf{k})e^{i\mathbf{k}\cdot(\mathbf{x}-\mathbf{x}')}d^{3}\mathbf{k}\right)e^{-i\omega t}d\omega.$$
$$(85)$$

Eq. (85) gives an explicit expression for the wavefield in the spacetime domain and it has significant implications for seismological applications, for instance, it can be used to synthesize three-component seismogram envelops for realistic earthquake sources [3, 176]. However, exploring the solution goes beyond the scope of the current research. In this work, we only focus on the dispersion and attenuation behavior of a plane wave component. In a statistically isotropic medium, the dispersion behavior of a plane wave is independent of its propagation direction. Without loss of generality, we consider a plane wave propagating along the $x_3$ axis, and the wavevector $\mathbf{k}=[0,0,k]$. The coefficient matrix and the dispersion equations of longitudinal and transverse waves can be expressed explicitly as:

$$\left[P.S.\tilde{\Gamma}_{ge}(\mathbf{k})\right]^{-1}-\langle\Xi_{gc}\Xi_{de}\rangle\left[\frac{1}{8\pi^{3}}\iiint_{V(\mathbf{s})}P.S.\tilde{\Gamma}_{cd}(\mathbf{s})\tilde{P}(\mathbf{k}-\mathbf{s})d^{3}\mathbf{s}\right]=\begin{bmatrix} M_{11} & 0 & 0 & 0 & 0 & 0 & 0 & M_{18} & 0 \\ 0 & M_{22} & 0 & 0 & 0 & 0 & M_{27} & 0 & 0 \\ 0 & 0 & M_{33} & M_{34} & M_{35} & M_{36} & 0 & 0 & 0 \\ 0 & 0 & M_{34} & M_{44} & M_{45} & M_{46} & 0 & 0 & 0 \\ 0 & 0 & M_{35} & M_{45} & M_{55} & M_{56} & 0 & 0 & 0 \\ 0 & 0 & M_{36} & M_{46} & M_{56} & M_{66} & 0 & 0 & 0 \\ 0 & M_{27} & 0 & 0 & 0 & 0 & M_{77} & 0 & 0 \\ M_{18} & 0 & 0 & 0 & 0 & 0 & 0 & M_{88} & 0 \\ 0 & 0 & 0 & 0 & 0 & 0 & 0 & 0 & M_{99} \end{bmatrix}. \quad (86)$$

The non-vanishing elements of the matrix $\mathbf{M}$ are given by:

$$M_{11}=\mu(k^{2}-k_{T}^{2})-K_{44}k^{2}-\omega^{4}(\rho_{1}-\rho_{2})^{2}f_{1}f_{2}\Sigma_{11},\quad M_{33}=(\lambda+2\mu)(k^{2}-k_{L}^{2})-K_{11}k^{2}-\omega^{4}(\rho_{1}-\rho_{2})^{2}f_{1}f_{2}\Sigma_{33}, \quad (87a)$$

$$M_{34}=-K_{12}ik-\omega^{2}f_{1}f_{2}(\rho_{1}-\rho_{2})[(\Xi_{11}^{(1)}-\Xi_{11}^{(2)})+(\Xi_{12}^{(1)}-\Xi_{12}^{(2)})]\Sigma_{34}-\omega^{2}f_{1}f_{2}(\rho_{1}-\rho_{2})(\Xi_{12}^{(1)}-\Xi_{12}^{(2)})\Sigma_{36}, \quad (87b)$$

$$M_{36}=-K_{11}ik-2\omega^{2}f_{1}f_{2}(\rho_{1}-\rho_{2})(\Xi_{12}^{(1)}-\Xi_{12}^{(2)})]\Sigma_{34}-\omega^{2}f_{1}f_{2}(\rho_{1}-\rho_{2})(\Xi_{11}^{(1)}-\Xi_{11}^{(2)})\Sigma_{36}, \quad (87c)$$

$$M_{18}=-K_{44}ik-\omega^{2}f_{1}f_{2}(\rho_{1}-\rho_{2})(\Xi_{44}^{(1)}-\Xi_{44}^{(2)})]\Sigma_{18}, \quad (87d)$$

$$M_{44} = K_{11} - f_1f_2[(\Xi_{11}^{(1)} - \Xi_{11}^{(2)})^2 + (\Xi_{12}^{(1)} - \Xi_{12}^{(2)})^2]\Sigma_{44} - 2f_1f_2(\Xi_{11}^{(1)} - \Xi_{11}^{(2)})(\Xi_{12}^{(1)} - \Xi_{12}^{(2)})\Sigma_{45}$$
$$- 2f_1f_2[(\Xi_{11}^{(1)} - \Xi_{11}^{(2)})(\Xi_{12}^{(1)} - \Xi_{12}^{(2)}) + (\Xi_{12}^{(1)} - \Xi_{12}^{(2)})^2]\Sigma_{46} - f_1f_2(\Xi_{12}^{(1)} - \Xi_{12}^{(2)})^2\Sigma_{66},$$
(87e)

$$M_{45} = K_{12} - 2f_1f_2(\Xi_{11}^{(1)} - \Xi_{11}^{(2)})(\Xi_{12}^{(1)} - \Xi_{12}^{(2)})\Sigma_{44} - f_1f_2[(\Xi_{11}^{(1)} - \Xi_{11}^{(2)})^2 + (\Xi_{12}^{(1)} - \Xi_{12}^{(2)})^2]\Sigma_{45}$$
$$- 2f_1f_2[(\Xi_{11}^{(1)} - \Xi_{11}^{(2)})(\Xi_{12}^{(1)} - \Xi_{12}^{(2)}) + (\Xi_{12}^{(1)} - \Xi_{12}^{(2)})^2]\Sigma_{46} - f_1f_2(\Xi_{12}^{(1)} - \Xi_{12}^{(2)})^2\Sigma_{66},$$
(87f)

$$M_{46} = K_{12} - f_1f_2[(\Xi_{11}^{(1)} - \Xi_{11}^{(2)})(\Xi_{12}^{(1)} - \Xi_{12}^{(2)}) + (\Xi_{12}^{(1)} - \Xi_{12}^{(2)})^2]\Sigma_{44} - f_1f_2[(\Xi_{11}^{(1)} - \Xi_{11}^{(2)})(\Xi_{12}^{(1)} - \Xi_{12}^{(2)}) + (\Xi_{12}^{(1)} - \Xi_{12}^{(2)})^2]\Sigma_{45}$$
$$- f_1f_2[(\Xi_{11}^{(1)} - \Xi_{11}^{(2)})^2 + (\Xi_{11}^{(1)} - \Xi_{11}^{(2)})(\Xi_{12}^{(1)} - \Xi_{12}^{(2)}) + 2(\Xi_{12}^{(1)} - \Xi_{12}^{(2)})^2]\Sigma_{46} - f_1f_2(\Xi_{11}^{(1)} - \Xi_{11}^{(2)})(\Xi_{12}^{(1)} - \Xi_{12}^{(2)})\Sigma_{66},$$
(87g)

$$M_{66} = K_{11} - 2f_1f_2(\Xi_{12}^{(1)} - \Xi_{12}^{(2)})^2\Sigma_{44} - 2f_1f_2(\Xi_{12}^{(1)} - \Xi_{12}^{(2)})^2\Sigma_{45} - 4f_1f_2(\Xi_{11}^{(1)} - \Xi_{11}^{(2)})(\Xi_{12}^{(1)} - \Xi_{12}^{(2)})\Sigma_{46} - f_1f_2(\Xi_{11}^{(1)} - \Xi_{11}^{(2)})^2\Sigma_{66},$$
(87h)

$$M_{77} = K_{44} - f_1f_2(\Xi_{44}^{(1)} - \Xi_{44}^{(2)})^2\Sigma_{77}, \quad M_{99} = K_{44} - f_1f_2(\Xi_{44}^{(1)} - \Xi_{44}^{(2)})^2\Sigma_{99},$$
(87i)

$$M_{22} = M_{11}, \quad M_{88} = M_{77}, \quad M_{35} = M_{34}, \quad M_{27} = M_{18}, \quad M_{56} = M_{46}, \quad M_{55} = M_{44},$$
(87j)

where

$$K_{11} = \frac{3(\lambda + 6\mu)(\lambda + 2\mu)}{3\lambda + 8\mu}, \quad K_{12} = \frac{3(\lambda + \mu)(\lambda + 2\mu)}{3\lambda + 8\mu}, \quad K_{44} = \frac{15(\lambda + 2\mu)\mu}{2(3\lambda + 8\mu)}, \quad K_{11} = K_{12} + 2K_{44}.$$
(88)

$$\Sigma_{11}(\mathbf{k}) = \frac{1}{8\pi^3}\iiint_{V(\mathbf{s})}\tilde{G}_{11}(\mathbf{s})\tilde{P}(\mathbf{k}-\mathbf{s})d^3\mathbf{s}, \quad \Sigma_{18}(\mathbf{k}) = \frac{1}{8\pi^3}\iiint_{V(\mathbf{s})}[is_3\tilde{G}_{11}(\mathbf{s}) + is_1\tilde{G}_{13}(\mathbf{s})]\tilde{P}(\mathbf{k}-\mathbf{s})d^3\mathbf{s},$$
(89a)

$$\Sigma_{33}(\mathbf{k}) = \frac{1}{8\pi^3}\iiint_{V(\mathbf{s})}\tilde{G}_{33}(\mathbf{s})\tilde{P}(\mathbf{k}-\mathbf{s})d^3\mathbf{s}, \quad \Sigma_{34}(\mathbf{k}) = \frac{1}{8\pi^3}\iiint_{V(\mathbf{s})}is_1\tilde{G}_{13}(\mathbf{s})\tilde{P}(\mathbf{k}-\mathbf{s})d^3\mathbf{s},$$
(89b)

$$\Sigma_{36}(\mathbf{k}) = \frac{1}{8\pi^3}\iiint_{V(\mathbf{s})}is_3\tilde{G}_{33}(\mathbf{s})\tilde{P}(\mathbf{k}-\mathbf{s})d^3\mathbf{s}, \quad \Sigma_{44}(\mathbf{k}) = S_{1111} - \frac{1}{8\pi^3}\iiint_{V(\mathbf{s})}s_1^2\tilde{G}_{11}(\mathbf{s})\tilde{P}(\mathbf{k}-\mathbf{s})d^3\mathbf{s},$$
(89c)

$$\Sigma_{45}(\mathbf{k}) = S_{1221} - \frac{1}{8\pi^3}\iiint_{V(\mathbf{s})}s_1s_2\tilde{G}_{12}(\mathbf{s})\tilde{P}(\mathbf{k}-\mathbf{s})d^3\mathbf{s}, \quad \Sigma_{46}(\mathbf{k}) = S_{1221} - \frac{1}{8\pi^3}\iiint_{V(\mathbf{s})}s_1s_3\tilde{G}_{13}(\mathbf{s})\tilde{P}(\mathbf{k}-\mathbf{s})d^3\mathbf{s},$$
(89d)

$$\Sigma_{66}(\mathbf{k}) = S_{1111} - \frac{1}{8\pi^3}\iiint_{V(\mathbf{s})}s_3^2\tilde{G}_{33}(\mathbf{s})\tilde{P}(\mathbf{k}-\mathbf{s})d^3\mathbf{s}, \quad \Sigma_{77}(\mathbf{k}) = 4S_{2233} - \frac{1}{8\pi^3}\iiint_{V(\mathbf{s})}[s_2^2\tilde{G}_{33}(\mathbf{s}) + s_3^2\tilde{G}_{22}(\mathbf{s}) + 2s_2s_3\tilde{G}_{23}(\mathbf{s})]\tilde{P}(\mathbf{k}-\mathbf{s})d^3\mathbf{s},$$
(89e)

$$\Sigma_{99}(\mathbf{k}) = 4S_{2233} - \frac{1}{8\pi^3}\iiint_{V(\mathbf{s})}[s_1^2\tilde{G}_{22}(\mathbf{s}) + s_2^2\tilde{G}_{11}(\mathbf{s}) + 2s_1s_2\tilde{G}_{12}(\mathbf{s})]\tilde{P}(\mathbf{k}-\mathbf{s})d^3\mathbf{s},$$
(89f)

$$\Sigma_{22} = \Sigma_{11}, \quad \Sigma_{27} = \Sigma_{18}, \quad \Sigma_{35} = \Sigma_{34}, \quad \Sigma_{55} = \Sigma_{44}, \quad \Sigma_{56} = \Sigma_{46}, \quad \Sigma_{88} = \Sigma_{77}.$$
(89g)

The dispersion equation for longitudinal coherent waves is:

$$M_{33}(M_{44}M_{66} + M_{45}M_{66} - 2M_{46}^2) - M_{44}M_{36}^2 - 2M_{34}^2M_{66} - M_{45}M_{36}^2 + 4M_{34}M_{36}M_{46} = 0.$$
(90)

The dispersion equation for transverse coherent waves is:

$$M_{11}M_{88} - M_{18}^2 = 0.$$
(91)

The solution to the dispersion equations is a complex propagation constant, $k = \mathrm{Re}(k) + i\mathrm{Im}(k)$, where $\mathrm{Re}(k) = \omega/V$, $\mathrm{Im}(k) = \alpha$, $\omega$ is the circular frequency, $V$ is the coherent wave velocity, and $\alpha$ is the attenuation coefficient of the coherent wave. The propagation characteristics of the coherent wave is fully described by the complex wavenumber.

The integrals appeared in $\Sigma_{ij}(\mathbf{k})$ contain singular points: $s = k_T$ and $s = k_L$. The correct definition of these integrals is discussed in Sheng [112] and Calvet and Margerin [12], where they gave the explicit expression by splitting the homogeneous Green's functions into the real part and the imaginary part. The real part of Green's functions is defined as the Cauchy principal value at the singular points while the imaginary part is given by the Dirac-delta function. The sign of the imaginary part is determined by the causality of elastic waves, see Sheng [112]. In the numerical implementation, these integrals are broken into two or three parts, and the general forms are

$$\int_0^{+\infty}\frac{F(s)ds}{(s^2-k_T^2)} := P.V.\int_0^B\frac{F(s)ds}{(s^2-k_T^2)} + \int_B^{+\infty}\frac{F(s)ds}{(s^2-k_T^2)} + \int_0^B i\delta(s^2-k_T^2)F(s)ds,$$
(92)

$$\int_0^{+\infty}\frac{F(s)ds}{(s^2-k_T^2)(s^2-k_L^2)} := P.V.\int_0^A\frac{F(s)ds}{(s^2-k_T^2)(s^2-k_L^2)} + P.V.\int_A^B\frac{F(s)ds}{(s^2-k_T^2)(s^2-k_L^2)}$$
$$+ \int_B^{+\infty}\frac{F(s)ds}{(s^2-k_T^2)(s^2-k_L^2)} + \int_0^A i\delta(s^2-k_T^2)\frac{F(s)ds}{(s^2-k_L^2)} + \int_A^{+\infty}i\delta(s^2-k_L^2)\frac{F(s)ds}{(s^2-k_T^2)},$$
(93)

where $k_L < A < k_T < B$ and $F(s)$ is a regular function, i.e., has no singularity along the positive real axis.

All the integrands in the above integrals decays proportionally to $1/s^2$ as $s \to +\infty$. Consequently, all the infinite integrals are convergent. The complex wavenumber is obtained by searching for the roots of the dispersion equations in the complex $k$-plane. The numerical algorithm is implemented on the platform Compaq Visual Fortran 6.6 for which the powerful IMSL numerical library is integrated. All the numerical calculations are carried out in terms of the dimensionless quantities, i.e., the fractional velocity variation, the adimensional attenuation and the dimensionless frequencies, as given by:

$$\delta \bar{V}_L = \frac{V - V_{0L}}{V_{0L}}, \quad \bar{\alpha}_L = \alpha_L d, \quad K_{0L} = k_{0L} d, \tag{94}$$

$$\delta \bar{V}_T = \frac{V - V_{0T}}{V_{0T}}, \quad \bar{\alpha}_T = \alpha_T d, \quad K_{0T} = k_{0T} d, \tag{95}$$

where $d$ is the average diameter (or characteristic dimension) of the heterogeneities, $d=2a$, $V_{0L}$ and $V_{0T}$ are the velocities of the longitudinal and transverse waves of the reference material, $k_{0L}$ and $k_{0T}$ are the nominal wavenumbers calculated using the quasi-static limits of the longitudinal and transverse wave velocities, i.e., $k_{0L}=\omega/V_{0L}$, $k_{0T}=\omega/V_{0T}$. Since $d$, $V_{0L}$ and $V_{0T}$ are constants, $K_{0L}$ and $K_{0T}$ can be viewed as dimensionless frequencies. $\alpha_L$ and $\alpha_T$ are the attenuation coefficients of longitudinal and transverse waves.

# III. COMPARISON WITH AN IMPROVED MULTIPLE SCATTERING MODEL FOR HETEROGENEOUS ELASTIC MEDIA WITH WEAK PROPERTY FLUCTUATION

Before applying the new model to a series of practical problems, it is meaningful to make a comprehensive comparison between the new model and an improved multiple scattering model for heterogeneous elastic media with weak property fluctuation, which was recently developed by the author based on Weaver's model [48]. For the convenience of subsequent discussion, we introduce two acronyms for the two models, where SFMS stands for the Strong-Fluctuation-Multiple-Scattering model developed in this work and WFMS stands for the Weak-Fluctuation-Multiple-Scattering model. We omit the rigorous development of the WFMS model and only list the major results here. Interested readers are referred to [129] for more details.

The dispersion equations for the longitudinal (L) and transverse (T) coherent waves are:

$$\bar{\rho}(k^2 \bar{V}_L^2 - \omega^2) + \tilde{M}_L(\mathbf{k}) = 0, \quad \bar{\rho}(k^2 \bar{V}_T^2 - \omega^2) + \tilde{M}_T(\mathbf{k}) = 0, \tag{96}$$

where $\bar{V}_L$ and $\bar{V}_T$ are the Voigt-average longitudinal and transverse velocities, $\tilde{M}_L(\mathbf{k})$ and $\tilde{M}_T(\mathbf{k})$ are the longitudinal and transverse mass operators, which are given by

$$\tilde{M}_L(\mathbf{k}) = \tilde{M}_{LL}(\mathbf{k}) + \tilde{M}_{LT}(\mathbf{k}), \quad \tilde{M}_T(\mathbf{k}) = \tilde{M}_{TL}(\mathbf{k}) + \tilde{M}_{TT}(\mathbf{k}), \tag{97}$$

where $\tilde{M}_{LL}(\mathbf{k})$, $\tilde{M}_{LT}(\mathbf{k})$, $\tilde{M}_{TL}(\mathbf{k})$ and $\tilde{M}_{TT}(\mathbf{k})$ represent the partial contributions from the L-L, L-T, T-L and T-T mode conversions. The mass operators are complex quantities and the real and imaginary parts are:

$$\text{Re}\,\tilde{M}_{LL}(\omega,k) = \frac{2 f_1 f_2}{\pi} P.V. \int_0^{+\infty} \frac{s^2 a^3 ds}{\bar{\rho}(s^2 \bar{V}_L^2 - \omega^2)} \int_{-1}^{+1} \Big\{ \omega^4 (\rho_1 - \rho_2)^2 x^2 + s^2 k^2 (\lambda_1 - \lambda_2)^2 + 4 s^2 k^2 (\mu_1 - \mu_2)^2 x^4$$
$$- 2\omega^2 k s (\rho_1 - \rho_2)(\lambda_1 - \lambda_2) x - 4\omega^2 k s (\rho_1 - \rho_2)(\mu_1 - \mu_2) x^3 + 4 k^2 s^2 (\lambda_1 - \lambda_2)(\mu_1 - \mu_2) x^2 \Big\} \tilde{\eta}(x) dx, \tag{98}$$

$$\text{Re}\,\tilde{M}_{LT}(\omega,k) = \frac{2 f_1 f_2}{\pi} P.V. \int_0^{+\infty} \frac{s^2 a^3 ds}{\bar{\rho}(s^2 \bar{V}_T^2 - \omega^2)} \int_{-1}^{+1} \Big\{ \omega^4 (\rho_1 - \rho_2)^2 (1-x^2) + 4 s^2 k^2 (\mu_1 - \mu_2)^2 (1-x^2) x^2$$
$$- 4\omega^2 k s (\rho_1 - \rho_2)(\mu_1 - \mu_2) x (1-x^2) \Big\} \tilde{\eta}(x) dx, \tag{99}$$

$$\text{Im}\,\tilde{M}_{LL}(\omega,k) = \frac{f_1 f_2 \omega^3 a^3}{\bar{\rho} \bar{V}_L^3} \int_{-1}^{+1} \Big\{ \omega^2 (\rho_1 - \rho_2)^2 x^2 + \frac{(\lambda_1 - \lambda_2)^2}{\bar{V}_L^2} k^2 + 4 \frac{(\mu_1 - \mu_2)^2}{\bar{V}_L^2} k^2 x^4 - 4 \frac{\omega}{\bar{V}_L} (\rho_1 - \rho_2)(\mu_1 - \mu_2) k x^3$$
$$- 2 \frac{\omega}{\bar{V}_L} (\rho_1 - \rho_2)(\lambda_1 - \lambda_2) k x + 4 \frac{(\lambda_1 - \lambda_2)(\mu_1 - \mu_2)}{\bar{V}_L^2} k^2 x^2 \Big\} \tilde{\eta}^{LL}(x) dx, \tag{100}$$

$$\operatorname{Im}\tilde{M}_{LT}(\omega,k) = \frac{f_1 f_2 \omega^3 a^3}{\bar{\rho}\bar{V}_T^3}\int_{-1}^{+1}\left\{\omega^2(\rho_1-\rho_2)^2(1-x^2)+4\frac{(\mu_1-\mu_2)^2}{\bar{V}_T^2}k^2x^2(1-x^2)\right.$$
$$\left.-4\frac{\omega}{\bar{V}_T}(\rho_1-\rho_2)(\mu_1-\mu_2)(1-x^2)kx\right\}\tilde{\eta}^{LT}(x)dx,\tag{101}$$

$$\operatorname{Re}\tilde{M}_{TL}(\omega,k) = \frac{f_1 f_2}{\pi}P.V.\int_0^{+\infty}\frac{s^2 a^3 ds}{\bar{\rho}(s^2\bar{V}_L^2-\omega^2)}\int_{-1}^{+1}\left\{\omega^4(\rho_1-\rho_2)^2(1-x^2)+4s^2k^2(\mu_1-\mu_2)^2(1-x^2)x^2\right.$$
$$\left.-4\omega^2 ks(\rho_1-\rho_2)(\mu_1-\mu_2)x(1-x^2)\right\}\tilde{\eta}(x)dx,\tag{102}$$

$$\operatorname{Re}\tilde{M}_{TT}(\omega,k) = \frac{f_1 f_2}{\pi}P.V.\int_0^{+\infty}\frac{s^2 a^3 ds}{\bar{\rho}(s^2\bar{V}_T^2-\omega^2)}\int_{-1}^{+1}\left\{\omega^4(\rho_1-\rho_2)^2(1+x^2)+s^2k^2(\mu_1-\mu_2)^2(1-3x^2+4x^4)\right.$$
$$\left.-4\omega^2 ks(\rho_1-\rho_2)(\mu_1-\mu_2)x^3\right\}\tilde{\eta}(x)dx,\tag{103}$$

$$\operatorname{Im}\tilde{M}_{TL}(\omega,k) = \frac{f_1 f_2 \omega^3 a^3}{2\bar{\rho}\bar{V}_L^3}\int_{-1}^{+1}\left\{\omega^2(\rho_1-\rho_2)^2(1-x^2)+4\frac{(\mu_1-\mu_2)^2}{\bar{V}_L^2}k^2x^2(1-x^2)-4\frac{\omega}{\bar{V}_L}k(\rho_1-\rho_2)(\mu_1-\mu_2)x(1-x^2)\right\}\tilde{\eta}^{TL}(x)dx,\tag{104}$$

$$\operatorname{Im}\tilde{M}_{TT}(\omega,k) = \frac{f_1 f_2 \omega^3 a^3}{2\bar{\rho}\bar{V}_T^3}\int_{-1}^{+1}\left\{\omega^2(\rho_1-\rho_2)^2(1+x^2)+\frac{(\mu_1-\mu_2)^2}{\bar{V}_T^2}k^2(1-3x^2+4x^4)-4\frac{\omega}{\bar{V}_T}k(\rho_1-\rho_2)(\mu_1-\mu_2)x^3\right\}\tilde{\eta}^{TT}(x)dx,\tag{105}$$

where

$$\tilde{\eta}(x)=\frac{1}{(1+a^2k^2+a^2s^2-2a^2ksx)^2},\quad \tilde{\eta}^{LL}(x)=\tilde{\eta}^{TL}(x)=\frac{1}{(1+a^2s^2+a^2k_L^2-2a^2sk_Lx)^2},$$
$$\tilde{\eta}^{LT}(x)=\tilde{\eta}^{TT}(x)=\frac{1}{(1+a^2s^2+a^2k_T^2-2a^2sk_Tx)^2},\quad k_L=\frac{\omega}{\bar{V}_L},\quad k_T=\frac{\omega}{\bar{V}_T}.\tag{106}$$

The Voigt average velocities $\bar{V}_L$ and $\bar{V}_T$ are defined in terms of the Voigt average material properties as:

$$\bar{V}_L=\sqrt{\frac{\bar{\lambda}+2\bar{\mu}}{\bar{\rho}}},\ \bar{V}_T=\sqrt{\frac{\bar{\mu}}{\bar{\rho}}},\ \bar{\lambda}=f_1\lambda_1+f_2\lambda_2,\ \bar{\mu}=f_1\mu_1+f_2\mu_2,\ \bar{\rho}=f_1\rho_1+f_2\rho_2,\tag{107}$$

where $\lambda_1$, $\lambda_2$, $\mu_1$, $\mu_2$, $\rho_1$, $\rho_2$, $f_1$ and $f_2$ are the Lamé constants, density, and volume fraction of the two component materials.

These expressions differ from that given in [12] in that all the cross-terms of mass density and elastic moduli have a minus sign instead of a plus sign. It is exactly due to this "small" change that the results are dramatically different from that given in [12]. It is worth mentioning that the improved model is developed by introducing the concept of exterior product, which has been extensively used in Differential Geometry [157]. It provides an ideal tool for dealing with integrals with multiple independent variables. As pointed out by Calvet and Margerin, the spectral function obtained in [12] gives two peaks in the high frequency domain for Media D and E, but there should be only one longitudinal mode for Media D and E and one transverse mode for Medium E. Contrarily, the spectral functions of Media D and E obtained using the new formula gives spectral curves exactly the same as predicted, i.e., one peak for longitudinal waves in Media D and E and one peak for transverse waves in Medium E in the whole frequency range, which perfectly resolves the raised question. For more details please see [129].

## A. Instability of the WFMS model in predicting the dispersion behavior of weak-property-fluctuation elastic media

To keep consistent with [12], we use the same properties and the same numbering system for materials A-E, as shown in Tab. 3. Materials F and G are introduced additionally for the purpose of comparing the stability of the two models. The materials considered in this section are all equal-phased, i.e., the volume fraction of the two phases are equal, $f_1=f_2=50\%$. As a quantitative indicator of the property fluctuation, the fractional velocity fluctuations of each component phase, defined by $\delta V_{L1}/V_{0L}=(V_{L1}-V_{0L})/V_{0L}$ etc. are shown in the table.

Table 3. Material Properties of Media A-G.

| Medium | $\rho_1$ (kg/m³) | $\rho_2$ (kg/m³) | $\lambda_1$ (GPa) | $\lambda_2$ (GPa) | $\mu_1$ (GPa) | $\mu_2$ (GPa) | $V_{L1}$ (m/s) | $V_{L2}$ (m/s) | $V_{T1}$ (m/s) | $V_{T2}$ (m/s) |
|---|---|---|---|---|---|---|---|---|---|---|
| A | 11000 | 11000 | 800 | 800 | 380 | 340 | 11908.74 | 11599.37 | 5877.54 | 5559.59 |
| B | 11000 | 11000 | 880 | 800 | 380 | 380 | 12210.28 | 11908.74 | 5877.54 | 5877.54 |
| C | 11000 | 11000 | 800 | 840 | 340 | 380 | 11599.37 | 12060.45 | 5559.59 | 5877.54 |
| D | 11000 | 11000 | 880 | 800 | 340 | 380 | 11908.74 | 11908.74 | 5559.59 | 5877.54 |
| E | 11550 | 11000 | 840 | 800 | 399 | 380 | 11908.74 | 11908.74 | 5877.54 | 5877.54 |

| | | | | | | | | | |
|---|---|---|---|---|---|---|---|---|---|
| F | 11000 | 10500 | 880 | 800 | 380 | 340 | 12210.28 | 11872.34 | 5877.54 | 5690.43 |
| G | 11000 | 11500 | 880 | 880 | 380 | 370 | 12210.28 | 11868.85 | 5877.54 | 5672.21 |

Table 4. Reference velocities of longitudinal and transverse waves of Media A-G.

| Medium | $V_{0L}$ (m/s) | $V_{0T}$ (m/s) | $\bar{V}_L$ (m/s) | $\bar{V}_T$ (m/s) | $\delta V_{L1}/V_{0L}$ | $\delta V_{L2}/V_{0L}$ | $\delta V_{T1}/V_{0T}$ | $\delta V_{T2}/V_{0T}$ |
|---|---|---|---|---|---|---|---|---|
| A | 11751.97 | 5716.68 | 11755.08 | 5720.78 | 0.01334 | -0.01298 | 0.02814 | -0.02748 |
| B | 12056.68 | 5877.54 | 12060.45 | 5877.54 | 0.01274 | -0.01227 | -3.17×10⁻⁷ | -3.17×10⁻⁷ |
| C | 11826.75 | 5716.69 | 11832.16 | 5720.78 | -0.01923 | 0.01976 | -0.02748 | 0.02814 |
| D | 11904.39 | 5716.70 | 11908.74 | 5720.78 | 0.00037 | 0.00037 | -0.02748 | 0.02813 |
| E | 11906.59 | 5876.73 | 11908.74 | 5877.54 | 0.00018 | 0.00018 | 0.00014 | 0.00014 |
| F | 12036.73 | 5782.79 | 12046.42 | 5786.91 | 0.01442 | -0.01366 | 0.01638 | -0.01597 |
| G | 12036.80 | 5773.27 | 12036.98 | 5773.50 | 0.01441 | -0.01395 | 0.01806 | -0.01751 |

At this point it is necessary to highlight the features of each material: the shear modulus $\mu$ is the only different property of the two constituent materials of medium A, while the Lamé constant $\lambda$ is the only different property for medium B, both the Lamé constants $\lambda$ and $\mu$ are different for medium C. As is seen from Tab. 3, the two component materials of both Media A and C have different longitudinal and transverse wave velocities, while the component phases of Medium B have different longitudinal velocities but identical transverse velocity. The two components of Medium D have a unique combination of Lamé constants such that their longitudinal velocities are the same, while their transverse velocities are still different. Medium E has a more special combination of material properties for which both the longitudinal and the transverse velocities of the two phases are equal. Medium F and G are general materials for which the two component materials have different elastic constants and densities.

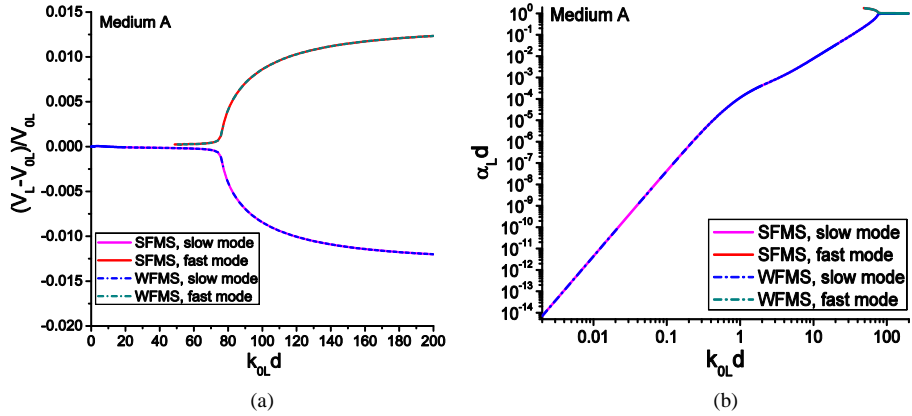

FIG. 8. Comparison of the velocity (a) and attenuation (b) of medium A calculated using SFMS and WFMS model.

Figure 8 shows the longitudinal wave dispersion and attenuation of Medium A, which are calculated from Eq. (88) and Eq. (92) for SFMS and WFMS models, respectively. From the results, we can see the velocity and attenuation calculated from the WFMS model show excellent agreement with that obtained using the SFMS model. The longitudinal velocity starts from the quasi-static limit $V_{0L}$, and then slightly decreases as the frequency increases. Meanwhile, the dimensionless attenuation $\alpha_L d$ increases nonlinearly from zero to 0.1. As the dimensionless frequency $k_{0L}d$ approaches 50, a second, faster mode begins to appear, but with much larger attenuation and slight positive dispersion. As the frequency increases further, the velocities of the two modes quickly approach the upper and lower limits $V_{L1}$ and $V_{L2}$, respectively, and the attenuation coefficients reach a saturation value near unity. We can see the two models give accurate results in the whole frequency range, from the quasi-static region to the geometric domain. Other scattering models based on different approximations are only valid in a limited frequency range. For example, the Born approximation is only valid in a relatively low frequency domain, corresponding to $0 < k_{0L}d < 70$ for this case, while geometric approximation is only valid in high frequency range, corresponding to $k_{0L}d > 150$ for this case. In addition to dispersion, the new model can also give the accurate attenuation. All these unique features show that the multiple scattering models have incomparable advantages compared to other scattering models. The WFMS model has been widely used in characterization of polycrystalline alloys, for which only the elastic stiffness is a random variable. Previous calculations show the WFMS model can give stable prediction for dispersion and attenuation of coherent waves in polycrystals.

Here we will show that the WFMS model gives unstable prediction for the dispersion and attenuation of two-phase materials for which both the elastic stiffness and the density are random variables.

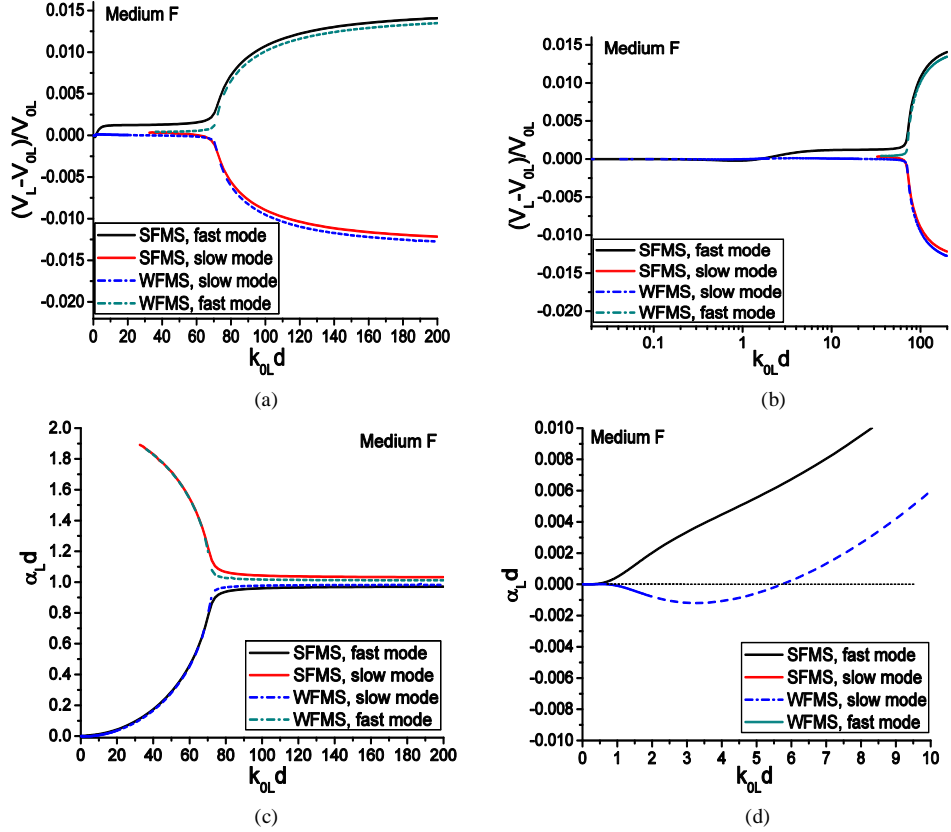

FIG. 9. Comparison of velocity (a) and attenuation (c) of medium F predicted by SFMS and WFMS models. (b) is a semi-logarithmic coordinate plot of (a) and (d) is an enlarged view of (c).

The dispersion and attenuation of longitudinal waves of Medium F calculated using the two models are shown in Fig. 9. It is seen from Figs. 9 (a) and (b) that the velocity predicted by the SFMS model first undergoes negative dispersion and then become positive dispersive, Contrary to Medium A, a second, slower mode appears in addition to the original, faster one at intermediate frequencies. At high frequencies, they approach the upper and lower limits of the two components, which is similar to Medium A. From these results, we can also observe that the dispersion given by the WFMS model have a similar pattern as Medium A, but it gives negative attenuation at dimensionless frequencies below 5.5, while the SFMS model gives positive attenuation in the whole frequency domain. Negative attenuation is physically impractical since it means the wave amplitude increases with propagation distance while no energy is input into the system. For velocity dispersion, differences in velocity up to 5% are observed between the results obtained by the two models. This example shows the performance of the WFMS model is unstable, in certain cases it gives wrong predictions.

Next, we examine the propagation characteristics of transverse waves predicted by the two models. The transverse dispersion and attenuation of Medium A are shown in Fig. 10. It is seen the dispersion curves calculated from the two models exhibit excellent agreement. However, the attenuation predicted by the WFMS model is negative in nearly the whole frequency range and form an image of the curves given by the SFMS model symmetric about the axis $\alpha_T d=0$. Fig. 10(c) shows the detailed variation the attenuation in frequency range (0, 1.0). In this range, the WFMS attenuation first increases with frequency. After reaching its positive maximum, it decreases with frequency and then becomes negative. This oscillation behavior prohibits any attempts to remedy the defect simply by reversing the sign of the attenuation.

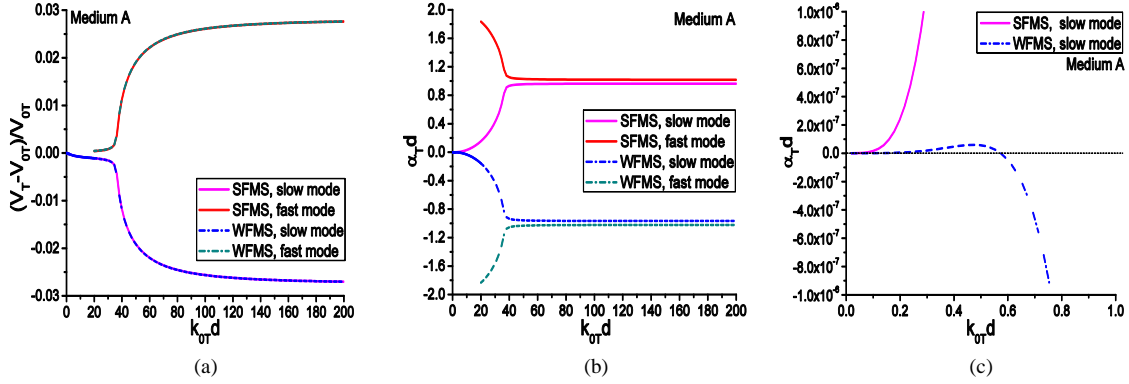

FIG. 10. Velocity and attenuation of medium A calculated using the SFMS and WFMS models.

As an additional example, Fig. 11 gives the dispersion and attenuation curves of transverse waves in Medium G. In this case the velocities predicted by the two models agree well at frequencies lower than 50, and the discrepancy becomes large at high frequencies. Moreover, the velocities calculated from the WFMS model do not converge to its geometric limits. Although it gives positive values, the WFMS model underestimates the attenuation of the slow mode and overestimates that of the fast mode in the whole frequency range.

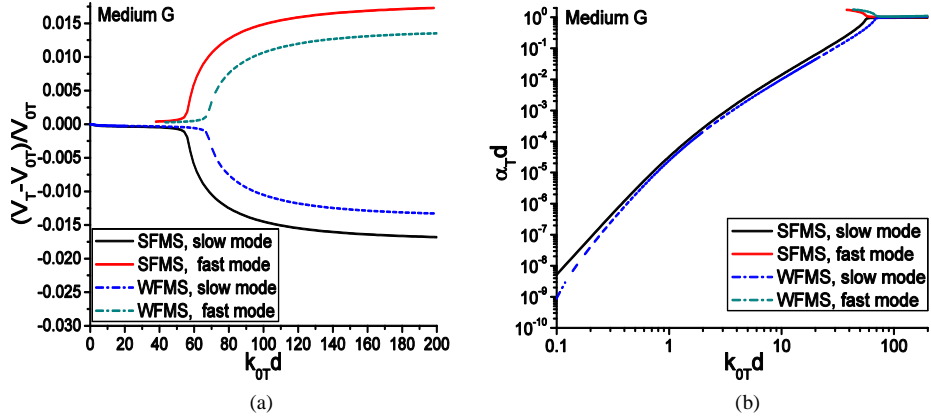

FIG. 11. Velocity and attenuation of medium E calculated using SFMS and WFMS model.

Through these examples, we can conclude that the WFMS model gives instable, and in certain cases wrong predictions of the velocity and attenuation. It is difficult to judge when the results converge to the correct solution. In all cases, the SFMS model gives stable and reasonable results in the whole frequency range. We are now in a position to examine the predictions given by the SFMS model for the propagation characteristics of coherent waves in Materials B-E, as discussed in [12]. For simplicity, the results calculated using the WFMS model are neglected here, and the results for longitudinal and transverse waves obtained using the SFMS model are presented in Figs. 12 and 13, respectively.

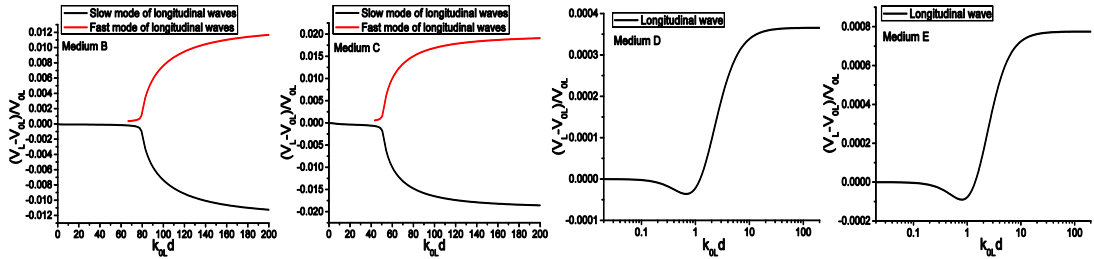

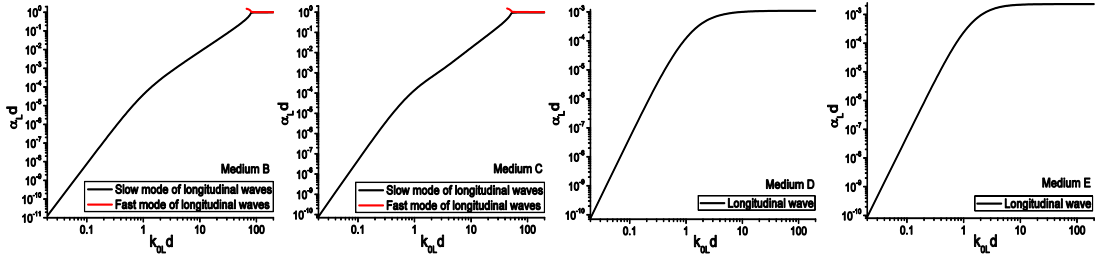

FIG. 12. Dispersion and attenuation of longitudinal waves in Media B-E calculated using the SFMS model.

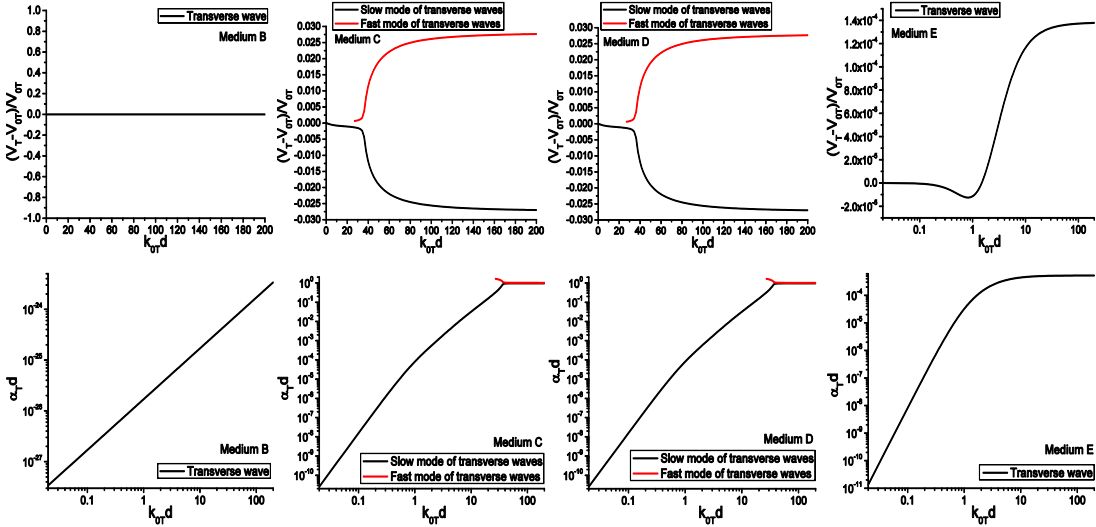

FIG. 13. Dispersion and attenuation of transverse waves in Media B-E calculated using the SFMS model.

From the results, we can see Medium B has one longitudinal mode at low frequencies and two at high frequencies, while only one transverse mode with negligible attenuation can be found in the whole frequency range. Both longitudinal and transverse waves in Medium C have one propagation mode at low frequencies and two at high frequencies. All these predictions are consistent with that given in [12]. Media D and E provide two subtle examples to examine the reliability of different models. For Medium D, as pointed out in [12], the two component materials have the same longitudinal velocity but different transverse velocities, so there should be one longitudinal mode in the whole frequency range while the transverse waves behave normally, i.e., one mode at low frequencies and two at high frequencies. Contrary to this presupposition, the spectral function in [12] gives two longitudinal modes at high frequencies. As discussed before, this is due to a mistake in the expression of the mass operators. Figure 13 shows results drastically different from those in [12], the SFMS model predicts one longitudinal mode in the whole frequency range, one transverse mode at low frequencies and two at high frequencies, which perfectly confirms the original judgment. Different from that of Medium D, both the longitudinal and transverse waves of the two phases in Medium E have the same velocity. Once again, the predictions given by SFMS model show one longitudinal and transverse mode in the whole frequency range. It is also noted that when there is only one mode, the relative variation of velocities is very small, less than 0.001, which can be regarded as zero. Meanwhile, the dimensionless attenuation is also very small, normally less than $10^{-3}$. All these predictions are in excellent agreement with the conjectures in [12].

## B. Failure of the WFMS model in predicting the dispersion behaviors of strong-property-fluctuation elastic media

In this section, we examine the velocity and attenuation of strong-property-fluctuation materials predicted by the two models. Cortical bone is a typical strong-property-fluctuation material, for which the two component phases: solid bone frame and water/marrow saturated pores, have drastically different material properties, as shown in Tab. 4. As a simplified mechanical model, cortical bone is assumed to be comprised of a homogeneous isotropic solid and water-saturated spherical pores. The pores are uniformly distributed in the solid frame, so the cortical bone is a statistically homogeneous medium.

Table 4. Material properties of the component phases of cortical bone.

| Material | $\rho$ (kg/m$^3$) | $C_{11}$ (GPa) | $C_{12}$ (GPa) | $C_{66}$ (GPa) | $V_L$ (m/s) | $V_T$ (m/s) |
|---|---|---|---|---|---|---|
| Bone | 1850 | 29.60 | 17.60 | 6.00 | 4000.00 | 1800.90 |
| Water | 1000 | 2.37 | 2.37 | 0 | 1539.48 | 0 |

The reference velocities for cortical bones with different porosity are summarized in Tab. 5 and plotted in Fig. 14.

Table 5. Material properties of the reference medium of cortical bone with different porosity.

| Porosity | $\rho$ (kg/m$^3$) | $\lambda$ (GPa) | $\mu$ (GPa) | $V_{0L}$ (m/s) | $V_{0T}$ (m/s) | $\bar{V}_L$ (m/s) | $\bar{V}_T$ (m/s) |
|---|---|---|---|---|---|---|---|
| $f_{pore}$=1% | 1841.5 | 17.13 | 5.89 | 3962.22 | 1788.43 | 3990.74 | 1796.00 |
| $f_{pore}$=2% | 1833.0 | 16.67 | 5.78 | 3924.41 | 1775.75 | 3981.37 | 1791.05 |
| $f_{pore}$=3% | 1824.5 | 16.22 | 5.67 | 3886.58 | 1762.87 | 3971.89 | 1786.03 |
| $f_{pore}$=4% | 1816.0 | 15.78 | 5.56 | 3848.74 | 1749.76 | 3962.29 | 1780.96 |
| $f_{pore}$=5% | 1807.5 | 15.35 | 5.45 | 3810.88 | 1736.44 | 3952.59 | 1775.82 |
| $f_{pore}$=6% | 1799.0 | 14.93 | 5.34 | 3773.02 | 1722.88 | 3942.77 | 1770.61 |
| $f_{pore}$=7% | 1790.5 | 14.52 | 5.23 | 3735.16 | 1709.09 | 3932.83 | 1765.35 |
| $f_{pore}$=8% | 1782.0 | 14.11 | 5.12 | 3696.54 | 1695.04 | 3922.77 | 1760.01 |
| $f_{pore}$=9% | 1773.5 | 13.72 | 5.01 | 3658.68 | 1680.75 | 3912.58 | 1754.61 |
| $f_{pore}$=10% | 1765.0 | 13.34 | 4.90 | 3620.84 | 1666.19 | 3902.28 | 1749.14 |
| $f_{pore}$=15% | 1722.5 | 11.55 | 4.34 | 3427.03 | 1587.32 | 3848.77 | 1720.70 |
| $f_{pore}$=20% | 1680.0 | 9.98 | 3.79 | 3233.01 | 1501.98 | 3791.75 | 1690.31 |
| $f_{pore}$=25% | 1637.5 | 8.60 | 3.24 | 3034.66 | 1406.64 | 3730.83 | 1657.74 |
| $f_{pore}$=30% | 1595.0 | 7.41 | 2.69 | 2831.75 | 1298.66 | 3665.56 | 1622.72 |
| $f_{pore}$=35% | 1552.5 | 6.41 | 2.16 | 2628.96 | 1179.54 | 3595.44 | 1584.95 |
| $f_{pore}$=40% | 1510.0 | 5.57 | 1.64 | 2420.94 | 1042.16 | 3519.86 | 1544.06 |
| $f_{pore}$=45% | 1467.5 | 4.89 | 1.15 | 2213.48 | 885.24 | 3438.09 | 1499.57 |
| $f_{pore}$=50% | 1425.0 | 4.36 | 0.71 | 2013.99 | 705.87 | 3349.26 | 1450.95 |

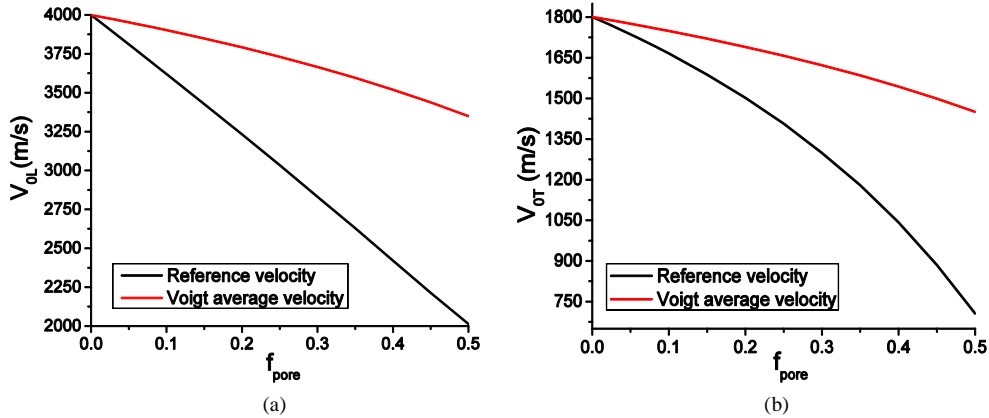

(a) (b)

FIG. 14. Quasi-static limit of the longitudinal and transverse wave velocities.

From Fig. 14 we see the Voigt average velocity systematically overestimates the quasi-static limits of both longitudinal and transverse waves, and the differences increase dramatically as the porosity increases. These results qualitatively agree with the predictions given by the self-consistent approach [49-50].

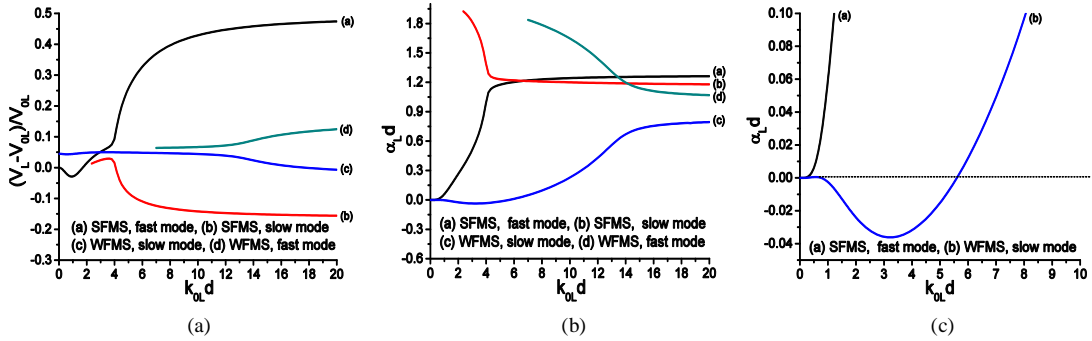

(a) (b) (c)

FIG. 15. Longitudinal velocity and attenuation of cortical bone calculated using the SFMS and WFMS models.

Figure 15 shows the longitudinal dispersion and attenuation of cortical bone with 10% porosity. According to the prediction given by the SFMS model, the longitudinal velocity starts from the quasi-static limit and then decreases sharply as the frequency increases. This is the famous phenomenon known as the anomalous negative dispersion, which has been observed in experiments and captured in numerical simulations [70, 73, 89]. After assuming its minimum, the longitudinal velocity increases quickly. At a dimensionless frequency around $k_{0L}d$=2, an additional slow mode begins to appear and its velocity slowly increases with frequency. At the dimensionless frequency $k_{0L}d$=4.5, the velocity of the fast mode approaches the upper bond, i.e., the velocity of pure bone, and then the two modes quickly separate and the velocity of the fast mode goes beyond the upper limit. This phenomenon could not be explained in the context of scattering, since the fastest propagation speed of a scattered wave is the longitudinal velocity in pure bone. This does not mean that the new model is invalid, instead, it tells us that the coherent wave disappears. The rationale behind this explanation is as follows: As the dimensionless frequency increases, the wavelength becomes shorter and shorter. When the wavelength is comparable to the size of a single scatterer, the wave is able to discern the structure of individual scatterer and the interaction of the scatterers with the wave gets significantly enhanced. Large property contrast further strengthens the wave-scatterer interaction. As a consequence, the incoherent component occupies a large proportion of the total random wave field. Thus, when the frequency goes beyond a certain limit, the wave losses coherence. In our model, this critical frequency is identified as the frequency at which the coherent wave velocity goes beyond the faster velocity of the two component phases. The disappearance of coherent waves is systematically studied in Sheng [112]. The loss of coherence is also reported in Sato [3, pp. 58-59], where the experimental studies of the coherent wave propagation and dismiss are presented. In seismology, this frequency range is called the saturated scattering regime by Wu [1]. Wu proposed a seismic ray explanation of this phenomenon: In this regime, the wave amplitude fluctuations are saturated and rays split into numerous microrays and interfere with each other, no coherent waves exist. In Fig. 15, no similarities can be observed between the predictions given by the two models. The WFMS model gives wrong initial guess, cannot reveal the negative dispersion, and unable to predict the frequency at which the coherent waves disappear. Moreover, the WFMS model gives negative attenuation at frequencies below $k_{0L}d$=5.5.

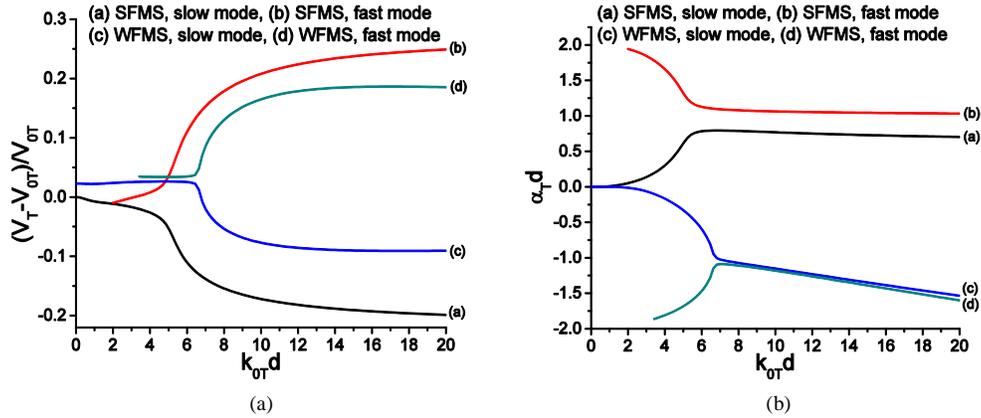

FIG. 16. Transverse velocity and attenuation of cortical bone calculated using the SFMS and WFMS models.

The transverse wave dispersion and attenuation are depicted in Fig. 16. According to the results calculated from the SFMS model, the velocity of transverse waves starts from the quasi-static limit and decreases monotonically as the frequency increases. At the dimensionless frequency around $k_{0T}d$=2, an additional faster mode appears. This mode has nearly the same velocity as that of the original mode but with much larger attenuation, so it is very difficult to observe in experiments. As the frequency continues to increase, the fast mode undergoes positive dispersion while the velocity of the slow mode still decreases. When the dimensionless frequency $k_{0T}d$ reaches 5, the velocity of the fast wave exceeds the transverse velocity of pure bone, which indicates the disappearance of coherent transverse waves. Once again, the WFMS model fails. It gives positive dispersion at low frequencies and negative attenuation in the whole frequency range.

Through the above discussion we can draw the following major conclusions:

1) The performance of the WFMS model is unstable, in certain cases it gives results with large errors or physically impractical results, such as negative attenuation, while the SFMS model gives very accurate results and the performance is extremely robust;

2) For two-phase media with a unique combination of mass density and Lamé constants such that both phases have the same longitudinal and/or transverse wave velocity, the SFMS model gives only one longitudinal and/or transverse propagation mode in the whole frequency range, the question raised in [12] has been resolved perfectly;

3) The coexistence of two bulk modes at intermediate to high frequency range is a dynamic bifurcation phenomenon, which universally exists in nearly all types of heterogeneous materials, it is not an artefact and the model has no deficiency in this sense;

4) The Voigt average velocity always overestimates the quasi-static velocity for heterogeneous media. This error becomes unacceptable for strong-property-fluctuation materials;

5) The WFMS model completely fails for strong-property-fluctuation materials, while the SFMS model still can give accurate prediction of the dispersion and attenuation behaviors of the coherent waves, it can also indicate the disappearance of the coherent waves by giving a velocity exceeding the upper bond of the two component phases;

6) The spectral function approach developed in [12] is unable to reveal the accurate dispersion and attenuation in the transition frequency range, and cannot distinguish between negative attenuation and positive attenuation, while the new model can accurately capture all these characteristics in the whole frequency range.

## IV. PRACTICAL APPLICATIONS OF THE NEW MODEL

In this section, we perform comprehensive studies on the propagation behavior of multiple scattered elastic waves in a series of heterogeneous media. As mentioned in the Introduction, seismology, ultrasonic nondestructive evaluation and bone quantitative ultrasound are three major fields in which multiple scattering theory is extensively used. Most materials of practical importance, including the heterogeneous lithosphere, particulate composite materials, poroelastic materials, polycrystals with strongly anisotropic crystallites, and human cortical bone exhibit heterogeneities with strong property fluctuation. Moreover, the central frequency of practical elastic waves ranges from 0.01 Hz to 50 MHz or even higher, which leads to ratios of wavelength to characteristic dimension of heterogeneities covering several orders of magnitude, from about 0.01 for seismic waves in the lithosphere to 100 or even larger for ultrasonic waves in fine-grain polycrystals or nanoparticulate composite materials. Therefore, accurate and quantitative characterization and reliable inversion of microstructures require powerful multiple scattering theories that applicable to strong-property-fluctuation materials and valid in the whole frequency range. The new model exactly meets these requirements since it gives up the weak-property-fluctuation approximation and does not introduce any approximation for the frequency. The new model provides a unified theoretical framework which establishes accurate quantitative relationship between the wave dispersion and attenuation characteristics and the microstructure features and is naturally suitable for extracting the microstructural information, such as the average size of the inclusions, the volume fraction of each component, the mass density and elastic stiffness fluctuation. To show its power in practical applications, we consider four examples: scattering and attenuation of seismic waves propagating in the Earth's lithosphere, scattering of ultrasonic waves in the porous aluminum and two-phase alloy, and scattering of ultrasound in human cortical bone. From the point of view of material composition, these examples cover nearly all types of two-phased materials, including the solid-solid, solid-vacuum and solid-fluid combinations.

### A. Applications in seismology

The planet Earth exhibits different degrees of heterogeneity from the inner core [13], to the mantle [15, 32, 158], and finally to the lithosphere [3, 8, 158-160]. The apparent attenuation of major arrivals and the existence of coda waves in nearly all the recorded seismograms are direct evidences of the existence of such inhomogeneities. However, due to lack of knowledge of wave propagation in such highly inhomogeneous media, the classical seismology is based on the multilayered model, in which each layer is treated as laterally homogeneous and vertically varying materials as a natural result of the action of gravitation. With the development of modern geophysics and seismology, more and more seismologists realize that the Earth also exhibits considerable lateral heterogeneity and tend to analyze the measured seismograms using scattering theories based on the random medium model [3, 8-9, 11-15, 178-179]. The prominent seismologist Aki first noticed the importance of coda waves and suggested that coda waves contain rich information about the statistical property of the heterogeneous lithosphere [7]. Aki [5-6], Wu [10] and Sato [3] also proposed scattering models to explain the apparent attenuation of the direct longitudinal and transverse arrivals. Up to date, scattering and attenuation of seismic waves have become the central topic of seismology [1-4]. A brief literature review shows that most previous studies are based on certain approximations, such as the Born approximation [3], the Rytov approximation [3], the single scattering approximation [18], the scalar wave approximation [17, 161], or the weak-fluctuation approximation [2, 16]. As pointed out in the Introduction, these approximations put severe restrictions on the application of these models. Consequently, a complete understanding of the propagation behavior of seismic waves is still missing. Many fundamental aspects of seismic waves are still poorly understood, some of which are listed below:

1. Is the attenuation of seismic waves mainly caused by scattering or by other intrinsic mechanisms?
2. Is the single scattering approximation enough to explain the apparent attenuation?
3. Can the Kramers-Kronig relation be used to explain the measured velocity and attenuation?
4. Does the Mohorovičić discontinuity really exist?
5. Is the multi-layered model appropriate to explain the seismic data?
6. Is the classical ray theory applicable to the explanation of seismic data?
7. What are the scattering dispersion and Q-factor in the whole frequency range?

With these questions in mind, we first conduct an in-depth investigation on the fundamental propagation behavior of longitudinal and transverse seismic waves in the Earth's lithosphere based on a realistic random medium model. The numerical results calculated from various combinations of material properties (density and elastic moduli) of the real rocks in the lithosphere will provide us important insights into all these questions. Through comparison of the new results with that observed in local, regional and global earthquakes, we will give novel and consistent answers to a series of longstanding problems.

Evidence from amplitude and phase fluctuation observed by the large scale seismic arrays LASA and NORSAR show that in the lithosphere, there exists a velocity fluctuation about 2% to 10% and the correlation distance of the fluctuation is about 10-20 km [5, 8, 160]. Moreover, it is also discovered that the velocity fluctuations are remarkably uniform. Based on these observations, the lithosphere is frequently modeled as a statistically homogeneous and isotropic random medium. Two-point correlation function is an important characteristic function to characterize various heterogeneous media. Up to now, random medium model with Gaussian, exponential and Von Karman correlation functions have all been adopted to describe the heterogeneities in the Earth [3, 5, 159-160, 177]. However, it is reported that the Gaussian correlation function is too smooth to represent the actual inhomogeneities in the Earth. Transverse coherence function (TCF) and angular coherence function (ACF) analysis measured by the LASA seismic array reveal [8] that the exponential or the Von Karman-type correlation functions fit the seismic data better. For the above reasons, in the present work we model the heterogeneous lithosphere as an exponential-type random medium. To model the velocity fluctuation, we consider a two-phase continuum composed of a hard phase with relatively large longitudinal and transverse velocities and a soft phase with relatively low velocities. Since the velocity fluctuation in the lithosphere is rather uniform, we assume that the two phases have equal volume fraction, i.e., $f_1=f_2=50\%$.

Dispersion and attenuation are two prominent features of waves propagating in a heterogeneous medium. However, researches concerning the dispersion of seismic waves are rather limited. This is partially because of the irregular surface of the Earth which significantly disturbs the measurement of dispersion. Another reason is that most seismic signals lie in the high frequency range, or even in the geometric range, in which the seismic waves have very small dispersion. The majority of existing studies are concentrated on the attenuation of direct arrivals and coda waves. It is well known that scattering attenuation and intrinsic attenuation are two major physical mechanisms for the overall attenuation of seismic signals. Intrinsic attenuation transfers elastic energy into heat and thus causes energy loss of the total wavefield. Contrarily, scattering cause magnitude decrease of the coherent waves by exciting coda waves, while the elastic energy of the total wavefield is conserved. The relative contribution of the two mechanisms to the overall seismic attenuation is an open problem in seismology. Nevertheless, many seismologists believe that the scattering attenuation contribute the major part of the total attenuation [3]. It was also pointed out in [9] that multiple scattering theory is necessary to separate the contributions from the two mechanisms. In this work, we assume that the scattering attenuation is the dominant mechanism and all the materials are considered as purely elastic materials. The answer to this question will be drawn after obtaining the exact numerical solution and comparing the numerical results with measured seismic data.

Seismologists usually quantify the attenuation of seismic waves by using the inverse quality factor $Q^{-1}$ instead of the attenuation coefficient $\alpha$. For simplicity, we call the inverse Q factor as Q-factor directly. The quality factor is defined as the ratio of the energy dissipated in a cycle of vibration to the total energy supplied. Carcione and Cavallini [162] conducted a rigorous derivation of the Q factor based on this definition and gives the following expression:

$$Q^{-1} = \frac{\mathrm{Im}(k^2)}{\mathrm{Re}(k^2)}, \tag{108}$$

where $k$ is the wavenumber of the longitudinal or transverse waves.

In textbooks of seismology [2, 3], the quality factor is also defined by the asymptotic expression of a spherically outgoing body wave:

$$u(r,f) = \frac{1}{r}\exp\left(-\frac{\pi f}{QV}r\right), \tag{109}$$

where $Q$ is the quality factor, $f$ is the frequency, $V$ is the velocity, $r$ is the distance between the source and the field point, and the prefactor $1/r$ accounts for the geometric spreading.

Considering the definition of attenuation coefficient $\alpha$:

$$u(r,f) = \frac{1}{r}\exp(-\alpha r), \tag{110}$$

we have the following identity:

$$\alpha = \frac{\pi f}{QV}, \tag{111}$$

or equivalently:

$$Q^{-1} = \frac{\alpha V}{\pi f}. \tag{112}$$

On the other hand, $k$ can be split into real and imaginary parts as:

$$k = \frac{2\pi f}{V} + i\alpha. \tag{113}$$

Substitution of (113) into (108) yields:

$$Q^{-1} = \frac{2\alpha \frac{2\pi f}{V}}{\left(\frac{2\pi f}{V}\right)^2 - \alpha^2}. \tag{114}$$

When the attenuation coefficient is small compared to the real part of $k$, terms of second and higher order in $\alpha$ can be neglected, thus we have

$$Q^{-1} \approx 2\alpha \frac{V}{2\pi f} = \frac{\alpha V}{\pi f}, \quad \text{when} \quad \frac{\alpha V}{2\pi f} \ll 1. \tag{115}$$

Comparing Eq. (115) with Eq. (112), we see the two definitions are equivalent. Since the definition in (108) is more accurate, we use this definition in subsequent calculations.

The ratio of the longitudinal Q-factor to the transverse Q-factor can be obtained from Eq. (115), given by:

$$\frac{Q_L^{-1}}{Q_T^{-1}} \approx \frac{\alpha_L V_L}{\alpha_T V_T}, \quad \text{when} \quad \frac{\alpha V}{2\pi f} \ll 1. \tag{116}$$

Numerical calculations in Sec. III show that in the geometric regime, the attenuation coefficients of the longitudinal and transverse waves in weak-property fluctuation media satisfy the following relation:

$$\alpha_L d \approx \alpha_T d \approx 1, \tag{117}$$

which leads to $\alpha_L \approx \alpha_T$, consequently we obtain:

$$\frac{Q_L^{-1}}{Q_T^{-1}} \approx \frac{V_L}{V_T}, \quad \text{when} \quad f \to \infty. \tag{118}$$

Equation (118) gives a rough estimation of the Q-factor ratio for seismic waves in the geometric regime. It tells us that in the geometric regime the Q-factor ratio is frequency-independent and solely determined by the wave velocities. This conclusion will be verified through numerical calculations for a series of material combinations, as shown in the following examples.

Velocity fluctuation can be caused either by density fluctuation or by elastic modulus fluctuation. In the following numerical examples, we study the effects of the two types of properties separately. Seismological and geophysical studies show that the density and velocity of rocks in the Earth's lithosphere have a linear relationship, known as Birch's law [3]. Although the specific coefficients appeared in the Birch law strongly depend on the average atomic weight of the rocks, there is a general conclusion that rocks with larger density also have greater longitudinal and transverse velocities. Geophysicists have proposed different velocity models for the substructure in the lithosphere using various seismic tomography techniques [163-165]. It is generally acknowledged that the longitudinal velocity in the lithosphere varies from 5.8 to 8.3 km/s, while the transverse wave velocity varies from 3.5 to 4.7 km/s. In the subsequent calculations, we choose the material properties following these general guidelines. We first study the effects of modulus fluctuation. The material properties of Media I-IV are listed in Tab. 6. It is noted that the material properties of Phase 1 do not change while the elastic moduli of Phase 2 increase gradually, which results in a velocity fluctuation from ±2.5% for Medium I to about ±6% for Medium IV.

Table 6. Material Properties and longitudinal velocities of Media I-IV.

| Medium | $\rho_1$ (kg/m$^3$) | $\rho_2$ (kg/m$^3$) | $\lambda_1$ (GPa) | $\lambda_2$ (GPa) | $\mu_1$ (GPa) | $\mu_2$ (GPa) | $V_{0L}$ (m/s) | $V_{L1}$ (m/s) | $V_{L2}$ (m/s) | $\delta V_{L1}/V_{0L}$ | $\delta V_{L2}/V_{0L}$ |
|---|---|---|---|---|---|---|---|---|---|---|---|
| I | 3000 | 3100 | 50 | 60 | 45 | 50 | 7003.94 | 6831.30 | 7184.21 | -0.0250 | 0.0260 |
| II | 3000 | 3100 | 50 | 60 | 45 | 53 | 7068.83 | 6831.30 | 7317.68 | -0.0336 | 0.0352 |
| III | 3000 | 3100 | 50 | 65 | 45 | 55 | 7161.58 | 6831.30 | 7513.43 | -0.0461 | 0.0491 |
| IV | 3000 | 3100 | 50 | 65 | 45 | 60 | 7261.76 | 6831.30 | 7725.12 | -0.0593 | 0.0638 |

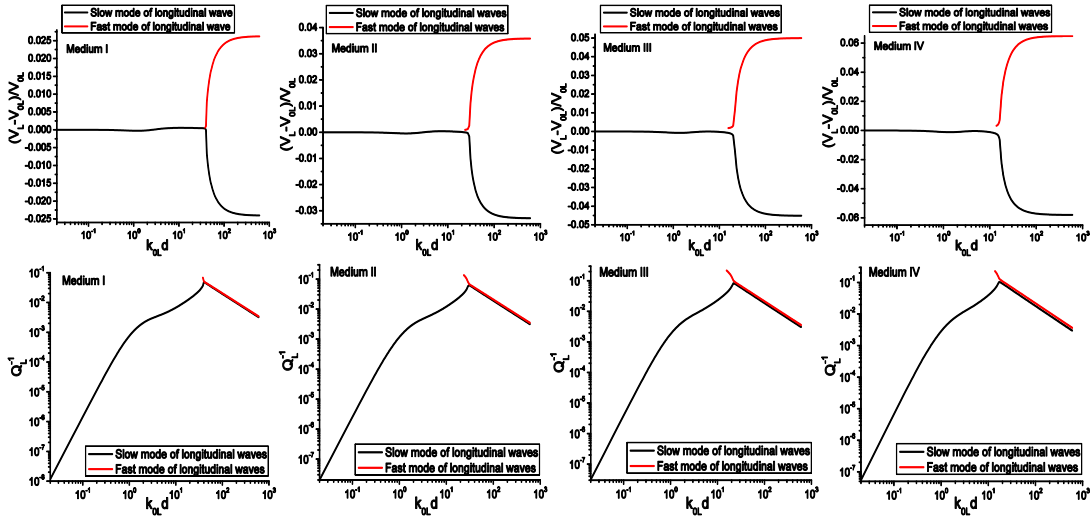

FIG. 17. Effects of elastic moduli fluctuation on longitudinal wave dispersion (first row) and Q factor (second row).

The longitudinal wave dispersion and Q factors of Media I-IV are depicted in Fig. 17. Similar to the examples discussed in Section III, there exists only one propagation mode with very small dispersion at low and intermediate frequencies. At a critical frequency, an additional, faster mode begins to appear. As the frequency continues to increase, the velocity of the fast (slow) mode quickly increases (decreases) and then approaches its geometric limit, i.e., the longitudinal velocity of the fast (slow) phase. At high frequencies, the dispersions of both the fast and slow modes are very small. As a consequence, seismic signals at relatively low frequencies and at high frequencies can propagate for a long distance while maintaining it waveform nearly unchanged. To show the detailed variation at very low frequency and very high frequency, the Q factors are plotted in a double-logarithmic coordinate system, as shown in the second row in Fig.17. At low frequency $k_{0L}d < 1$, the attenuation increases from zero following a power law. In the intermediate frequency range, the Q-factor increases with frequency nonlinearly, then it assumes its maximum soon after the faster mode appears. Then the attenuation decreases with frequency following a negative power law. The Q-factor of the fast mode is very large at its emergence and then decreases dramatically. After the slow mode reaching its summit, it decreases following a negative power law similar to that of the slow mode but with a slightly larger value. With the increase of velocity fluctuation, the frequency of the attenuation peak decreases significantly, from $k_{0L}d$ =41.60 for Medium I to $k_{0L}d$ =17.12 for Medium IV. The value of the peak increases from about 0.047 for Medium I to 0.10 for Medium IV. The Q-factor in the geometric regime $k_{0L}d > 100$ decreases from 0.01 to 0.001.

Table 7. Material Properties and transverse velocities of Media I-IV.

| Medium | $\rho_1$ (kg/m$^3$) | $\rho_2$ (kg/m$^3$) | $\lambda_1$ (GPa) | $\lambda_2$ (GPa) | $\mu_1$ (GPa) | $\mu_2$ (GPa) | $V_{0T}$ (m/s) | $V_{T1}$ (m/s) | $V_{T2}$ (m/s) | $\delta V_{T1}/V_{0T}$ | $\delta V_{T2}/V_{0T}$ |
|---|---|---|---|---|---|---|---|---|---|---|---|
| I | 3000 | 3100 | 50 | 60 | 45 | 50 | 3943.71 | 3872.98 | 4016.10 | -0.0180 | 0.0180 |
| II | 3000 | 3100 | 50 | 60 | 45 | 53 | 4001.69 | 3872.98 | 4134.82 | -0.0322 | 0.0333 |

| | | | | | | | | | | |
|---|---|---|---|---|---|---|---|---|---|---|
| III | 3000 | 3100 | 50 | 65 | 45 | 55 | 4039.03 | 3872.98 | 4212.12 | -0.0411 | 0.0429 |
| IV | 3000 | 3100 | 50 | 65 | 45 | 60 | 4128.13 | 3872.98 | 4399.41 | -0.0618 | 0.0657 |

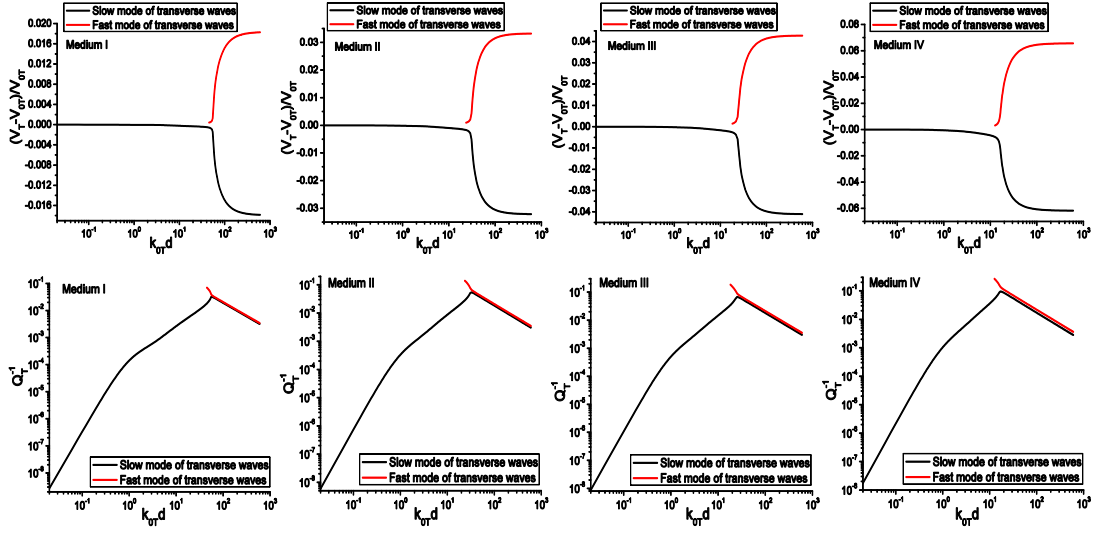

FIG. 18. Effects of elastic moduli fluctuation on transverse wave dispersion (first row) and Q factor (second row).

Figure 18 shows the dispersion and attenuation of transverse waves. It is seen that the dispersion and Q-factors follow a rather similar variation tendency as the longitudinal waves, i.e., there is only one propagation mode at low frequency and two at intermediate to high frequency, and there also exists a peak in the transverse Q-factor curves. To compare the relative magnitude in the same frequency scale, the Q-factors of longitudinal and transverse waves and their ratios are plotted in Fig. 19. The subscripts SL, FL, ST and FT stand for the slow longitudinal mode, the fast longitudinal mode, the slow transverse mode and the fast transverse mode, respectively. For convenience of subsequent discussion, we denote the frequencies at which the Q-factors of the longitudinal and transverse waves assume their peaks by $K_{crL}$ and $K_{crT}$, respectively.

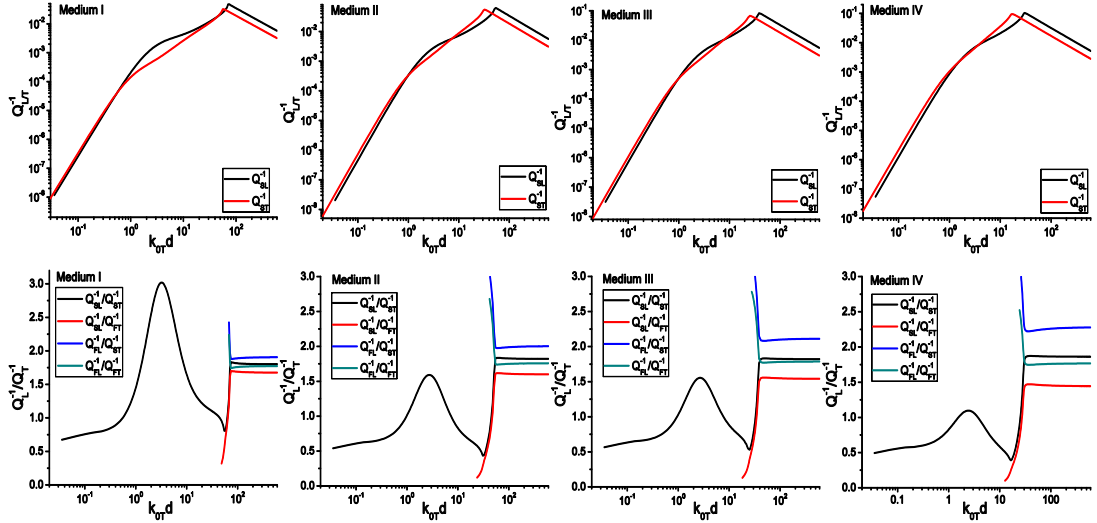

FIG. 19. Comparison of longitudinal and transverse Q-factors (first row) and
the ratios of longitudinal to transverse Q-factors (second row) of Media I-IV.

It is observed that $K_{crL}$ is always greater than $K_{crT}$. At low frequencies, the longitudinal Q-factor is smaller than that of the transverse waves. At frequencies higher than $k_{crL}$, longitudinal Q-factors are always greater than transverse Q-factors, and both of them strictly follow a negative power law. The ratios of the longitudinal to transverse Q-factors have a very complicated profile. At low frequencies, the ratio is relatively small, near 0.5. As the frequency increases, the ratio increases gradually and then reaches a peak. As the frequency continues to increase, the ratio gradually decreases back to less than unity and then decreases sharply to its minimum at $K_{crT}$. After reaching its minimum, the Q-factors increases monotonically and reaches a stable value between 1.5 and 2.5 at $K_{crL}$. Since there are two sets of longitudinal and transverse waves, we have four different Q-factor ratios in the high frequency range. The most prominent feature in the high frequency regime $k_{0L}d > K_{crL}$ is that the ratio keeps constant at four different values. The differences among the four values increase notably as the velocity fluctuations increase from Medium I to Medium IV. It is also worth noting that the peak of the Q-factor ratio decreases from 3 to 1 as the velocity fluctuations increase from Medium I to Medium IV. Table 8 gives the Q-factor ratios in the geometric range and the velocity ratios of the two component phases. It is seen the ratios roughly agrees with the approximate relation given by Eq. (118). As the velocity fluctuation increases from Medium I to Medium IV, the relative error increases from 3.24% to 14.6%.

Table 8. The geometric limits of Q-factor ratios and the velocity ratios of component materials of Media I-IV.

| Medium | $Q_{SL}^{-1}/Q_{ST}^{-1}$ | $Q_{FL}^{-1}/Q_{ST}^{-1}$ | $Q_{SL}^{-1}/Q_{FT}^{-1}$ | $Q_{FL}^{-1}/Q_{FT}^{-1}$ | $V_{L1}/V_{T1}$ | $V_{L2}/V_{T1}$ | $V_{L1}/V_{T2}$ | $V_{L2}/V_{T2}$ |
|---|---|---|---|---|---|---|---|---|
| I | 1.80 | 1.91 | 1.68 | 1.77 | 1.76 | 1.85 | 1.70 | 1.79 |
| II | 1.82 | 2.00 | 1.60 | 1.76 | 1.76 | 1.89 | 1.65 | 1.77 |
| III | 1.82 | 2.11 | 1.54 | 1.79 | 1.76 | 1.94 | 1.62 | 1.78 |
| IV | 1.86 | 2.28 | 1.44 | 1.77 | 1.76 | 1.99 | 1.55 | 1.76 |

Now we study the effects of density fluctuations on the dispersion and attenuation characteristics. For this case, we enforce the velocity of the two component phases be constant, and vary the density and the elastic moduli simultaneously. The material properties of Media V-VIII are given in Tab. 9.

Table 9. Material Properties and longitudinal velocities of Media V-VIII.

| Medium | $\rho_1$ (kg/m³) | $\rho_2$ (kg/m³) | $\lambda_1$ (GPa) | $\lambda_2$ (GPa) | $\mu_1$ (GPa) | $\mu_2$ (GPa) | $V_{0L}$ (m/s) | $V_{L1}$ (m/s) | $V_{L2}$ (m/s) | $\delta V_{L1}/V_{0L}$ | $\delta V_{L2}/V_{0L}$ |
|---|---|---|---|---|---|---|---|---|---|---|---|
| V | 3000 | 3100 | 73.50 | 99.20 | 36.75 | 49.60 | 7474.46 | 7000.00 | 8000.00 | -0.0635 | 0.0703 |
| VI | 3000 | 3300 | 73.50 | 105.60 | 36.75 | 52.80 | 7466.91 | 7000.00 | 8000.00 | -0.0625 | 0.0714 |
| VII | 3000 | 3400 | 73.50 | 108.80 | 36.75 | 54.40 | 7461.77 | 7000.00 | 8000.00 | -0.0619 | 0.0721 |
| VIII | 3000 | 3500 | 73.50 | 112.00 | 36.75 | 56.00 | 7455.85 | 7000.00 | 8000.00 | -0.0611 | 0.0730 |

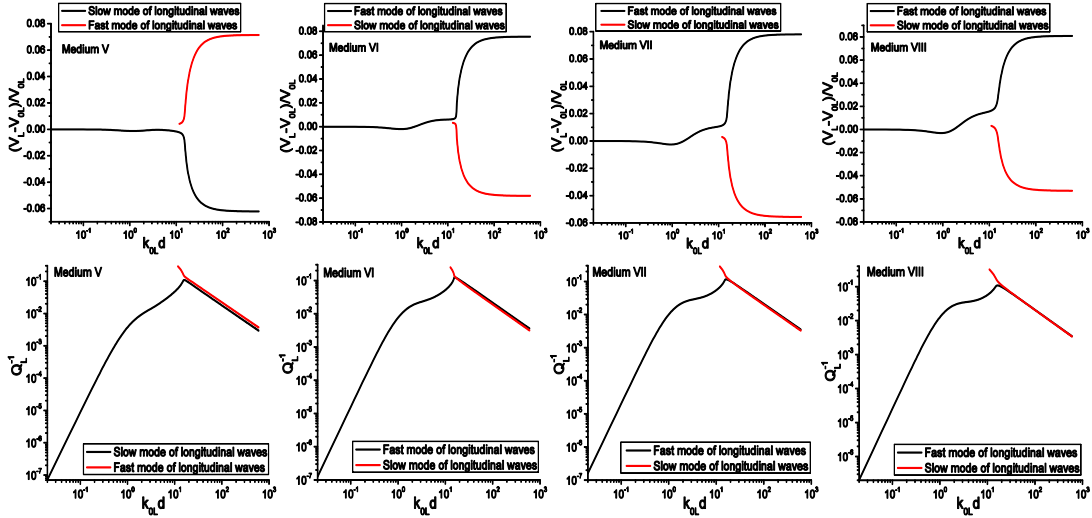

FIG. 20. Longitudinal wave dispersions and Q factors of Media V-VIII.

Figure 20 shows the velocity and Q-factors of the longitudinal waves. As can be seen, as the density fluctuation increases from Medium V to Medium VIII, the dispersion of the longitudinal waves in the intermediate frequency range ($1 < k_{0L}d < 20$) experiences a remarkable increase. Moreover, the structure of the dispersion curves is also changed significantly: For Medium V, the low frequency branch is connected to the slow mode at high frequencies, while for Media VI-VIII, the low frequency branch is connected to the high-velocity branch in the high frequency regime. Changing the density and elastic moduli fluctuation while keeping the velocities of the two component phases constant has less prominent effects on the critical frequencies $K_{crL}$ and the peak values of the Q-factors.

Table 10. Material properties and transverse velocities of Media V-VIII.

| Medium | $\rho_1$ (kg/m³) | $\rho_2$ (kg/m³) | $\lambda_1$ (GPa) | $\lambda_2$ (GPa) | $\mu_1$ (GPa) | $\mu_2$ (GPa) | $V_{0T}$ (m/s) | $V_{T1}$ (m/s) | $V_{T2}$ (m/s) | $\delta V_{T1}/V_{0T}$ | $\delta V_{T2}/V_{0T}$ |
|---|---|---|---|---|---|---|---|---|---|---|---|

| V | 3000 | 3100 | 73.50 | 99.20 | 36.75 | 49.60 | 3742.80 | 3500.00 | 4000.00 | -0.0649 | 0.0687 |
| VI | 3000 | 3300 | 73.50 | 105.60 | 36.75 | 52.80 | 3741.57 | 3500.00 | 4000.00 | -0.0646 | 0.0691 |
| VII | 3000 | 3400 | 73.50 | 108.80 | 36.75 | 54.40 | 3740.38 | 3500.00 | 4000.00 | -0.0643 | 0.0694 |
| VIII | 3000 | 3500 | 73.50 | 112.00 | 36.75 | 56.00 | 3738.85 | 3500.00 | 4000.00 | -0.0639 | 0.0699 |

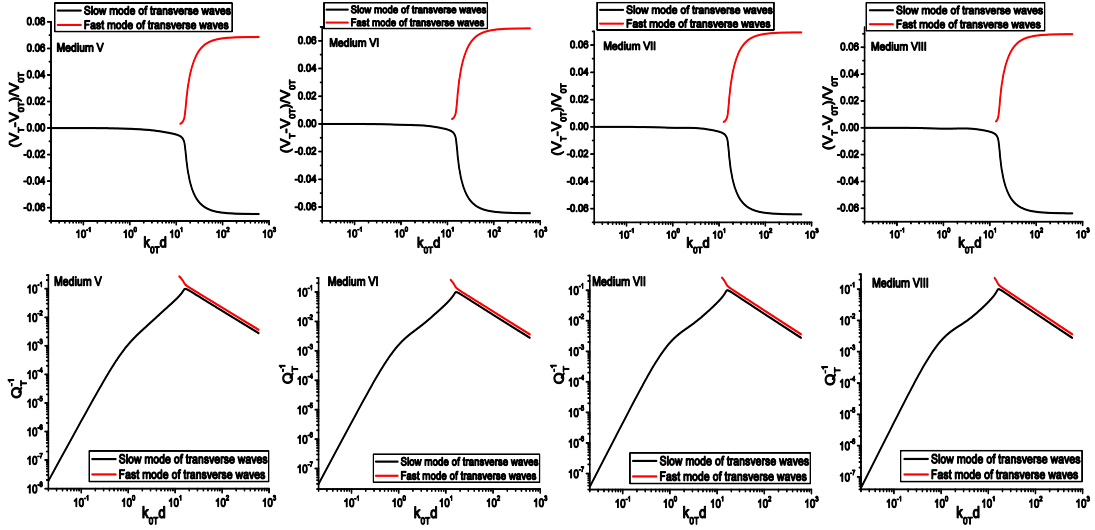

FIG. 21. Transverse wave dispersion and Q factor of Media V-VIII.

The dispersion and Q-factors of transverse waves of Media V-VIII are shown in Fig. 21. Although the overall tendency of velocity and Q-factors are similar to that of longitudinal waves, it is noteworthy that the structure of the dispersion curves does not change as the material property fluctuation changes. For all the cases, the low frequency branch is connected to the branch of the slow mode.

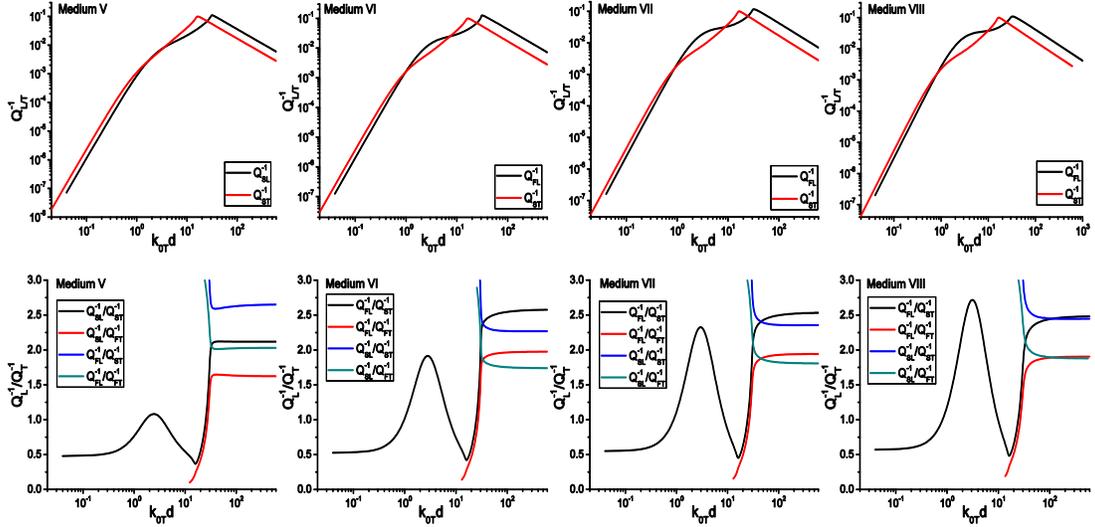

FIG. 22. Comparison of longitudinal and transverse inverse Q-factors (first row) of media with different density fluctuation and ratio of longitudinal to transverse Q-factors (second row).

To show their relative changes, the longitudinal and transverse Q-factors and their ratios are plotted in the same frequency scale, as shown in Fig. 22. As can be seen, the fluctuation in density significantly changes the relative magnitude of the longitudinal and transverse Q-factors in the whole frequency range. The ratio starts from about 0.5 and then increases gradually until reaching its maximum, and then decreases rapidly to a minimum about 0.5 at $K_{crT}$. The maximum value of the ratio is strongly dependent on the property fluctuations. As we can see in the figures, it reaches 1 for Medium V and 2.75 for Medium VIII. At frequencies greater than $K_{crL}$, the Q-factor ratios of the four combinations approach four different high-frequency limits. To validate the approximate relation given by Eq. (118), the Q-factor ratios and the velocity ratios are listed in Tab. 11. It is seen that the relation is roughly satisfied within an average error about 15%.

Table 11. The geometric limits of Q-factor ratios and velocity ratios of component materials of Media V-VIII.

| Medium | $Q_{SL}^{-1}/Q_{ST}^{-1}$ | $Q_{FL}^{-1}/Q_{ST}^{-1}$ | $Q_{SL}^{-1}/Q_{FT}^{-1}$ | $Q_{FL}^{-1}/Q_{FT}^{-1}$ | $V_{L1}/V_{T1}$ | $V_{L2}/V_{T1}$ | $V_{L1}/V_{T2}$ | $V_{L2}/V_{T2}$ |
|---|---|---|---|---|---|---|---|---|
| V | 2.12 | 2.65 | 1.62 | 2.03 | 2.00 | 2.29 | 1.75 | 2.00 |
| VI | 2.27 | 2.57 | 1.74 | 1.98 | 2.00 | 2.29 | 1.75 | 2.00 |
| VII | 2.36 | 2.53 | 1.81 | 1.94 | 2.00 | 2.29 | 1.75 | 2.00 |
| VIII | 2.45 | 2.48 | 1.88 | 1.91 | 2.00 | 2.29 | 1.75 | 2.00 |

The materials considered in the above examples exhibit various degrees of property fluctuations, which lead to a velocity fluctuation ranging from ± 2.5% to ± 7%. In the following examples, we increase the property fluctuation of the component phases such that the fractional velocity fluctuation reaches up to ± 15%. We consider a series of media with properties chosen randomly, for which the longitudinal velocity of the slow phase ranges from 5700 m/s to 6200 m/s, the longitudinal velocity of the fast phase varies from 7800 m/s to 8300 m/s, the transverse velocity of the slow phase ranges from 3300 m/s to 3600 m/s, and the transverse velocity of the fast phase varies from 4200 m/s to 4600 m/s. The material parameters are listed in Tabs. 12 and 13. We need to point out that the Medium XVI has much larger velocities and will be used as an exemplar model for the next section.

Table 12. Material Properties of Media IX-XVI.

| Medium | $\rho_1$ (kg/m³) | $\rho_2$ (kg/m³) | $\lambda_1$ (GPa) | $\lambda_2$ (GPa) | $\mu_1$ (GPa) | $\mu_2$ (GPa) |
|---|---|---|---|---|---|---|
| IX | 3000 | 3100 | 42.00 | 75.00 | 35.00 | 65.00 |
| X | 3000 | 3300 | 42.00 | 72.00 | 35.00 | 67.00 |
| XI | 3000 | 3500 | 42.00 | 85.00 | 35.00 | 70.00 |
| XII | 3000 | 3500 | 40.00 | 100.00 | 35.00 | 65.00 |
| XIII | 3000 | 3600 | 40.00 | 115.00 | 35.00 | 67.00 |
| XIV | 3000 | 3200 | 29.16 | 53.77 | 39.42 | 58.78 |
| XV | 3800 | 4000 | 170.16 | 181.14 | 81.91 | 123.46 |
| XVI | 3000 | 3300 | 30.42 | 70.67 | 33.87 | 67.12 |

Table 13. Longitudinal and transverse velocities of Media IX-XVI.

| Medium | $V_{0L}$ (m/s) | $V_{L1}$ (m/s) | $V_{L2}$ (m/s) | $\delta V_{L1}/V_{0L}$ | $\delta V_{L2}/V_{0L}$ | $V_{0T}$ (m/s) | $V_{T1}$ (m/s) | $V_{T2}$ (m/s) | $\delta V_{T1}/V_{0T}$ | $\delta V_{T2}/V_{0T}$ |
|---|---|---|---|---|---|---|---|---|---|---|
| IX | 7036.44 | 6110.10 | 8131.96 | -0.1317 | 0.1557 | 3957.52 | 3415.65 | 4579.05 | -0.1369 | 0.1571 |
| X | 6933.59 | 6110.10 | 7900.90 | -0.1188 | 0.1395 | 3923.90 | 3415.65 | 4505.89 | -0.1295 | 0.1483 |
| XI | 6971.26 | 6110.10 | 8017.84 | -0.1235 | 0.1501 | 3906.41 | 3415.65 | 4472.14 | -0.1256 | 0.1448 |
| XII | 6960.73 | 6055.30 | 8106.43 | -0.1301 | 0.1646 | 3834.46 | 3415.65 | 4309.46 | -0.1092 | 0.1239 |
| XIII | 7028.99 | 6055.30 | 8316.65 | -0.1385 | 0.1832 | 3834.91 | 3415.65 | 4314.06 | -0.1093 | 0.1249 |
| XIV | 6619.90 | 6000.00 | 7317.07 | -0.0936 | 0.1053 | 3940.97 | 3625.00 | 4285.71 | -0.0802 | 0.0875 |
| XV | 9836.66 | 9375.00 | 10344.83 | -0.0469 | 0.0517 | 5080.99 | 4642.86 | 5555.56 | -0.0862 | 0.0934 |
| XVI | 6696.80 | 5720.00 | 7880.00 | -0.1459 | 0.1767 | 3892.83 | 3360.00 | 4510.00 | -0.1369 | 0.1585 |

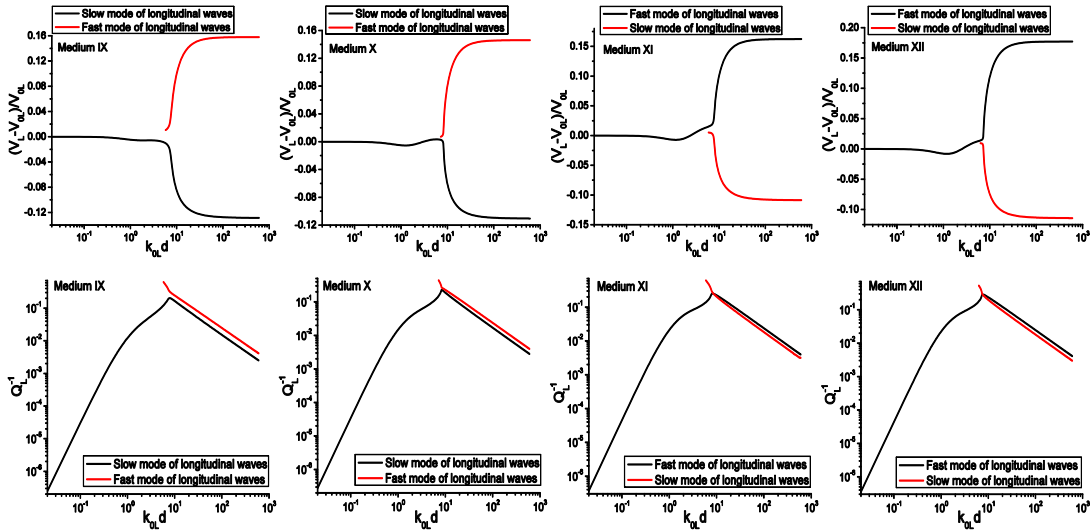

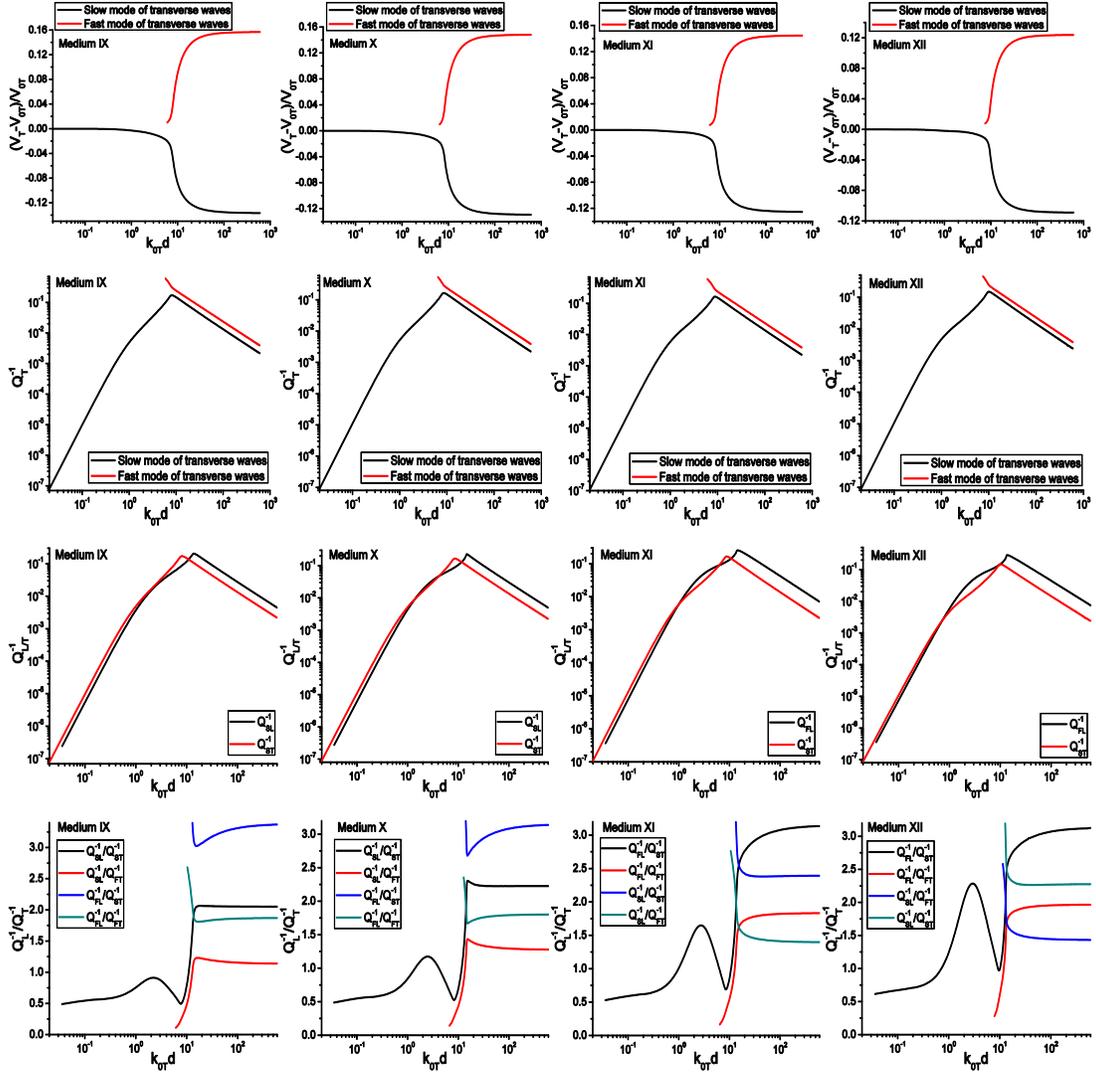

FIG. 23 Dispersion, Q-factors and Q-factor ratios of longitudinal and transverse waves in Media IX-XII

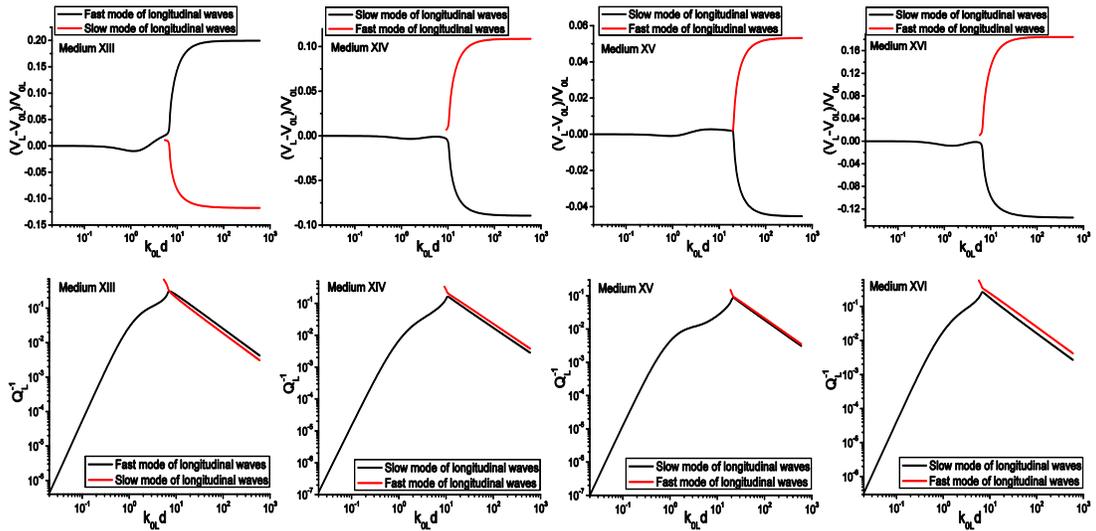

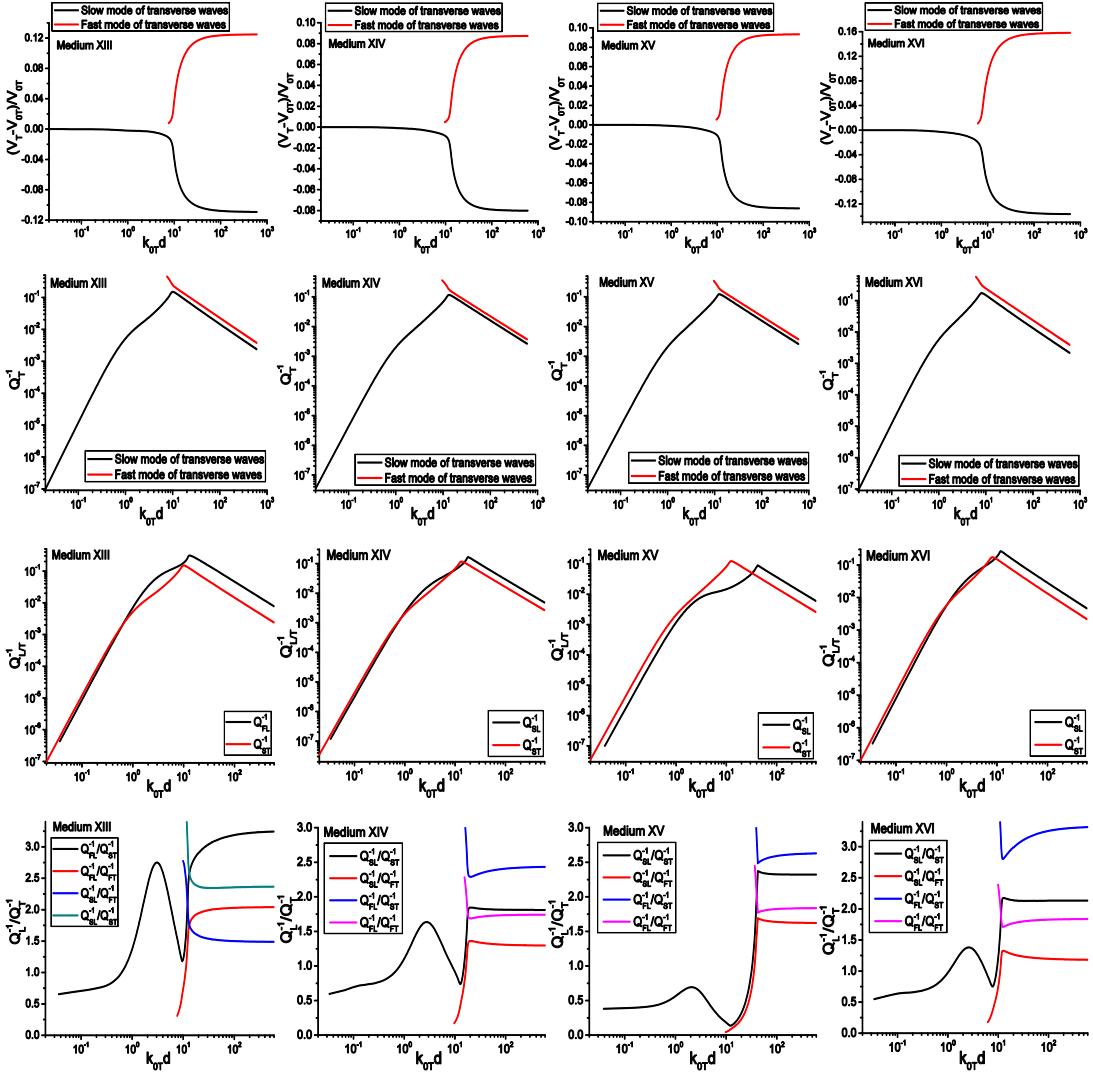

FIG. 24 Dispersion, Q-factors and Q-factor ratios of longitudinal and transverse waves in Media XIII-XVI

The longitudinal and transverse wave dispersion, Q-factors and the Q-factor ratios of Media IX-XVI are presented in Figs. 23 and 24. From the numerical results, we can draw the following conclusions: In the low to intermediate frequency range, i.e., $0 < k_{0L}d < 8$ for longitudinal waves and $0 < k_{0T}d < 8$ for transverse waves, there is only one longitudinal and transverse mode. The velocities of these propagation modes are nearly the same as that of the homogeneous reference media. The dispersion of these modes at low frequency $k_{0L}d < 1$ or $k_{0T}d < 1$ is negligible, while in the frequency range $1 < k_{0L}d, k_{0T}d < 8$ the dispersion increases slightly but still very small. In the low frequency range $k_{0L}d < 1$ or $k_{0T}d < 1$, the Q-factors of these media increase from $10^{-7}$ to 0.01 following a power law. In the frequency range $1 < k_{0L}d, k_{0T}d < 8$, the Q-factors increase with frequency following a complicated nonlinear law, and the relative magnitude of the longitudinal and transverse Q-factors strongly depend on the material properties. We can also find that the Q-factors increase from 0.01 to 0.05 in this range. In the moderately high frequency range $8 < k_{0L}d, k_{0T}d < 30$, an additional longitudinal mode and transverse mode starts to appear, while the dispersions of both modes are extremely large. The velocities of the two sets of longitudinal and transverse waves approach that of the fast and slow phases, respectively. The Q-factors in this range first increase from 0.05 to a peak value ranging from 0.1 to 0.2, and then decrease to about 0.05 following an inverse power law. In the high frequency range $k_{0L}d, k_{0T}d > 30$, the velocities of both modes of longitudinal and transverse waves asymptotically approach their geometric limits and the dispersion is nearly negligible. The Q-factors in this range decrease from 0.05 to 0.001 or even smaller following a negative power law. The figures in which the longitudinal and transverse Q-factors plotted in the same scale show that the Q-factor of longitudinal waves is always smaller than that of the transverse waves at low frequencies $k_{0L}d, k_{0T}d < 1$, and greater than the transverse Q-factors in the high frequency range $k_{0L}d$, $k_{0T}d > 30$. The critical frequency at which the longitudinal Q-factors reaching their summits is always greater than that of the transverse Q-factors, i.e. $K_{crL} > K_{crT}$. The Q-factor ratios exhibit a rather complicated pattern. At low frequencies $k_{0L}d$, $k_{0T}d < 1$, the Q-factor ratios increase gradually from about 0.5. As the frequency continues to increase, the Q-factor ratios increase rapidly and reaches a summit ranging from 1.0 to 3.0 at $k_{0T}d = 2\sim3$. After assuming its summit, the Q-factors decrease rapidly until reaching the minimum at $K_{crT}$. The minimum value is also strongly dependent on the material properties, which normally ranges from 0.5 to 1.2. At frequencies higher than $K_{crT}$, The Q-factor increases rapidly again. Since in the high frequency range, there are two sets of longitudinal and transverse waves, there are four different combination of Q-factor ratios. As is evident from the figures, the high-frequency limits of the ratios vary from 1.1 to 3.2, for which the range is much larger than that of Media I-VIII. It is worth noting that the high-frequency limits of the four ratios have the following relation: $Q_{FL}^{-1} > Q_{SL}^{-1} > Q_{FT}^{-1} > Q_{ST}^{-1}$. This conclusion informs us that for high frequency seismic waves and at large epicentral distances, the amplitude of the longitudinal waves is smaller than that of transverse waves. Moreover, the slow transverse wave has the largest amplitude and carries the largest amount of energy, which is consistent with observations in real earthquakes. Table 14 shows the geometric limits of Q-factor ratios and velocity ratios of component materials of Media IX-XVI. It is seen the discrepancies between the numerical results and the approximation formula Eq. (118) are relatively large compared with that for Media I-VIII, especially for the discrepancy between $Q_{FL}^{-1}/Q_{ST}^{-1}$ and $V_{L2}/V_{T1}$. This is because the fractional velocity fluctuations for Media IX-XVI are larger than that for Media I-VIII.

Table 14. The geometric limits of Q-factor ratios and velocity ratios of component materials of Media IX-XVI.

| Medium | $Q_{SL}^{-1}/Q_{ST}^{-1}$ | $Q_{FL}^{-1}/Q_{ST}^{-1}$ | $Q_{SL}^{-1}/Q_{FT}^{-1}$ | $Q_{FL}^{-1}/Q_{FT}^{-1}$ | $V_{L1}/V_{T1}$ | $V_{L2}/V_{T1}$ | $V_{L1}/V_{T2}$ | $V_{L2}/V_{T2}$ |
|---|---|---|---|---|---|---|---|---|
| IX | 2.050 | 3.370 | 1.140 | 1.870 | 1.789 | 2.381 | 1.334 | 1.776 |
| X | 2.227 | 3.139 | 1.278 | 1.802 | 1.789 | 2.313 | 1.356 | 1.753 |
| XI | 2.393 | 3.136 | 1.398 | 1.832 | 1.789 | 2.347 | 1.366 | 1.793 |
| XII | 2.275 | 3.126 | 1.433 | 1.966 | 1.773 | 2.373 | 1.405 | 1.881 |
| XIII | 2.366 | 3.244 | 1.488 | 2.040 | 1.773 | 2.435 | 1.404 | 1.928 |
| XIV | 1.810 | 2.433 | 1.296 | 1.742 | 1.655 | 2.019 | 1.400 | 1.707 |
| XV | 2.322 | 2.630 | 1.622 | 1.837 | 2.019 | 2.228 | 1.688 | 1.862 |
| XVI | 2.134 | 3.315 | 1.182 | 1.837 | 1.702 | 2.345 | 1.268 | 1.747 |

Through the above numerical examples, we see the dispersion and attenuation are very sensitive to the material property fluctuations. Both density and elastic moduli have significant influence on the propagation characteristics of seismic waves. All these theoretical results help us gain important insight into the dispersion and attenuation characteristics of seismic waves measured in local, regional and global earthquakes. In subsequent sections, the theoretical results will be applied to explain seismic data collected worldwide.

*1. Explanation of the observed Q-factors and Q-factor ratios*

Attenuation of the direct arrivals of the longitudinal and transverse waves has been the central topic of nemerious seismological studies ever since the beginning of modern seismology, interested readers please see the manograph by Sato and collaborators [3] and references therein. The longitudinal and transverse Q-factors and their ratios of a series of local and regional earthquakes occurred worldwide are summarized in monograph [3]. For the convenience of subsequent diacussion, the results are presented in Fig. 25.

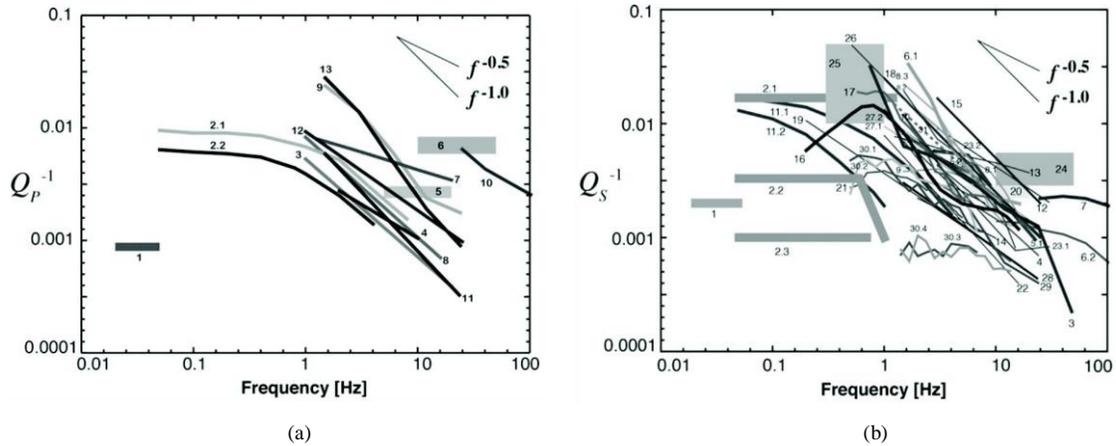

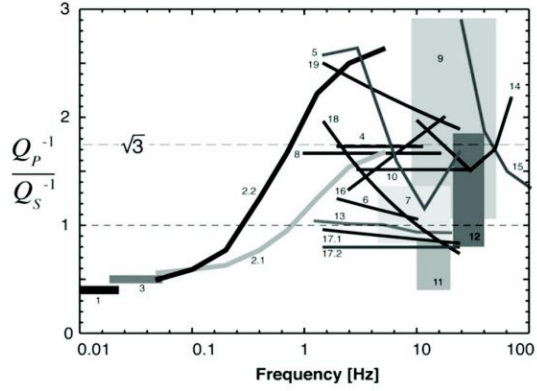

(c)

FIG. 25. Q-factors of longitudinal (a) and transverse (b) waves, and their ratios (c) measured in local, regional and global earthquakes [3].

It needs to mention that the notations $Q_P^{-1}$ and $Q_S^{-1}$ denote the Q-factors of primary and secondary waves, corresponding to the Q-factors of the longitudinal and transverse waves in this work. Before comparing the measured Q-factors with the theoretical predictions, we first give a rough estimation of the dimensionless frequency to which the measured curves correspond. As pointed out before, the average velocities of the longitudinal and transverse waves are about 7 km/s and 4 km/s, and the majority of the measured curves lie between 1 to 100 Hz. The correlation length, i.e., the characteristic dimension of the rocks in the lithosphere lies between several hundred meters to tens of kilometers [5, 8], here we use $d =10$ km for the estimation. Calculation using the above data gives $k_{0L}d$ = 9-900 and $k_{0T}d$ = 16-1600, thus we can see most seismic signals lie in the intermediate to high frequency ranges. In this regime, both the longitudinal and the transverse Q-factors increase from about 0.001, reaching their peaks lying between 0.05 to 0.2, and then decrease monotonically to about 0.001. In the high frequency regime, the Q-factor curves follows an inverse power law, i.e., decreasing quasi-linearly in the double-logarithmic coordinate system. Comparing the theoretical predictions with the measured curves, we see all these theoretical predictions show excellent agreement with that measured in local and regional events. Both the longitudinal and transverse Q-factors decrease following an inverse power law in the frequency band $1 < f < 100$ Hz. Meanwhile, the magnitude of the Q-factors lies in the range of 0.1 to 0.001, which is also consistent with the theoretical calculations. In addition, it is observed that the Curve 16 exhibits a sharp peak at around 1 Hz. This curve partially confirms Aki's conjecture that there should exist a peak in the Q-factor curves in the intermediate frequency range [3]. Extending all the measured curves to the intermediate frequencies $0.05 < f < 1$Hz, we can infer that the peaks of the longitudinal and transverse Q-factor curves lie between 0.1 and 0.2. Our theoretical predictions, as shown in Figs. 17-24, along with this reasonable inference fully support Aki's conjecture. At this point, we are ready to give an explanation on why the attenuation peaks are rarely captured in practical measurements. From the theoretical calculation, we can see the peak occurs at a critical frequency at which the dispersion curve branches. Correspondingly, the waves with a central frequency near the critical frequency have both very large dispersion and attenuation. As a consequence, the wavepackets in this range quickly broaden and decay as the epicentral distance increases, and finally dismissed in the ambient noises. Therefore, it is an extremely challenging task for seismic recording systems to identify weak signals near the critical frequency. We also note that the curves numbered 1, 2.1, 2.2 and 2.3 lie in the low to intermediate frequency regimes. Due to lack of accurate data, they are marked with thick lines. It was pointed out that attenuation at low frequency is extremely difficult to be measured accurately [3].

The measured Q-factor ratios are shown in Fig. 25(c). At first glance, we may find it is difficult to extract any law by which these curves abide. After careful inspection, we find the curves can be classified into six categories according to their variation tendencies: Low frequency stage: Curves 1, 3; Ascent stage: Curves 2.1, 2.2, 16; Descent stage: Curves 5, 15, 18; Turning stage: Curves 5, 14, Plateau stage: Curves 4, 6, 8, 10, 13, 17.1, 17.2, 19. Surprisingly, we can establish a rough correspondence between these curves and the theoretical curves: Curves 1 and 3 correspond to the low frequency range in which the ratio is nearly constant and lies near 0.5; Curves 2.1, 2.2 and 16 correspond to the ascendant stage in which the ratio raises from 0.5 to 1.5, 2 or 3, depending on the specific material property fluctuation; Curves 5, 18 correspond to the descendant stage; Curves 5, 14 correspond to the turning stage in which the ratio decreases to its minimum and then increases; Curves 4, 6, 8, 10, 13, 17.1, 17.2 and 19 correspond to the plateau stage in which the ratio lies between 1 and 3. These curves correspond to the high frequency range, also called the saturated scattering region, in which the ratios are stabilized at four different constants. The shaded region 9 also cover the range of high-frequency limits of the four Q-factor ratios. It need to mention the Curves 17.1 and 17.2 are for the ratios of longitudinal waves to the SV and SH components of transverse waves, respectively, which are less than unity. However, if recalculated for the complete transverse waves, it is highly possible that it will give a value greater than unity. This is because at high frequencies, the attenuation of longitudinal waves is larger than that of transverse waves. The existence of Q-factor ratios with magnitudes less than 1.5 or near unity in the high frequency range strongly suggests that the lithosphere is a heterogeneous medium with strong property fluctuation, as evident in Tab. 14. Through the comparison we see the new model can give an accurate quantitative prediction of the longitudinal and transverse attenuations and their ratios. The rich variety of the tendency of the Q-factor ratios is also perfectly captured in the new model. At this point we need to mention the scattering theory proposed in [10] based on the Born approximation predicts that the Q-factor curve increases monotonically with frequency, which is in contradiction with the measured seismic attenuation. We also note the curves in the high-frequency regime, for which the magnitude of the ratio is a constant near unity, along with the curves in the descendent stage and the turning stage, cannot be explained using the travel-time corrected Born approximation [3].

At last, we try to give a rough estimate of the correlation lengths of the portion of the lithosphere that corresponds to each curve of the Q-factor ratios in Fig. 25(c). Although the information for the velocities is missing, existing seismic data recorded in numerous seismic events reveal that the reference velocities of the transverse waves lie in the range from 3800 to 4000 m/s. For the purpose of a primary evaluation, we use $V_{0T}$ = 3900 m/s. From Figs. 23-24, we can observe that the peak of the Q-factor ratios for most medium models occurs at $K_{0T}$ = 2-3, here we use $K_{0TMAX}$ = 2.5. The minima of the Q-factor ratios occur in the range $8 < K_{0T} < 15$, here we choose $K_{0TMIN}$ = 10. The high-frequency regime begins roughly from $K_{0T}$ = 20, where the Q-factor ratios either have a very small slope or keep constant. Based on these preliminary knowledge, we can find that the curves 2.1 and 2.2 lie in the range $0 < K_{0T} < 2.5$, and at $f \approx 5$ Hz they reach the maximum value. Substitution of the above data into the following relation: $2\pi f$ (= 5 Hz) d/$V_{0T}$ = 2.5, we obtain d ≈ 310 m. The curves 5 and 18 lie in the adimensional frequency range $2 < K_{0T} < 20$, while the corresponding frequency range is $1 < f < 30$ Hz. Solving the equations: $2\pi f$ (=1 Hz) d/$V_{0T}$ = 1 and $2\pi f$ (=30 Hz) d/$V_{0T}$ = 20, we get d = 414~621 m. The curves 4, 8, 10, 13, 17.1, 17.2, and 19 all lie in the high-frequency range, i.e., $K_{0T} \geq 20$, while the corresponding frequency range is $f > 1$~2 Hz. Substituting these data into the relation: $2\pi f$ (=1 or 2 Hz) d/$V_{0T}$ = 20, we get d ≥ 12.42 km or d ≥ 6.21 km, respectively. We can see the estimates of the correlation lengths (characteristic dimension of the inhomogeneities) given by the new multiple scattering theory exactly lie in the range determined by other approaches, such as the transverse coherence function (TCF) and angular coherence function (ACF) analysis [8].

## *2. Does the Mohorovičić discontinuity really exist?*

On October 8, 1909, an earthquake with the macroseismic intensity VIII °MCS struck Pokupsko, 40 km southeast of Zagreb, Croatia [166]. Mohorovičić collected the seismograms recorded by numerous European seismological stations and plotted the travel time curves, as shown in Fig. 26(a). He immediately recognized that there are two sets of longitudinal and transverse waves. To clearly indicate these phases, we replot the travel time curves in Fig. 26(b). From the travel-time curve, we can see four different arrivals, denoted by Pn, Pg, Sn and Sg, respectively. The velocities of the four phases can be roughly estimated from the curves, $V_{Pn}$ = 8230 m/s, $V_{Pg}$ = 5680 m/s, $V_{Sn}$ = 4640 m/s, $V_{Sg}$ = 3310 m/s. It is also noted that the Pn and Sn phases begin to appear from the epicentral distance d = 300 km and propagate a long distance even longer than d = 2600 km, the Pg and Sg phases exist in the range from the epicenter to an epicentral distance d = 1500 km.

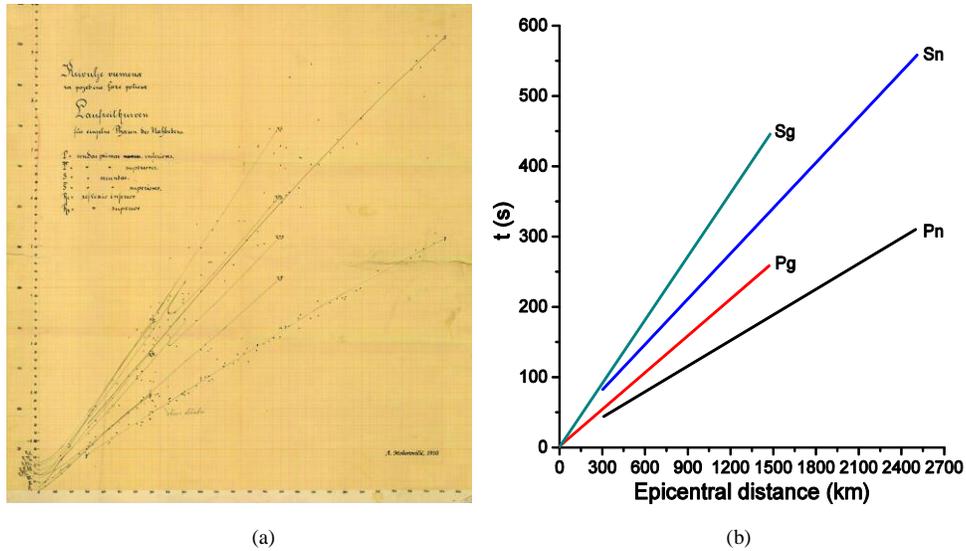

FIG. 26. Mohorovičić travel-time curve (a) (Courtesy of the Department of Geophysics, Faculty of Science, Zagreb) [166] and (b) best fitting curves

Since then, more and more seismograms recorded in the local, regional and global earthquakes show that two sets of longitudinal and transverse waves can be observed. Figure 27 shows the travel-time curves of the Pn, Pg, Sn and Sg phases recorded by observation stations across the Eurasian continent, including the eastern, central and northern Russia, and China [167-168]. The velocities of the four phases can be evaluated from the curves as: $V_{Pn}$ = 8330 m/s, $V_{Pg}$ = 6250 m/s, $V_{Sn}$ = 4630 m/s, $V_{Sg}$ = 3520 m/s. Different from the travel-time curves shown in Fig. 26, all phases propagate a long distance greater than 1500 km. The Sg phase propagates an exceedingly large distance, even longer than 2500 km. It is worth noting that both the Pn and Pg phases are difficult to discriminate in the range from 0 to 500 km, and start to separate from each other at distances greater than 500 km. Similar phenomena can be observed for the Sn and Sg phases.

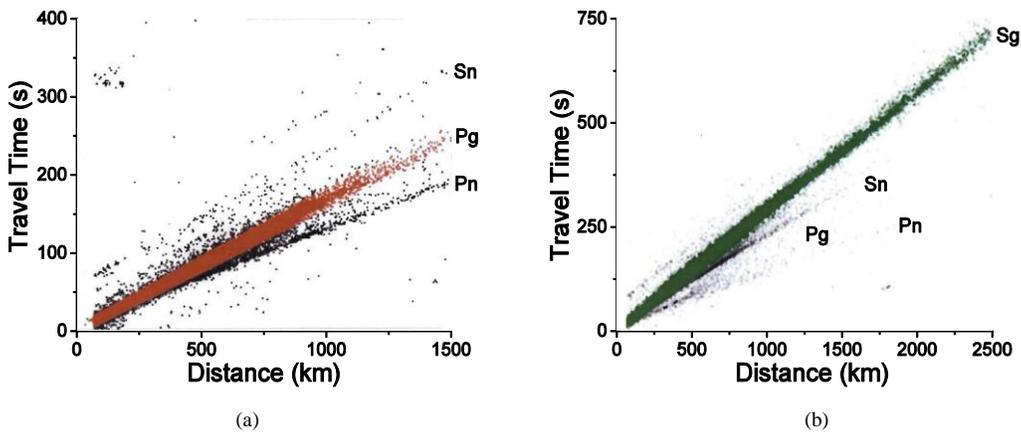

FIG. 27. Travel time-distance plots of the Pg, Pn, Sg, Sn phases in Eurasia [167-168]. For Pg, 1453318 traveltimes from 3709 stations and 407131 events are recorded, and for Sg, 1049125 traveltimes from 3084 stations and 266751 events are recorded.

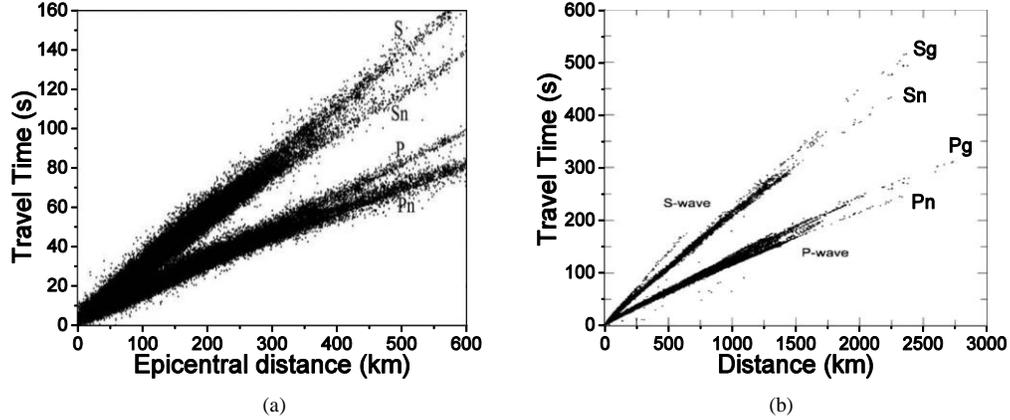

(a)                                                        (b)

FIG. 28. Travel time-distance plot of the Pn, Pg, Sn and Sg phases in (a) Southwest China [169] (139462 Pn and Pg traveltimes and 102184 Sn and Sg traveltimes from 15316 events) and (b) Japan [170] (207000 arrival times from 7743 shallow and deep earthquakes and 34148 arrival times from 333 teleseismic events recorded by over 1000 seismic stations).

Figure 28(a) show the travel-time curves of 15316 earthquakes occurred in southwest China, which is mainly a tectonically active highland region, including southeast Tibet, Sichuan and Yunnan provinces [169]. It provides a textbook example for the four seismic phases in which the four seismic phases can be explicitly identified. At distances less than 350 km, there is only one longitudinal and one transverse wavefront can be observed. Starting from the epicentral distance d = 350 km, the wavefronts of P waves and S-waves begin to split. The velocities of the four phases can be estimated from the curves as: $V_{Pn}$ = 7320 m/s, $V_{Pg}$ = 6000 m/s, $V_{Sn}$ = 4300 m/s, $V_{Sg}$ = 3600 m/s. All the four phases extended to more than 600 km. Figure 28(b) presents the travel-time curves collected from 7743 shallow and deep earthquakes and 333 teleseismic events recorded by over 1000 seismic stations on the Japan Islands [170]. The velocities of the P- and S-waves are evaluated from the curves as: $V_{Pn}$ = 10300 m/s, $V_{Pg}$ = 9400 m/s, $V_{Sn}$ = 5500 m/s, $V_{Sg}$ = 4600 m/s, which are much higher than that of other portions in the world. We also note all the four phases travel a long distance up to 2500 km.

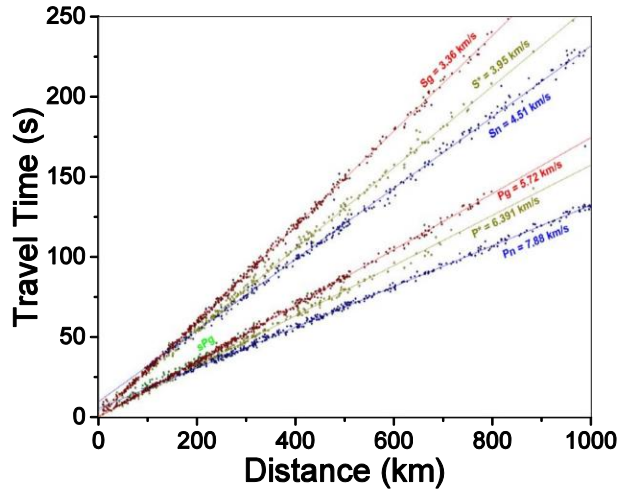

FIG. 29. Travel-time curves of the Pn, P*, Pg, Sn, S* and Sg phases in eastern Russia [171].
The traveltimes are recorded from 92 Sakhalin earthquakes.

Figure 29 shows the travel-time curves and their linear regression obtained from 92 Sakhalin earthquakes in eastern Russia [171]. The most prominent feature is that in addition to the conventional four phases Pn, Pg, Sn, and Sg, two novel phases denoted by P* and S* are clearly observed. The velocities of the six phases are extracted from the linear regression as: $V_{Pn}$ = 7880 m/s, $V_{P*}$ = 6391 m/s, $V_{Pg}$ = 5720 m/s, $V_{Sn}$ = 4510 m/s, $V_{S*}$ = 3950 m/s, $V_{Sg}$ = 3360 m/s. It is seen that the velocities of the P* (S*) phase lies between the Pn and Pg (Sn and Sg) phases.

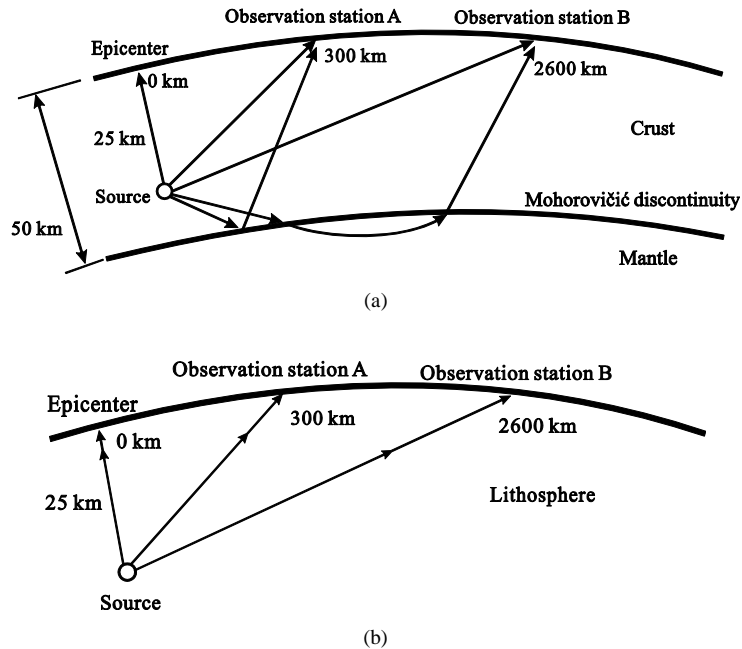

FIG. 30. Schematics for Mohorovičić's explanation (a) and new explanation based on the SFMS theory (b) for the observed two sets of longitudinal and transverse waves.

Mohorovičić denied the possibility that the two sets of longitudinal and transverse waves are emanated from the same source and traverse the same path. Instead, he proposed an explanation based on a layered model of the lithosphere.

**Mohorovičić's explanation**: There exists an interface about 50 km beneath the Earth's surface, which separates two layers of elastic materials (rocks) with dramatically different properties. The upper layer is relatively soft and has a longitudinal velocity about 6.0 km/s, the lower material is relatively hard and has a longitudinal velocity about 8.0 km/s. He further assumed that the wave velocity increases with depth following an exponential law, of which he thought as a natural result of the gravitation. When a shallow earthquake occurs, the seismic waves propagate to the observation station along two paths. One propagates upward and arrives at the station directly. The other propagates downward, when impinging at the discontinuity, it gets refracted and portion of the wave propagates in the hard medium. As the velocity of the material increases with depth, the wave gradually reflected back and gets refracted again at the discontinuity. After that, the wave propagates in the upper medium again and arrives at the observation station. A schematic diagram for the paths of the seismic rays are given in Fig. 30 (a).

As similar phenomena are observed in the travel-time curves extracted from numerous local, regional and teleseismic earthquakes, as shown in Figs. 27-29, seismologists infer that there must be a global discontinuity existing in the lithosphere. In honor of its discoverer, this discontinuity is named the Mohorovičić discontinuity. Since then, the two-layered model of the lithosphere is accepted worldwide. The top layer with relatively low velocity is called the crust, and the P- and S-waves propagating in this layer are denoted by Pg and Sg, respectively, while the bottom layer with relatively high velocity is called the mantle, and the waves propagating in this layer are denoted by Pn and Sn [164]. The Mohorovičić discontinuity is identified as the interface between the crust and the mantle. In 1925, Victor Conrad first identified a new set of longitudinal and transverse waves, denoted by P* and S*, when he analyzed the eastern Alps earthquake in 1923 [130, 172-173]. Since the new phases propagate at an intermediate velocity between that of the upper crust and mantle, Conrad postulated that the waves propagate in the lower crust and there must be an additional interface which separates the upper and lower crust. The new phases are also observed worldwide as shown in Fig. 29. Consequently, the interface between the upper and lower crust is named as the Conrad discontinuity. Later on, additional phases are identified in local and regional earthquakes and correspondingly, new discontinuities such as the Riel Discontinuity and the Fortsch discontinuity are identified. Currently, the multi-layered model of the lithosphere is widely used in the seismological community for seismic tomography and inversion for the velocity structure [164-165].

Despite its success in explaining the coexistence of two or more sets of body waves, disputes about the multi-layered model along with the existence of the Mohorovičić discontinuity and the Conrad discontinuity never end in the scientific community [130, 132, 173].

Modern seismology improves the estimation of the velocity distribution and predicts the velocities of primary seismic waves (P-waves) immediately above the Moho are about 6.7–7.2 km/s, and below the Moho they are about 7.6–8.6 km/s [164]. One naturally raises the question: what is the physical mechanism that causes such an abrupt change in velocity, phase transition or change in chemical composition? It is pointed out that phase transition induced by temperature increase results in a transition zone with gradually varying material properties instead of a sharp interface [130]. Thus, geophysicists suggest that the Moho marks a change in chemical composition. From the knowledge of petrology, we know the longitudinal velocity immediately above the Moho is consistent with that of basalt, while the velocity immediately below the Moho is similar to those of peridotite or dunite. Nevertheless, any believable sorting mechanism gives an irregular surface instead of an even interface [130, 132]. Consequently, both the mechanisms encounter difficulties in explaining such an even and sharp discontinuity. Moreover, as discussed above, the amplitudes of all these waves have obvious attenuation besides the geometric spreading attenuation. The theories based on the geometric ray and the layered model can hardly give a reasonable explanation for the apparent attenuation. The seismic waves penetrating into the mantle have a rather complicated path, and the amplitude attenuation in each layer, of which the mechanical properties vary exponentially with depth, is extremely difficult to evaluate. Finally, there are a series of coda waves which are also decaying and may last for tens of minutes. The multi-layered model encounters significant difficulties in explaining these stochastic coda waves. Through the above discussion, we see a number of phenomena are extremely difficult to explain in the framework of the multi-layered model. Contrarily, all these questions can be answered consistently and quantitatively in the framework of the new multiple scattering theory.

**New explanation**: Modern petrological study reveals that the lithosphere is composed of a variety types of rocks, such as granite, basalt, limestone, sandstone, shale, and serpentinite. Each type of rocks has its unique mass density and elastic moduli. Consequently, the lithosphere is a heterogeneous medium with certain degree of velocity fluctuation. As a result of billion years of geological activities, including collisions among tectonic plates, subduction of sea slabs, volcanic activities, etc., there are active mass exchanges in the lithosphere, both horizontally and vertically, which lead to a statistically uniform distribution of the inhomogeneities. The fluctuation in materials properties results in a fast longitudinal velocity lying between 7.6-8.3 km/s and a slow longitudinal velocity lying between 5.7-6.3 km/s. The characteristic size of the inhomogeneities ranges from several hundred meters to tens of kilometers. When an earthquake occurs, a seismic signal with a duration from several seconds to a few minutes is launched. Depending on the source mechanism, the seismic waves may be composed of waves with different frequencies, ranging from low frequency, intermediate frequency to high frequency. The wave packet gets multiply scattered during its travel from the source to the observation stations. As shown in the numerical examples given in Figs. 23-24, both the wave components at low or intermediate frequencies and at high frequencies have small dispersion. As a result, the wave packets in these frequency ranges can propagate for a long distance. Contrarily, wave packets with a frequency near the critical frequency at which the dispersion curves branches have both large dispersion and attenuation. Consequently, these wave packets can only propagate a short distance and then dismissed in the background noises. At observation stations near the epicenter, waves at all frequencies are mixed and the steady multiple-scattered waves have not been established. Consequently, there is only one wave packet can be observed in the first two hundred kilometers. After propagating for a certain distance (about two hundred kilometers), stable multiple scattered waves are established. If the center frequency of the seismic wave lies in the high-frequency regime, or even in the geometric regime, there exist two sets of longitudinal and transverse waves, of which the fast longitudinal (Pn) and transverse wave (Sn) speeds approach those of the hard material and the slow longitudinal (Pg) and transverse (Sg) wave speeds approach those of the soft material. Meanwhile, the dispersions of all the waves are very small, so all the wave packages can propagate for a long distance. A schematic diagram of the seismic rays is shown in Fig. 30(b). If the wave packet emanated from the source contains both low- or intermediate-frequency components and high-frequency components, the original wave packet will split into three branches, as shown in Fig. 29. The phase velocities of the P* and S* phases lie near that of the homogeneous reference medium. In real measurements, the P* and S* phases may be contaminated by the P-coda waves or S-coda waves, so the recorded signal is a superposition of low- or intermedia-frequency phase signal and the high-frequency noises. By comparing the velocities of Media IX-XVI and that evaluated from Figs. 26-29, we can establish a rough correspondence between the virtual material models and the lithospheric regions: Media IX-XIII correspond to Figs. 26-27, Medium XIV corresponds to Fig. 28(a), Medium XV corresponds to Fig. 28(b), and Medium XVI corresponds to Fig. 29. With these correspondences in mind, we can try to give a consistent explanation for both the dispersion and attenuation characteristics of the measured data. Fig. 25 shows that most seismic signals decrease with frequency following an inverse power law, which indicates that the signals lie in the high-frequency regime. Meanwhile, the majority of the observed travel-time curves exhibit four branches corresponding to the Pn, Pg, Sn and Sg phases, respectively, which also indicates that most seismic signals lie in the high-frequency regime. By adjusting the properties of the component phases, the Q-factor ratios show a variety of different patterns which are sufficient to cover all the values and variations of measured Q-factor ratios, as shown in Fig. 25(c). The travel-time curves in Fig. 28(b) exhibit high-velocity, weak velocity fluctuations and low attenuation, which are related to the existence of the subducting Pacific and Philippine sea slabs. Correspondingly, these distinctive features are also fully captured in the numerical example for Medium XV. From Fig. 24, we can see Medium XV has high velocities, small velocity fluctuations and low attenuation (high-Q). Medium XVI gives a material model corresponding to Fig. 29, from which the P* and S* phases can be explicitly identified. Comparing the velocities in Fig. 29 and that shown in Fig. 24, we can find the velocities of all the phases calculated from Medium XV show excellent agreement with that measured in real seismic events.

In the new explanation, there needs not be any even interface that separates two or more layers with dramatically different material properties. Instead, it only requires that there exists a certain degree of property fluctuation and the fluctuation is statistically uniform, which are more likely to be realized in the heterogeneous Earth.

Through the above discussion we see the new model is capable of providing an accurate and consistent explanation to nearly all the observed seismic data. At this point we are ready to answer several long-standing problems raised in the seismology community, and draw a series of new conclusions:

1) The measured apparent attenuation for short-period seismic waves are mainly caused by scattering in the heterogeneous Earth, the portion contributed by anelasticity mechanisms is less important. This conclusion is also confirmed by the observation that seismic waves can propagates a very long distance (the seismic waves frequently run several times around the earth before their energy is completely dissipated) and last for a long time (about tens of minutes);

2) Multiple scattering theory is a necessary tool for the explanation of the attenuation of the major phases Pn, P*, Pg, Sn, S* and Sg, single scattering theory is insufficient;

3) Elastic waves are tensor waves in their nature, and the causality of the multiple scattered waves is secured by the proper choice of the imaginary part of the homogeneous Green's functions. The Kramers-Kronig relation, which was derived for scalar waves, cannot be applied to the scattered elastic waves [65-66, 133-134]. This is because the K-K relation always requires that there exists a single propagation mode in the whole frequency range. However, as shown in the numerical examples, most heterogeneous materials exhibit two propagation modes in the intermediate- to high-frequency range. Most importantly, for materials with extremely strong property-fluctuation, like bone and porous metals, there even does not exist any coherent wave at high frequencies;

4) All the direct arrivals and the coda waves form an integrate unity. Multiple scattering does not only cause dispersion and attenuation of the direct arrivals, but also induces split of the coherent wavefront and generates two or three sets of longitudinal and transverse modes. At frequencies far from the critical frequency (branch point of the dispersion curves), the dispersions of both longitudinal and transverse waves are extremely weak. Consequently, the wave packets can be maintained for a long distance. This unique propagation characteristic gives a new explanation to the Pn, P*, Pg, Sn, S* and Sg (or Lg) phases that observed worldwide [164]. Contrary to the explanation based on multi-layered model, it does not require the existence of any even and sharp discontinuity, such as the Mohorovičić discontinuity or the Conrad discontinuity;

5) Velocity dispersion and attenuation constitute a complete description of the propagation of the coherent multiple-scattered waves, they are simultaneously determined by the new model;

6) The multiple scattering theory suggests us give up the classic seismology, which is based on the assumption that the lithosphere is a multi-layered medium, and develop new seismology and new ray theory (or other high-frequency asymptotic approaches) based on the random medium model. This also motivates a generic statistical approach to classify the lithosphere based on the velocities, velocity fluctuations, and the correlation length (characteristic size) of the subsurface heterogeneities;

7) The existence of inhomogeneities that large enough to break the stochastic homogeneous assumption may change the waveform significantly. Thus, the model provides a new imaging mechanism for inversion of the large-scale structures in the lithosphere or beneath the lithosphere [146, 174, 180].

Finally, we need to stress that the theoretical predictions of the dispersion and attenuation are for the coherent waves. In an idealized case, the coherent waves are obtained by summing up the recorded waves from a number of observation stations that located concentrically about the seismic source. The quantities measured at a single observation station may be distorted by the near-receiver inhomogeneities, as pointed out by Wu [10]. The development of large scale seismic arrays, such as LASA, NORSAR, has significantly improved our ability to obtain such coherent waves by collecting large amounts of single-station signals. These seismic arrays along with modern signal processing techniques provide us a great opportunity to extract the complete set of coherent wave information, including the dispersion and attenuation of each phase. Based on the high-quality data, the new model can perform reliable inversion for the material property fluctuation and characteristic length of inhomogeneities with the aid of advanced optimization algorithms like the least-square method [175]. Furthermore, it is well known that mineral depositions can also influence the material property fluctuation, thus the inversion results can be used for exploring natural resources, such as hydrocarbone reservoirs and coal and iron mines.

## B. Applications in the ultrasonic nondestructive evaluation (NDE)

Porous metals have broad applications in aerospace industry [181-184] and biomedical engineering [185-187] due to its light weight and high strength. An additional prominent feature is that its mechanical properties are adjustable and can perfectly match those of biomedical materials like bone, so porous metals like porous Titanium are also used as bone substitution materials [186]. Studying the dispersion behavior of ultrasound in these materials is of vital importance for the nondestructive evaluation and characterization of porous metals [188-192]. Two-phase metals and particle reinforced composites are other categories of heterogeneous materials widely used in industry. These materials have unique physical and chemical properties, such as chemical stability, high strength, tunable mechanical properties [193-194]. The microstructure is the key factor that determines the material properties like fracture toughness, hardness, impact strength, yield strength, and tensile strength. Consequently, there is considerable interest in developing ultrasonic techniques to nondestructively characterize microstructures and finally determine these mechanical properties. It is well known that the ultrasonic attenuation and dispersion are very sensitive to the characteristic parameters of microstructures, like elastic moduli of the component phases, average sizes of the grains, so they provide important parameters for microstructural characterization. In general, the interpretation of frequency dependent velocity and attenuation data requires the use of multiple scattering theory. However, previous research on ultrasonic NDE has been mainly focused on characterization of metal polycrystals such as Titanium alloy, Nicole alloy or iron, which are generally regarded as weak scattering materials [42-50]. Very little effort is devoted to characterization of materials with strong property contrast, such as porous metals and two-phase alloys [188-192]. The model developed in this work exactly meet these requirements. As applications, we calculate the velocity dispersion and attenuation of porous aluminum with various porosities and Cu-Al alloys with different phase volume ratios.

### *1. Nondestructive evaluation of porous metals*

Porous aluminum is an extreme example of two-phase materials. The material properties used in the calculation are shown in Tab. 15,

Table 15. Material Properties of porous aluminum.

| Material | $\rho$ (kg/m$^3$) | $\lambda$ (GPa) | $\mu$ (GPa) | $V_L$ (m/s) | $V_T$ (m/s) |
|---|---|---|---|---|---|
| Al | 2720 | 50.35 | 25.94 | 6130.62 | 3088.17 |
| Vacuum | 0 | 0 | 0 | 0 | 0 |

The properties of the reference media for porous aluminum with porosities (pore volume fractions) varying from 5% to 30% are given in Tab. 16.

Table 16. Properties of the reference medium for porous aluminum.

| Volume fraction | $\rho$ (kg/m$^3$) | $\lambda$ (GPa) | $\mu$ (GPa) | $V_{0L}$ (m/s) | $V_{0T}$ (m/s) |
|---|---|---|---|---|---|
| $f_{pore}$=5% | 2584.0 | 42.34 | 23.49 | 5879.39 | 3015.20 |
| $f_{pore}$=10% | 2448.0 | 35.03 | 21.02 | 5610.96 | 2930.29 |
| $f_{pore}$=20% | 2176.0 | 22.45 | 15.99 | 5001.68 | 2711.17 |
| $f_{pore}$=30% | 1904.0 | 12.47 | 10.84 | 4235.00 | 2386.02 |

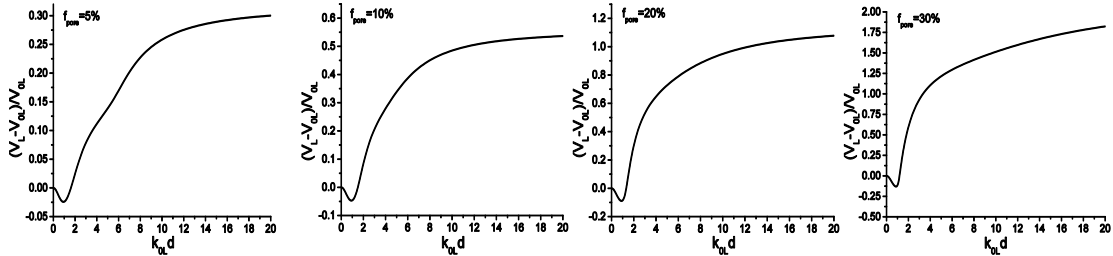

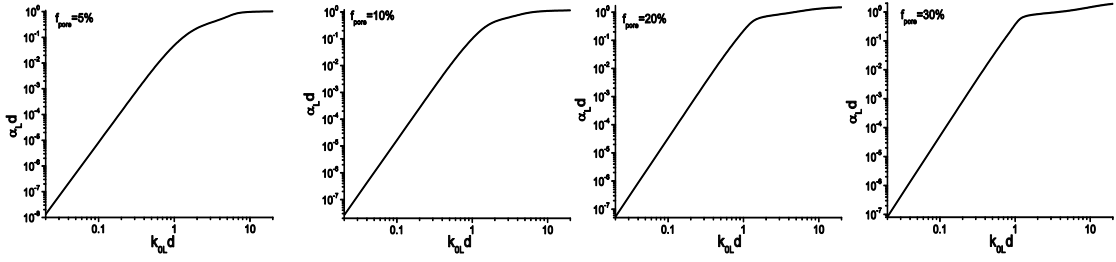

FIG. 31. Effects of porosity on dispersion and attenuation of longitudinal waves in porous aluminum.

The dispersion and attenuation of longitudinal waves and transverse waves in porous aluminum are shown in Figs. 31 and 32, respectively.

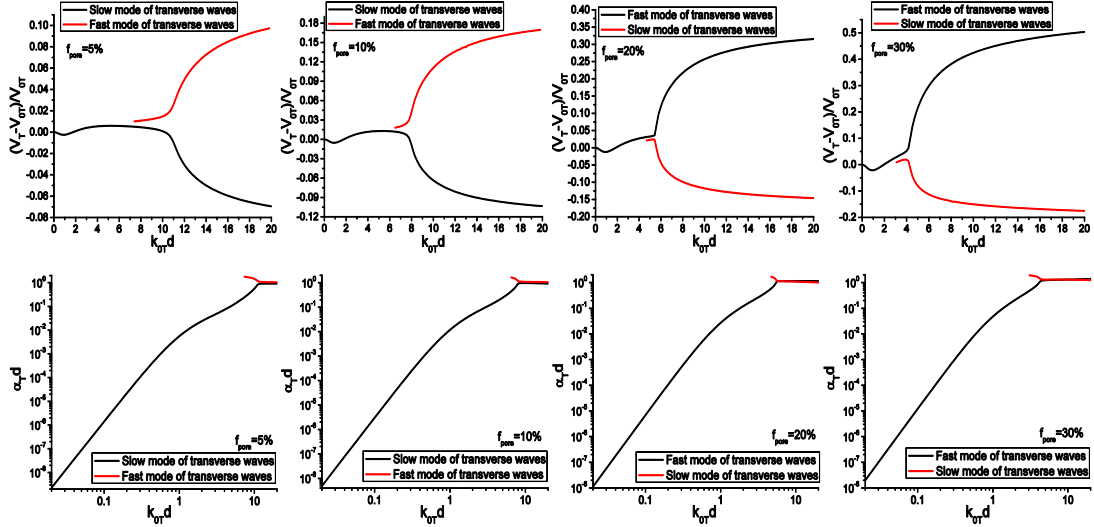

FIG. 32. Effects of porosity on dispersion and attenuation of transverse waves in porous aluminum.

From the results, we can find the following propagation characteristics of coherent waves in porous metals:

1) The longitudinal velocity first undergoes steep negative dispersion, after reaching the minimum value, it becomes positively dispersive and quickly goes beyond the longitudinal velocity of pure aluminum, which indicates the disappearance of the coherent waves. Only one longitudinal mode is found in the whole frequency range. This phenomenon tells us that the branch behavior of longitudinal coherent waves is strictly related to the elastic property of the component materials, or in other words, two longitudinal modes can only exist when both of the component phases are elastic materials. The dimensionless attenuation-frequency relation follows a power law at frequencies lower than 1. At dimensionless frequencies higher than 1, the attenuation approaches unity.

2) The coherent transverse waves first undergo negative dispersion with a magnitude smaller than that of the longitudinal waves, then become positively dispersive and approach the transverse wave velocity of pure aluminum. A second mode appears in the intermediate frequency range. The attenuation of the second mode is much larger than the first one at its initial appearance and then decreases quickly. At moderately high frequencies, the velocity of the fast mode quickly goes beyond the transverse velocity of pure aluminum, which indicates the disappearance of coherent waves. It is also noted that porous Al with low porosity support coherent transverse waves in a relatively wider frequency band.

### *2. Nondestructive evaluation of two-phase alloys*

Nondestructive characterization of microstructures in multiphase alloys by means of ultrasonic scattering has received extensive studies. The mechanical properties of frequently used metallic phases, including Fe, Al, Cu, Ti, or Nb all exhibit strong property fluctuation. Our model is naturally suitable for these types of materials. As an example, we consider a two-phase alloy made of Al and Cu [194]. The material properties of Al and Cu are given in Tab. 17. We consider Cu-Al alloys with different volume ratios. Tab. 18 lists the material properties of the reference media.

Table 17.    Material Properties of Cu and Al.

| Material | $\rho$ (kg/m$^3$) | $C_{11}$ (MPa) | $C_{12}$ (MPa) | $C_{66}$ (MPa) | $V_L$ (m/s) | $V_T$ (m/s) |
|---|---|---|---|---|---|---|
| Cu | 8490 | 134.45 | 60.45 | 37.00 | 3979.48 | 2087.60 |
| Al | 2720 | 102.23 | 50.35 | 25.94 | 6130.62 | 3088.17 |

Table 18. Material properties of the reference medium of Cu-Al alloys.

| Volume fraction | $\rho$ (kg/m$^3$) | $\lambda$ (MPa) | $\mu$ (MPa) | $V_{0L}$ (m/s) | $V_{0T}$ (m/s) | $\delta V_{LCu}/V_{0L}$ | $\delta V_{LAl}/V_{0L}$ | $\delta V_{TCu}/V_{0T}$ | $\delta V_{TAl}/V_{0T}$ |
|---|---|---|---|---|---|---|---|---|---|
| $f_{Cu}$=10% | 3297 | 51.24 | 26.88 | 5643.43 | 2855.33 | -0.29 | 0.09 | -0.27 | 0.08 |
| $f_{Cu}$=30% | 4451 | 53.10 | 28.87 | 4990.23 | 2546.82 | -0.20 | 0.23 | -0.18 | 0.21 |
| $f_{Cu}$=50% | 5605 | 55.06 | 31.01 | 4570.39 | 2352.10 | -0.13 | 0.34 | -0.11 | 0.31 |
| $f_{Cu}$=70% | 6759 | 57.13 | 33.30 | 4278.46 | 2219.49 | -0.07 | 0.43 | -0.06 | 0.39 |
| $f_{Cu}$=90% | 7913 | 59.32 | 35.73 | 4065.30 | 2124.92 | -0.02 | 0.51 | -0.02 | 0.45 |

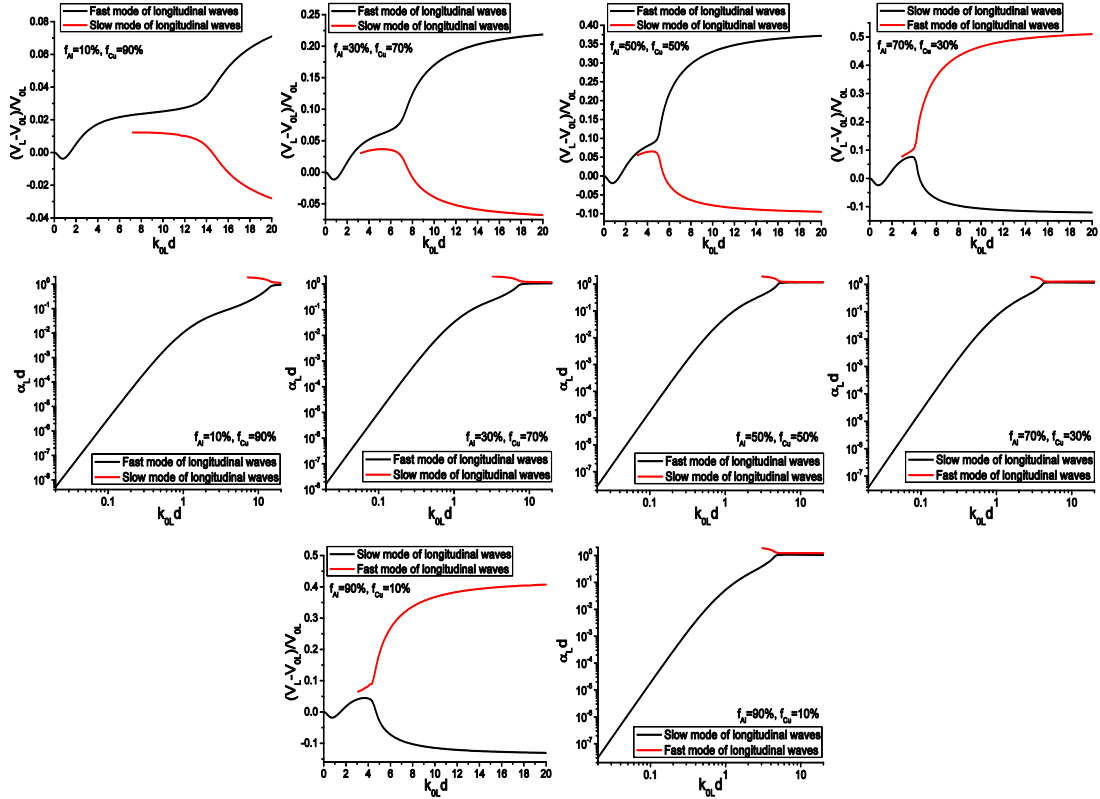

FIG. 33. Dispersion and attenuation of longitudinal waves in Cu-Al alloys with different volume ratios.

Figure 33 shows the dispersion and attenuation of longitudinal waves. From the numerical results, we can observe the following propagation characteristics: Starting from the quasi-static limit, the longitudinal waves first undergo negative dispersion. After reaching its minimum, the coherent waves become positively dispersive, then it approaches the longitudinal speed of the pure Al. In the intermediate frequency range, a second propagation mode starts to appear, its attenuation is much larger than the first mode and then decreases dramatically and approaches unity. At high frequencies, velocities of the two modes quickly approach the upper and lower bonds. Different from weak scattering media, the two bonds are different from the velocities of the two component phases. Once again, the coherent wave disappears when the velocity of the fast mode exceeds the upper bond of the two phases. One noteworthy phenomenon is that alloys in which Cu occupies a larger volume fraction support coherent wave propagation in a wider frequency range.

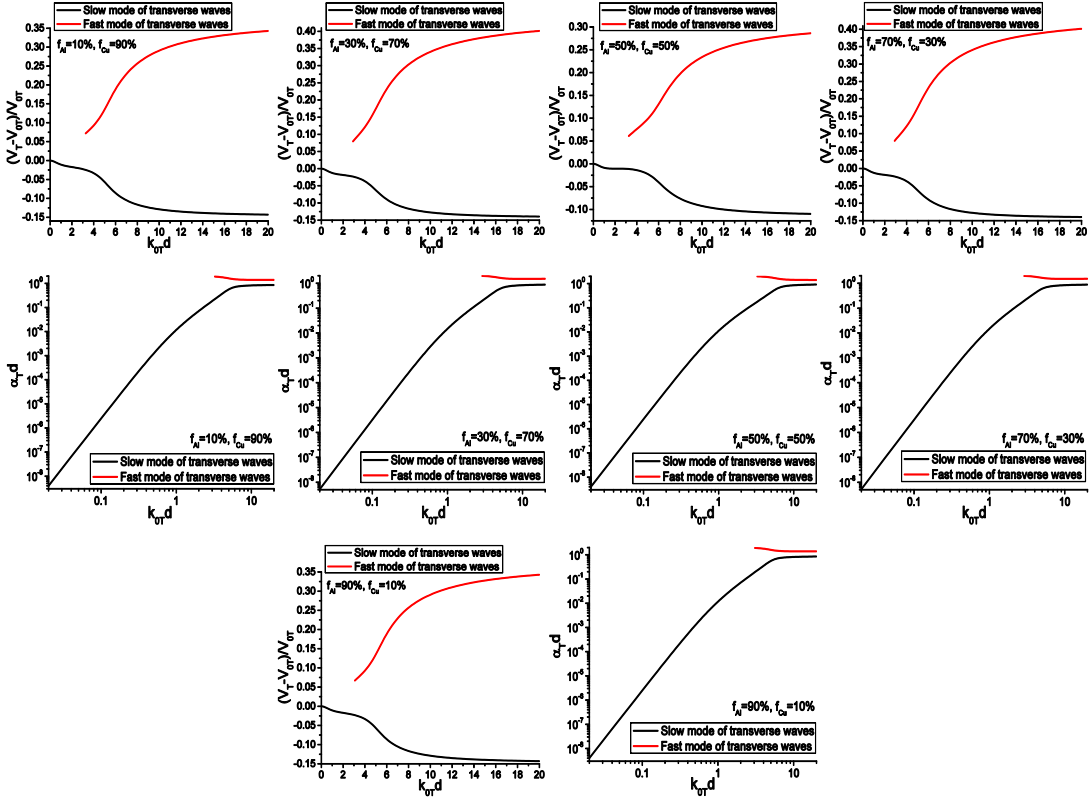

FIG. 34. Dispersion and attenuation of transverse waves in Cu-Al alloys with different volume ratios.

Figures 34 shows the dispersion and attenuation of transverse waves. From the numerical results, we can observe the following propagation characteristics: Starting from the quasi-static limit, the coherent transverse waves undergo negative dispersion monotonically, at intermediate frequency $k_{0T}d$=2-3, the wavelengths are comparable to the characteristic dimension of the inclusions, a second propagation mode with positive dispersion starts to appear. As the frequency increases, velocities of the two propagation modes quickly approach the upper and lower bonds. Similarly, when the fast mode goes beyond the upper bond of the component phases, it indicates vanish of the coherent waves. Different from longitudinal waves, the transverse waves at low frequencies have no positive dispersion regime.

From these numerical examples, we see that the new model can give an accurate and quantitative prediction of the dispersion and attenuation of heterogeneous medium regardless of whether the volume fraction of either phase is small. Both the effective static and dynamic modulus can be further derived from the dispersion and attenuation curves. The results have significant implications for microstructure characterization and theory-guided material design [207-208].

## C. Applications in bone quantitative characterization

Cortical bone is the major supporting structure of human body and most osteoporotic fractures are happened in it. Cortical bone is composed of compact bone and distributed pores saturated with marrow. The pore diameter is normally in the range of 20-300 micrometer and the volume fraction lies between 5% and 15% [70]. However, as a consequence of aging and continuous loss of minerals, the porosity of the middle-aged and elderly people may increase to 30% [78]. Early diagnosis and treatment of osteoporosis are essential for preventing bone fracture. Compared to microXRT or MRI, ultrasound provides an economic and noninvasive approach for the diagnosis of osteoporosis and monitoring the bone status. Different from MRI and XRT, which can only provide information about the porosity and average pore diameter, ultrasonic signals also carry rich information of the mechanical properties of bone, including the density and elastic stiffness. Therefore, quantitative ultrasound (QUS) provides a promising technique for predicting the bone mechanical strength. For this reason, bone quantitative ultrasound has become an active research field [70]. Multiple scattering theory of ultrasonic waves plays a central role in characterizing microarchitecture in bone. However, modeling the multiple scattering in cortical bones progresses very slowly. One reason is that cortical bone is an extremely complicated heterogeneous material with hierarchical microstructures covering a wide length scale. Another reason is that the wave-scatter interactions are very strong since wavelengths of ultrasound excited by most biomedical transducers are comparable to the characteristic dimension of microarchitectures in cortical bone. Reported experimental and numerical results are also full of controversy. The majority of the experimental research report that wave velocity is negatively dispersive [89], while some other researchers report that velocity have positive dispersion [73]. The interesting and diverse phenomena have attracted extensive attention from both academy and clinical community [70]. Due to the lack of a thorough understanding of the rich phenomena, quantitative ultrasound techniques for cortical bone characterization has been hampered for many years. It is generally realized the single scattering model and Biot's model are insufficient for characterization of bone samples. For instance, the single scattering model can predict dispersion and attenuation when the number density of pores is low, normally less than 5%, and the error increases as the porosity becomes larger. The Biot model cannot predict the anomalous negative dispersion and it involves several phenomenological parameters that difficult to be measured experimentally. In this section, we conduct comprehensive numerical calculation for cortical bones with different porosities and demonstrate the capability of the new model to quantitatively predict the scattering characteristics of cortical bones. In this work, real cortical bones are modeled as a poroelastic medium with spherical pores saturated by water. The dimensionless velocity variation and the attenuation are plotted versus the dimensionless frequency. The properties of the reference medium are listed in Tab. 5.

### *1. Dispersion and attenuation of longitudinal coherent waves*

The dispersion and attenuation of longitudinal waves are shown in Fig. 35.

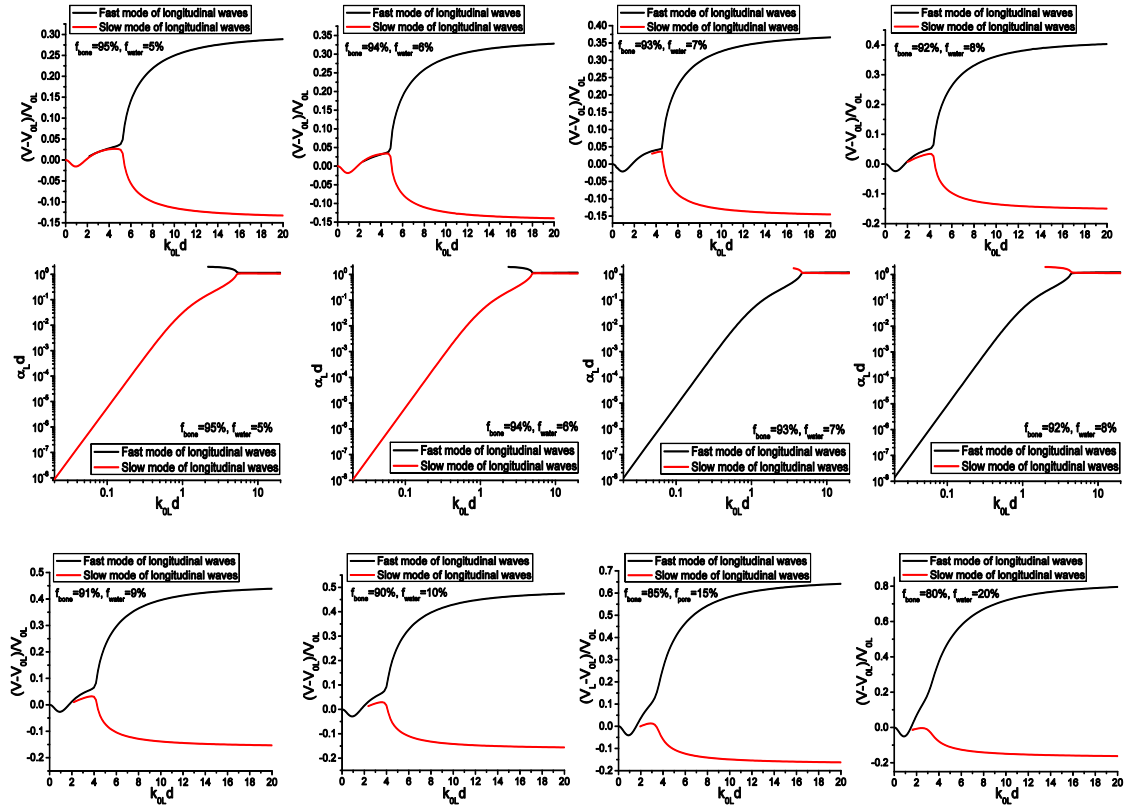

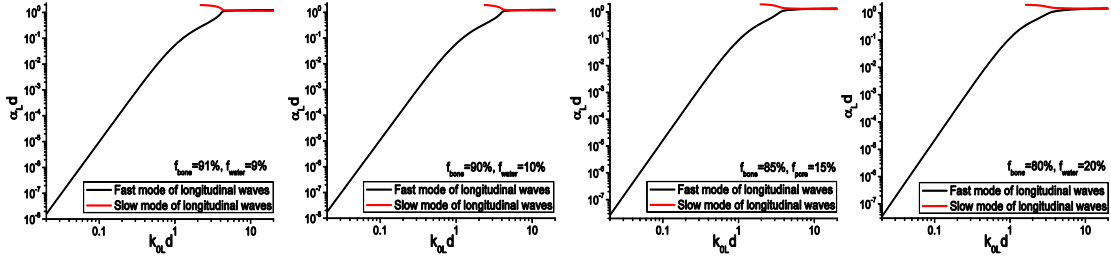

FIG. 35. Velocity and attenuation of coherent longitudinal waves in cortical bone with different porosities.

According to the characteristics of the dispersion curves, the frequency domain can be divided into four regions:

Region I ($0 < k_{0L}d < 2$): Starting from quasi-static limits, the wave undergoes negative dispersion. The velocity assumes its minimum at a dimensionless frequency near unity. The attenuation increases with frequency following a power law. There is only one propagation mode in this region;

Region II: After assuming its minimum, the velocity becomes positively dispersive. The attenuation increases with frequency in a general nonlinear manner. There is still only one propagation mode;

Region III: There are two propagation modes in this region, both have positive dispersion. The width of this range is strongly dependent on the volume fraction of the pores, the smaller the $f_{pore}$, the wider the Region III. At the emergence of the second mode, its velocity is nearly identical to that of the first mode, then the velocity difference becomes larger as the frequency continues to increases. We can also observe that the difference increases with porosity. This result tells us that the two propagation modes are more easily observed in bones with high-volume fractions. The attenuation of the second mode is very large and then decreases rapidly. When the pore volume fraction is smaller than 6%, the fast mode has larger attenuation, when $f_{pore}$ is greater than 7%, the slow mode has larger attenuation;

Region IV: The velocity of the fast mode quickly approaches the upper limit, and then goes beyond that of pure bone, which indicates that the coherent wave disappears. The slow mode becomes negatively dispersive and quickly approaches its lower limit.

The above results are presented in the dimensionless parameter space. However, in practical applications we need to analyze the dispersion behavior in the physical space. In doing so, the velocity and attenuation coefficient against the frequency $f$ are replotted in Figs. 36. Figure 36 shows the effects of porosity on the dispersion behavior by considering cortical bones with fixed pore diameter $d=100$ μm and various porosities, ranging from 1% to 10%.

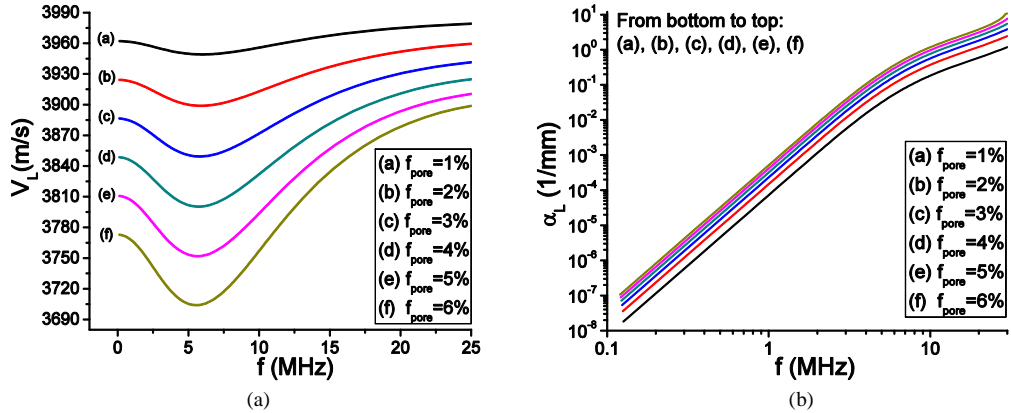

(a)         (b)

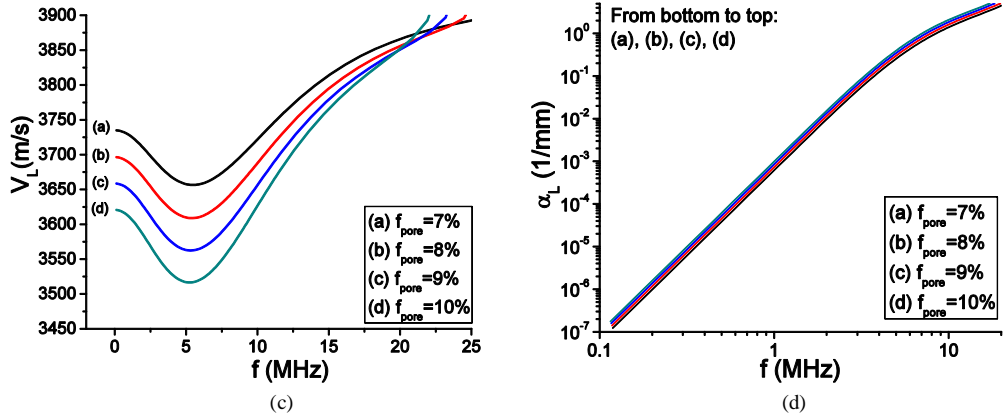

(c)           (d)

FIG. 36. Longitudinal dispersion (a), (c) and attenuation (b), (d) of cortical bone with different porosities: $d$=100 μm.

From Figs. 36(a) and (c) we observe that the velocity decreases as the porosity increases, and the negative dispersion in the frequency band [0, 5] MHz becomes more evident as the porosity increasing. It is also observed that the attenuation increases monotonically with porosity.

  The effects of pore diameter on dispersion and attenuation are shown in Fig. 37, where born samples with fixed porosity $f_{\text{pore}}$=10% but various pore diameters d = 50, 100, 150 and 200 μm are considered. It is observed the pore diameter does not change the quasi-static limit of the velocity. Instead, it changes the dispersion and attenuation following a scaling law, i.e., the dispersion curves are compressed horizontally as the pore diameter increases. As a consequence, bone samples with larger pore diameter have more obvious negative dispersion and becomes positively dispersive at a relatively low frequency.

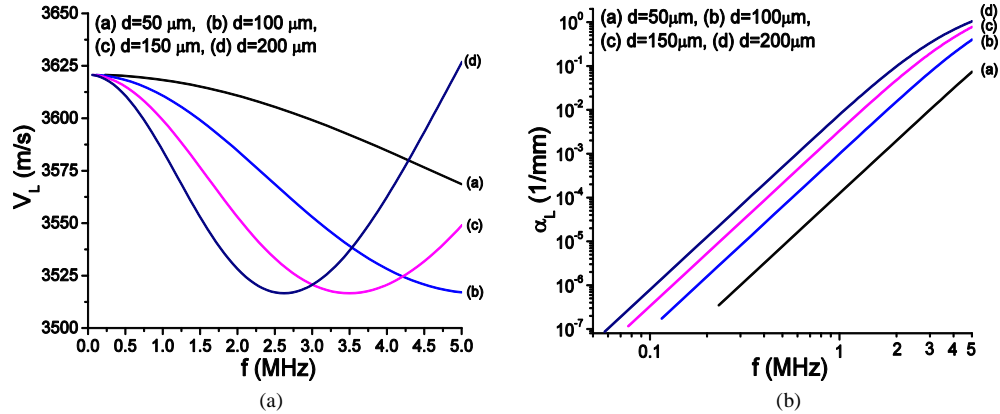

(a)           (b)

FIG. 37. Longitudinal dispersion (a) and attenuation (b) of cortical bone with different diameters: $f_{\text{pore}}$=10%.

The coexistence of two propagation modes is frequently mentioned in the literature but rarely quantified to the author's best knowledge. The new model allows us to estimate the condition under which two propagation modes can exist. It is seen from Fig. 35 that when the second mode just emerge, its velocity is extremely close to the first one. Meanwhile, the attenuation of the new mode is very large, so it is extremely difficult to discriminate the two modes. Indeed, the wave observed in experiments is a superposition of the two modes, and the dispersion behavior lies between the two. When the volume fraction of the pores is relatively large, reaching a value between 20% to 30%, the difference between the velocities of the two modes is relatively large. If the average pore size is also not very small, about 200~300 μm in diameter, it is possible to observe the two modes experimentally.

*2. Dispersion and attenuation of transverse coherent waves*

The dispersion and attenuation of transverse waves are shown in Fig. 38.

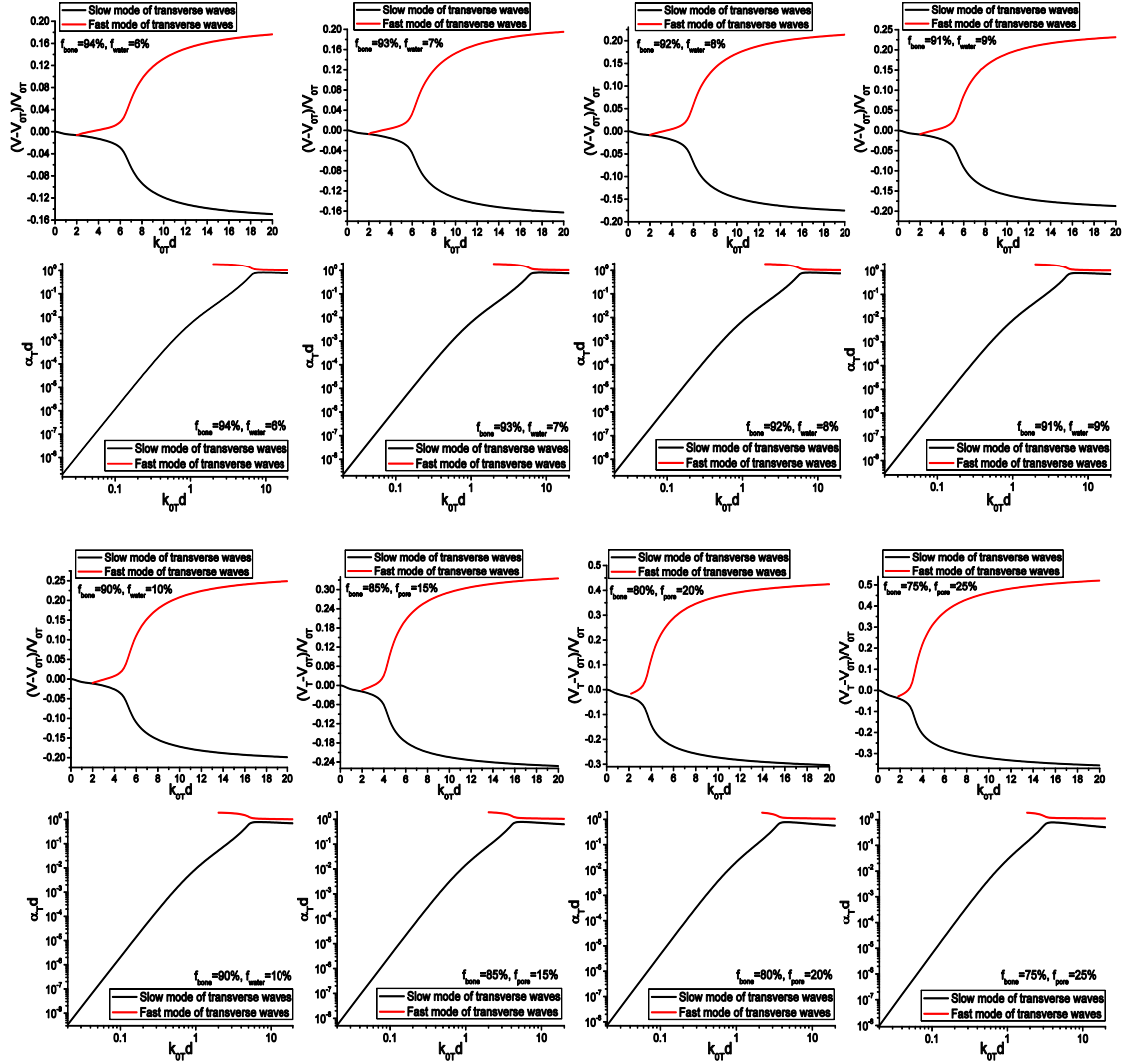

FIG. 38. Velocity and attenuation of coherent transverse waves in cortical bone with different porosities.

According to the characteristics of the dispersion curves, the frequency domain can be divided into three regions:

Region I ($0 < k_{0T}d < 2$): There is one propagation mode with negative dispersion;

Region II: There are two propagation modes in this region, one has positive dispersion with larger velocity. The width of this range is strongly dependent on the volume fraction of the pores, the smaller the $f_{pore}$, the broader the region II. The magnitude of the dispersion increases with porosity;

Region III: The velocity of the fast mode quickly goes beyond the upper bond of the two component materials, indicating the disappearance of the coherent wave. The slow mode keeps negatively dispersive and quickly approaches its lower limit. The existence of the slow mode needs to be verified by experiments or numerical simulation.

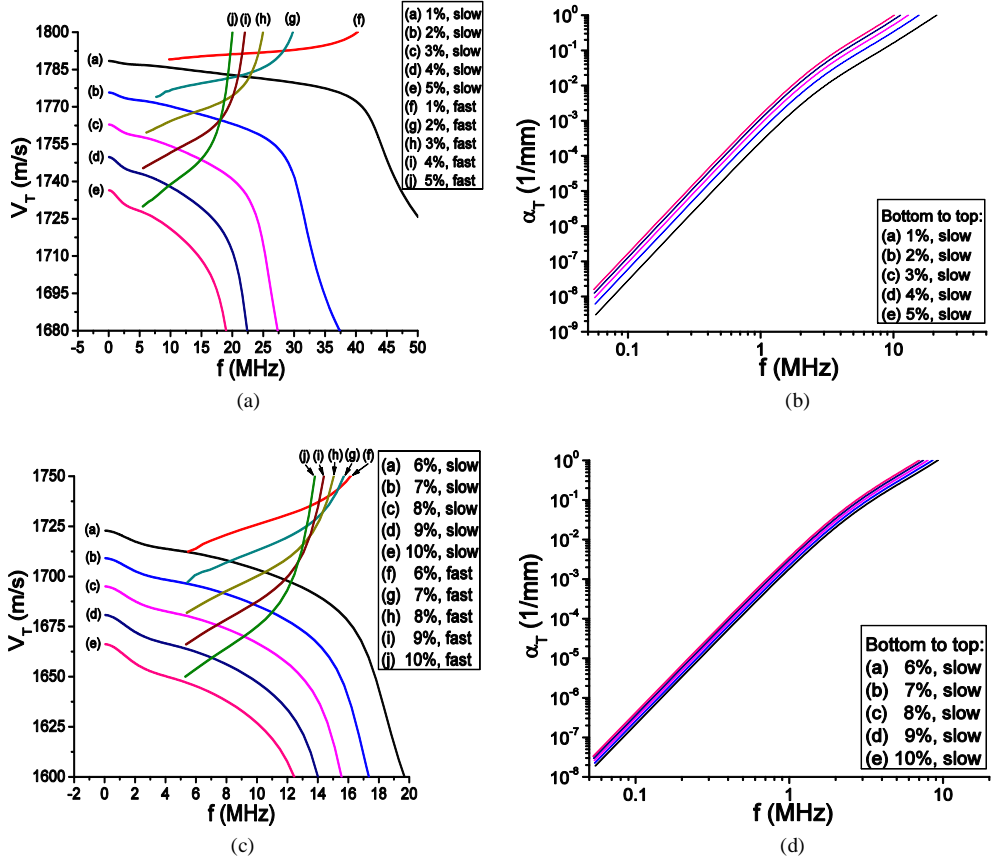

FIG. 39. Transverse dispersion (a), (c) and attenuation (b), (d) of cortical bone with different porosities: $d=100$ μm.

The transverse wave dispersion and attenuation of cortical bones with a fixed pore diameter and different porosities are shown in Fig. 39. Similar to longitudinal waves, the velocity decreases and the attenuation increases as the porosity increases.

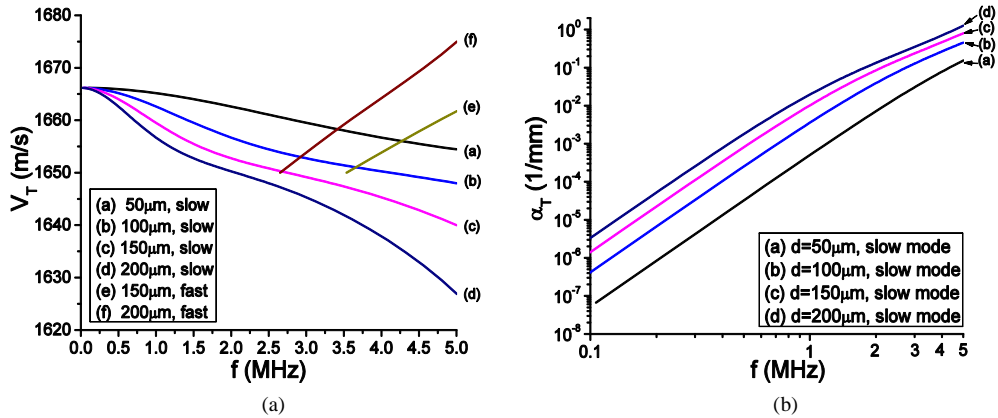

FIG. 40. Transverse dispersion (a) and attenuation (b) of cortical bone with different diameters: $f_{pore}=10\%$.

Figure 40 shows the effects of pore diameter on the dispersion and attenuation of transverse waves. They also follow a scaling law. As the pore diameter increasing, the velocity decreases more rapidly and the attenuation increases dramatically, though the quasi-static limit of the velocity keeps constant.

Through the foregoing examples we see the new model establishes an accurate quantitative relation between the wave propagation parameters and the bone microstructural information, and consequently. It provides a promising inversion model for quantitative bone characterization by using the two most important parameters: broadband ultrasonic attenuation (BUA) and speed of sound (SOS).

This work provides a completely independent theoretical framework for the prediction of dispersion behavior in fluid-saturated poroelastic materials. In contrast to Biot's model, the new model does not require any phenomenological parameters. Most importantly, the negative dispersion behavior, which is extremely difficult to predict by Biot's model, is perfectly captured in the new model. The new model supersedes the single scattering theories and discrete scattering models in that it is applicable to poroelastic materials with relatively large pore diameter and pore volumetric concentration.

# V. DISCUSSION

In the foregoing sections, we have demonstrated several practical applications of the new model, we are now at a position to discuss numerical and experimental validations of the model, and future work for possible extensions.

First of all, we need to point out that as a statistical model, providing an estimation of the statistical properties of the microstructure is sufficient for practical applications, such as clinical diagnosis or microstructure characterization. Accurate numerical and experimental verifications of this model are out of the scope of the current work. Nevertheless, it is meaningful to give a detailed discussion on how to conduct successful numerical and experimental verification for the new model. Up to date, significant efforts have been devoted to numerical simulation of wave scattering in heterogeneous materials using Finite-Difference-Time-Domain (FDTD) algorithms [29, 71, 195-198], finite element method (FEM) [67-68, 199-200] and spectral element method (SEM) [195]. Compared to experimental approaches, numerical simulation provides an efficient and flexible way to validate different theoretical models. The most attractive feature is its capability of generating finely controlled microstructures. Numerical models for heterogeneous materials with different types of spatial correlation functions [29], with specific material property combination, and with different realizations of random media can be created using advanced preprocessing techniques, such as the Voronoi tessellations [67-68, 155]. Moreover, time domain signals with arbitrary waveform and at any frequencies can also be generated, which enables the possibility of extracting the dispersion and attenuation features in a wide frequency band. Another advantage of numerical simulation is it provides a flexible approach to extract signals at any numerical nodes and at any sections at the post-processing stage, which is impractical in experiments. Numerical approaches do not introduce approximations for the scattering process, in other words, it includes all multiple scattering events and thus contains the complete information of the stochastic wavefield. Nevertheless, conventional numerical approaches suffer a number of drawbacks. For instance, discretization with large amount of meshes often introduces numerical dispersion and attenuation, especially for the case of high frequency signals [195]. Additionally, accurate description of the phase boundaries is extremely difficult, approximation to the geometry generates spurious diffraction [198]. Furthermore, the discontinuity of material properties cannot be accurately accounted for. The finite difference method frequently replaces the phase interfaces separating materials with sharp property contrast by an additional mesh with average properties of the adjacent materials [70]. Such homogenization scheme will inevitably change the scattering behavior of the inclusions. It is also noted the derivatives of the displacement is discontinuous at the phase boundaries, which may lead to numerical instability. Consequently, developing new numerical algorithms is necessary to conduct accurate full-frequency-range simulation of the multiple scattering phenomenon in strong-property-fluctuation materials. An ideal numerical simulation should satisfy the following conditions: (1) Negligible numerical dispersion and attenuation induced by mesh discretization; (2) The numerical algorithm should be pure, free of artificial excitation or dissipation; (3) The microstructure should be finely controlled, including the shape and size of the grains/pores, the distribution of the pores, no overlap of the inclusions should be introduced; (4) As the contrast of the density and elastic stiffness between different phases may be very large, strong scattering occurs at these phase boundaries, so in the simulation no approximation of the boundary conditions should be introduced; (5) Proper boundary conditions of the sample should be introduced to avoid undesired reflections, perfect-matching-layer gets in trouble when dealing with highly inhomogeneous media, symmetric boundary conditions, periodic boundary conditions are preferred; (6) An ideal source of signal for extracting the dispersion characteristics is a plane wave with infinite extent, finite sources should be avoided, (7) In the high frequency range, the discretization in the space and time domain, i.e., the time step $\Delta t$ and the grid spacing $\Delta x$ must be sufficient to capture the scatterer-wave interaction; (8) To facilitate the observation of the fast and slow modes, the sample should be large enough, so that each of the signals are completely separated. Additionally, high-quality signal processing techniques are also necessary for the analysis of its spectral content since these signals are highly attenuated.

Researchers have conducted comprehensive experimental studies on the velocity dispersion and attenuation of multiple scattered waves in polycrystalline metals [201-202], particulate composites [203-204] and cortical and trabecular bone [67, 72-73]. These works provide valuable data for evaluating the validity and accuracy of different theoretical models. However, the relevance of the data obtained in the previous experiments to the theoretical predictions in this work is questionable. Accurate experimental verification of the new model is still a very challenging task. A successful experimental verification should meet the following requirements: (1) The microstructure must be finely controlled, including the shape, size and distribution, especially that the spatial autocorrelation functions should match the theoretical hypothesis; (2) No microcracks, interfacial diffusion/debonding or air bubbles should be introduced in the sample; (3) Planar and/or focused transducers are frequently used in practical backscattering and transmission measurements, both of which have finite beam width. Consequently, the diffraction effects must be properly corrected before the dispersion and attenuation of a plane wave component can be extracted [40, 205-206]; (4) Sufficient propagation distance is required to establish a stable multiple scattering signal, so the sample should have a reasonable size. 3D printing is a promising technology for manufacturing such heterogeneous materials, but its capability still needs to verify.

The current work has laid the foundation of the multiple scattering theory for strong fluctuation materials. It can be extended in many aspects to incorporate more complicated microstructures. To the author's best knowledge, the following directions for future developments can be distinguished. First, it is reported that a number of heterogeneous materials have ellipsoidal inclusions. For example, Lotus-Type porous metals have slender cylindrical pores aligned in one direction [181]. High-temperature titanium alloys used in jet engine foils and disks also exhibit ellipsoidal grains due to certain thermomechanical procedures like rolling and casting [209]. To capture these microstructural information, a multiple scattering model for heterogeneous media with aligned ellipsoidal inclusions are in need. Second, the current model only considers materials with isotropic constituent materials. It is well known that polycrystalline alloys are composed of anisotropic crystallites with random or preferred orientations (textures). In order to take the crystallographic information into consideration, a multiple scattering model for alloys with anisotropic grains is necessary. Third, in this work we only considered random media with exponential correlation functions, which is appropriate for the description of inclusions with sharp boundaries. For heterogeneities with blurred boundary, Gaussian correlation function is the most proper choice. The Von-Karman type correlation function is adopted to describe inhomogeneous materials with self-similar microstructures. Thus, the current model can also be extended to study the scattering phenomena occurred in heterogeneous materials with different statistical characteristics. Fourth, this work only considers pure elastic materials, while the anelastic behaviors are completely neglected. Studying the relative contribution of scattering and viscoelasticity to the overall attenuation is important for a more accurate explanation of the measured seismic data. According to the theory of viscoelasticity, the effects of viscosity can be accounted for by introducing an imaginary part in the elastic constants. The work provides a theoretical framework naturally suitable for incorporating this type of complexities. Fifth, in the spirit of this work, a system of renormalized Bethe-Salpeter equation can also be developed, which further paves the way for the development of new radiative transfer theory. The new theory is anticipated to give a more accurate description of the energy diffusion in the regime where no coherent waves can exist [3, 48, 125].

# VI. CONCLUSION

This work provides an accurate and universal theoretical framework for investigation of elastic wave scattering in a wide variety of heterogeneous media, such as heterogeneous earth, metallic polycrystals, granular composite materials and biomedical materials like bone. The new model establishes accurate quantitative microstructure-property relations and fully supports characterization and inversion of microstructures by utilizing the dispersion and attenuation information. A series of numerical examples have been carried out to demonstrate its potential applications in seismology, ultrasonic nondestructive characterization, and bone quantitative ultrasound. Compared with current scattering theories, the new model possesses a series of advantages. The most prominent feature is that it is developed based on the accurate elastodynamic equations and the renormalized Dyson's equation, throughout the derivation no approximations are introduced except the first-order smoothing approximation. It naturally considers the mode conversion between longitudinal and transverse waves and the effects of phase boundaries. With the proper choice of the reference homogeneous medium and by using the Fourier transform technique, the new model is capable of predicting the dispersion and attenuation of coherent waves for heterogeneous media with arbitrary property fluctuation in the whole frequency range. Meanwhile, the input of the model is extremely simple. It does not require any phenomenological parameters as common effective medium models do. With the advanced numerical algorithms, exact dispersion and attenuation for a wide variety of medium models are obtained for the first time. Most importantly, a series of important discoveries have been made based on the results. It is shown that the negative dispersion phenomenon is a common feature of two-phase elastic materials with strong property fluctuation. The appearance of two sets of longitudinal and transverse propagation modes in the intermediate to high frequency range is a bifurcation phenomenon caused by the multiple scattering process. The exact condition under which the bifurcation occurs is accurately predicted by the new model. For weak-property-fluctuation materials, the velocities of the two modes at very high frequencies approach the geometric limits, while for strong-property-fluctuation materials the coherent waves disappear at relatively high frequencies, which is indicated by the model by giving a velocity lager than the upper bond of the constituent phases. At relatively low frequencies, both the dimensionless attenuation and Q-factors increase with frequency following a power law, while at high frequencies, the dimensionless attenuation approaches a constant value near unity and the Q-factor decreases following an inverse power law. Both the longitudinal and transverse Q-factors reach a summit at a critical intermediate frequency, which perfectly confirms Aki's conjecture. The curves of Q-factors and Q-factor ratios predicted by the model show excellent agreement with that collected from seismic data, showing that multiple scattering plays a dominant role in determining the attenuation of seismic waves. These new discoveries suggest us to interpret the seismic data in the new theoretical framework, and reexamine all the conclusions based on the classical multi-layered model, such as the existence of the Mohorovičić discontinuity. Finally, this work paves the way for developing more advanced models to characterize heterogeneous materials with more complicated microstructures.

## AUTHOR'S DECLARATION

This work is independently accomplished by the author and he takes full responsibility for the whole work.

## APPENDIX. CALCULATION OF THE SINGULARITY OF GREEN'S TENSOR

Green's functions are generally used to calculate the field induced by certain distributed sources, i.e., distributed forces for the case of elastodynamic problems. The resulting field is frequently expressed as integrals of spatial convolution type, as shown in Eq. (39). Different from the case of scalar wave scattering where only Green's function is introduced, the integral appeared in the elastic wave scattering also contains its first- and second-order derivatives. For the convenience of subsequent discussion, we list the elastodynamic Green's function and its derivatives explicitly:

$$G_{\alpha i}(\mathbf{x},\mathbf{x}',\omega) = \frac{F}{4\pi} \frac{\delta_{i\alpha}}{\mu} \frac{e^{ik_T|\mathbf{x}-\mathbf{x}'|}}{|\mathbf{x}-\mathbf{x}'|} - \frac{F}{4\pi\rho\omega^2} \frac{\partial^2}{\partial x_i \partial x_\alpha} \left[ \frac{e^{ik_L|\mathbf{x}-\mathbf{x}'|}}{|\mathbf{x}-\mathbf{x}'|} - \frac{e^{ik_T|\mathbf{x}-\mathbf{x}'|}}{|\mathbf{x}-\mathbf{x}'|} \right],$$
(A1)

$$G_{\alpha i,\beta}(\mathbf{x},\mathbf{x}',\omega) = \frac{F}{4\pi} \frac{\delta_{i\alpha}}{\mu} \frac{\partial}{\partial x_\beta} \frac{e^{ik_T|\mathbf{x}-\mathbf{x}'|}}{|\mathbf{x}-\mathbf{x}'|} - \frac{F}{4\pi\rho\omega^2} \frac{\partial^3}{\partial x_i \partial x_\alpha \partial x_\beta} \left[ \frac{e^{ik_L|\mathbf{x}-\mathbf{x}'|}}{|\mathbf{x}-\mathbf{x}'|} - \frac{e^{ik_T|\mathbf{x}-\mathbf{x}'|}}{|\mathbf{x}-\mathbf{x}'|} \right],$$
(A2)

$$G^0_{\alpha i,j\beta}(\mathbf{x},\mathbf{x}',\omega) = \frac{F}{4\pi} \frac{\delta_{i\alpha}}{\mu} \frac{\partial^2}{\partial x_j \partial x_\beta} \frac{e^{ik_T|\mathbf{x}-\mathbf{x}'|}}{|\mathbf{x}-\mathbf{x}'|} - \frac{F}{4\pi\rho\omega^2} \frac{\partial^4}{\partial x_i \partial x_\alpha \partial x_j \partial x_\beta} \left[ \frac{e^{ik_L|\mathbf{x}-\mathbf{x}'|}}{|\mathbf{x}-\mathbf{x}'|} - \frac{e^{ik_T|\mathbf{x}-\mathbf{x}'|}}{|\mathbf{x}-\mathbf{x}'|} \right].$$
(A3)

Consequently, three types of integrals are involved:

$$I_\mathrm{I} = \iiint_V G_{\alpha i}(\mathbf{x},\mathbf{x}',\omega) f(\mathbf{x}) dV, \quad I_\mathrm{II} = \iiint_V G_{\alpha i,\beta}(\mathbf{x},\mathbf{x}',\omega) f(\mathbf{x}) dV, \quad I_\mathrm{III} = \iiint_V G_{\alpha i,i\beta}(\mathbf{x},\mathbf{x}',\omega) f(\mathbf{x}) dV,$$
(A4)

where $f(\mathbf{x})$ represents a distributed body force and $V$ is the volume in which the body force is non-vanishing. In this work, the body force is assumed to be a regular function which has no singularities and is continuous in the whole volume.

As pointed out in Section II.2, when the field point lies in the source region, Green's function and its derivatives are ill-defined at $\mathbf{x} = \mathbf{x}'$. To properly define these integrals, one needs to introduce the concept of principal value of integrals. Mathematically the principal value integral is defined to be the integral over a modified domain obtained by excavating an infinitesimal volume surrounding the singular point. It is shown that the principal value of Type-I and Type-II integrals converges to the correct results, which can also be calculated using other accurate methods. However, when calculating Type-III integrals using the principal value approach, an additional term which does not dependent on the source volume appears. Interestingly, this term is dependent on the shape of the small exclusion volume, so the integral is called shape-dependent principal value integral, and denoted by P.S. Through comparison with the correct result obtained using other methods, we find it is exactly the remaining portion which is dependent on the shape and size of the source volume that corresponds to the correct results. Detailed calculation shows the additional term is a natural result of the δ-singularity of Green's tensor. In the following section, we only consider spherical inclusions and introduce three different but equivalent methods to calculate the singularity of Green's tensor.

## Method 1: Calculation of the singularity in the spatial domain

To calculate the principal value of the singular integrals, we assume that a body force of unit magnitude is uniformly distributed in a cylindrical volume with a radius $a$ and height $2b$. In accordance with the definition, a small spherical volume centered at the origin is excluded from the cylinder, as shown in Fig. A1.

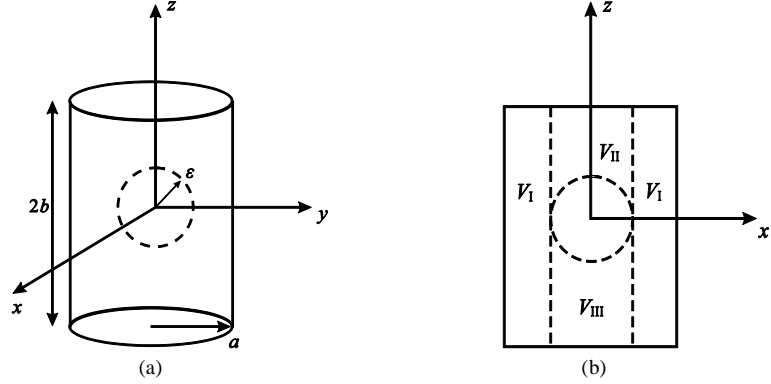

(a)        (b)

FIG. A1. Schematic diagram of the cylindrical coordinate and the integral domain:
(a) a cylindrical domain with a spherical exclusion, (b) integral domains I, II and III.

Without loss of generality, we assume that the origin of the coordinate system is located at $\mathbf{x}'$. The principal value of the integral can be simplified as:

$$P.S.I_{ij} = \lim_{\varepsilon \to 0} \iiint_{V-V_\varepsilon} \left[ \frac{F}{4\pi} \frac{\delta_{i\alpha}}{\mu} \frac{\partial^2}{\partial x_j \partial x_\beta} \frac{e^{ik_T|\mathbf{R}|}}{|\mathbf{R}|} - \frac{F}{4\pi\rho\omega^2} \frac{\partial^4}{\partial x_i \partial x_\alpha \partial x_j \partial x_\beta} \left( \frac{e^{ik_L|\mathbf{R}|}}{|\mathbf{R}|} - \frac{e^{ik_T|\mathbf{R}|}}{|\mathbf{R}|} \right) \right] dV, \quad (A5)$$

where $\varepsilon$ is the radius of the small exclusion, $\mathbf{R} = \mathbf{x} - \mathbf{x}'$ is the distance between the source point and the field point.

In the following discussion, we do not distinguish between the longitudinal and transverse waves and denote the wavenumber uniformly by $k$. When the fundamental results are obtained, we will attach the subscripts "L" or "T" to the wavenumbers of longitudinal and transverse waves again. For the convenience of the following calculation, we denote the spatial variables $x_1$, $x_2$ and $x_3$ as $x$, $y$ and $z$. From Eq. (A5) we know that the principal value of the integral involving $G_{33,33}$ is:

$$P.S.\iiint_V G_{33,33}(\mathbf{r}, 0, \omega) dV = \lim_{\varepsilon \to 0} \iiint_{V-V_\varepsilon} \left[ \frac{F}{4\pi} \frac{1}{\mu} \frac{\partial^2}{\partial z^2} \frac{e^{ik_T|\mathbf{R}|}}{|\mathbf{R}|} - \frac{F}{4\pi\rho\omega^2} \frac{\partial^4}{\partial z^4} \left( \frac{e^{ik_L|\mathbf{R}|}}{|\mathbf{R}|} - \frac{e^{ik_T|\mathbf{R}|}}{|\mathbf{R}|} \right) \right] dV, \quad (A6)$$

where the following integrals are involved:

$$I_{zz}(\varepsilon) = \iiint_{V-V_\varepsilon} \frac{\partial^2}{\partial z^2} \frac{e^{ik|\mathbf{R}|}}{|\mathbf{R}|} dV, \quad I_{zzzz}(\varepsilon) = \iiint_{V-V_\varepsilon} \frac{\partial^4}{\partial z^4} \frac{e^{ik|\mathbf{R}|}}{|\mathbf{R}|} dV. \quad (A7)$$

These integrals are most conveniently calculated in a cylindrical coordinate system shown in Fig. A1.

$$x = r\cos\theta, \quad y = r\sin\theta, \quad z = z, \quad |\mathbf{R}| = \sqrt{r^2 + z^2}. \quad (A8)$$

The integrals in Eq. (A7) are now expressed in the cylindrical coordinate system as:

$$I_{zz}(\varepsilon) = \iiint_{V-V_\varepsilon} \frac{\partial^2}{\partial z^2} \frac{e^{ik\sqrt{r^2+z^2}}}{\sqrt{r^2+z^2}} r dr d\theta dz, \quad I_{zzzz}(\varepsilon) = \iiint_{V-V_\varepsilon} \frac{\partial^4}{\partial z^4} \frac{e^{ik\sqrt{r^2+z^2}}}{\sqrt{r^2+z^2}} r dr d\theta dz. \quad (A9)$$

To further calculate these integrals, the integral domain is subdivided into three parts: $V_I$, $V_{II}$ and $V_{III}$. $V_I$ is the outer cylindrical shell formed by the inner cylindrical surface $r=\varepsilon$, outer cylindrical surface $r=a$ and the end surfaces $z=\pm b$, $V_{II}$ is enclosed by the inner cylindrical surface $r=\varepsilon$, top hemisphere and the top end $z=b$, $V_{III}$ is the symmetric counterpart of $V_{II}$, as shown in Fig. A1(b). Correspondingly, $I_{zz}(\varepsilon)$ can be divided into three parts:

$$I_{zz} = I_{zz}^{V_I} + I_{zz}^{V_{II}} + I_{zz}^{V_{III}}, \tag{A10}$$

where

$$I_{zz}^{V_I} = \int_\varepsilon^a r dr \int_0^{2\pi} d\theta \int_{-b}^b dz \frac{\partial^2}{\partial z^2} \frac{e^{ik\sqrt{r^2+z^2}}}{\sqrt{r^2+z^2}} = 4\pi b \left( \frac{e^{ik\sqrt{a^2+b^2}}}{\sqrt{a^2+b^2}} - \frac{e^{ik\sqrt{\varepsilon^2+b^2}}}{\sqrt{\varepsilon^2+b^2}} \right), \tag{A11}$$

$$I_{zz}^{V_{II}} = \int_0^\varepsilon r dr \int_0^{2\pi} d\theta \int_{\sqrt{\varepsilon^2-r^2}}^b dz \frac{\partial^2}{\partial z^2} \frac{e^{ik\sqrt{r^2+z^2}}}{\sqrt{r^2+z^2}} = \frac{2\pi}{3} e^{ik\varepsilon} - \frac{2\pi}{3} ik\varepsilon e^{ik\varepsilon} + 2\pi b \frac{e^{ik\sqrt{\varepsilon^2+b^2}}}{\sqrt{\varepsilon^2+b^2}} - 2\pi e^{ikb}. \tag{A12}$$

Considering the symmetry of the integrand and integral domain, we have $I_{zz}^{V_{III}} = I_{zz}^{V_{II}}$, thus

$$I_{zz}(\varepsilon) = I_{zz}^{V_I} + I_{zz}^{V_{II}} + I_{zz}^{V_{III}} = 4\pi b \left( \frac{e^{ik\sqrt{a^2+b^2}}}{\sqrt{a^2+b^2}} - \frac{e^{ik\sqrt{\varepsilon^2+b^2}}}{\sqrt{\varepsilon^2+b^2}} \right) + \frac{4\pi}{3} e^{ik\varepsilon} - \frac{4\pi}{3} ik\varepsilon e^{ik\varepsilon} + 4\pi b \frac{e^{ik\sqrt{\varepsilon^2+b^2}}}{\sqrt{\varepsilon^2+b^2}} - 4\pi e^{ikb}. \tag{A13}$$

The principal value of $I_{zz}(\varepsilon)$ is obtained by setting $\varepsilon$ approach zero:

$$P.S.I_{zz} = \lim_{\varepsilon \to 0} \left[ 4\pi b \left( \frac{e^{ik\sqrt{a^2+b^2}}}{\sqrt{a^2+b^2}} - \frac{e^{ik\sqrt{\varepsilon^2+b^2}}}{\sqrt{\varepsilon^2+b^2}} \right) + \frac{4\pi}{3} e^{ik\varepsilon} - \frac{4\pi}{3} ik\varepsilon e^{ik\varepsilon} + 4\pi b \frac{e^{ik\sqrt{\varepsilon^2+b^2}}}{\sqrt{\varepsilon^2+b^2}} - 4\pi e^{ikb} \right] = 4\pi \left( \frac{be^{ik\sqrt{a^2+b^2}}}{\sqrt{a^2+b^2}} - e^{ikb} \right) + \frac{4\pi}{3}. \tag{A14}$$

$I_{zzzz}(\varepsilon)$ can be calculated in the similar manner, we only present the major steps:

$$I_{zzzz}(\varepsilon) = I_{zzzz}^{V_I} + I_{zzzz}^{V_{II}} + I_{zzzz}^{V_{III}}, \tag{A15}$$

where

$$I_{zzzz}^{V_I} = 4\pi \int_\varepsilon^a \left[ -\frac{15b^3}{(r^2+b^2)^{\frac{7}{2}}} + \frac{15ikb^3}{(r^2+b^2)^3} + \frac{9b + 6k^2b^3}{(r^2+b^2)^{\frac{5}{2}}} - \frac{9ikb + ik^3b^3}{(r^2+b^2)^2} - \frac{3k^2b}{(r^2+b^2)^{\frac{3}{2}}} \right] e^{ik\sqrt{r^2+b^2}} r dr, \tag{A16}$$

$$I_{zzzz}^{V_{II}} = 2\pi \int_0^\varepsilon \left[ -\frac{15b^3}{(r^2+b^2)^{\frac{7}{2}}} + \frac{15ikb^3}{(r^2+b^2)^3} + \frac{9b + 6k^2b^3}{(r^2+b^2)^{\frac{5}{2}}} - \frac{9ikb + ik^3b^3}{(r^2+b^2)^2} - \frac{3k^2b}{(r^2+b^2)^{\frac{3}{2}}} \right] e^{ik\sqrt{r^2+b^2}} r dr + \frac{2\pi}{5} ik^3 \varepsilon e^{ik\varepsilon} - \frac{2\pi}{5} k^2 e^{ik\varepsilon}. \tag{A17}$$

Similarly, we can calculate $I_{zzzz}^{V_{III}}$, and get $I_{zzzz}^{V_{III}} = I_{zzzz}^{V_{II}}$, thus

$$I_{zzzz}(\varepsilon) = 4\pi \int_0^a \left[ -\frac{15b^3}{(r^2+b^2)^{\frac{7}{2}}} + \frac{15ikb^3}{(r^2+b^2)^3} + \frac{9b + 6k^2b^3}{(r^2+b^2)^{\frac{5}{2}}} - \frac{9ikb + ik^3b^3}{(r^2+b^2)^2} - \frac{3k^2b}{(r^2+b^2)^{\frac{3}{2}}} \right] e^{ik\sqrt{r^2+b^2}} r dr + \frac{4\pi}{5} ik^3 \varepsilon e^{ik\varepsilon} - \frac{4\pi}{5} k^2 e^{ik\varepsilon}, \tag{A18}$$

$$P.S.I_{zzzz} = \lim_{\varepsilon \to 0} I_{zzzz}(\varepsilon) = 4\pi \int_0^a \left[ -\frac{15b^3}{(r^2+b^2)^{\frac{7}{2}}} + \frac{15ikb^3}{(r^2+b^2)^3} + \frac{9b + 6k^2b^3}{(r^2+b^2)^{\frac{5}{2}}} - \frac{9ikb + ik^3b^3}{(r^2+b^2)^2} - \frac{3k^2b}{(r^2+b^2)^{\frac{3}{2}}} \right] e^{ik\sqrt{r^2+b^2}} r dr - \frac{4\pi}{5} k^2. \tag{A19}$$

Although the explicit expression of the first term on the righthand side is not available, we can find that the denominator is powers of $r^2 + b^2 \geq b^2$, and it does not bring any singularity, so only the second term have singularity.

Substituting from Eqs. (A14) and (A19) into Eq. (A6) and considering $F=1$ newton, we get:

$$P.S. \iiint_V G_{33,33}(\mathbf{r},0,\omega) dV = \frac{1}{\mu} \left( \frac{be^{ik_T\sqrt{a^2+b^2}}}{\sqrt{a^2+b^2}} - e^{ik_T b} \right)$$

$$- \int_0^a \left[ -\frac{15b^3}{(r^2+b^2)^{\frac{7}{2}}} + \frac{15ik_L b^3}{(r^2+b^2)^3} + \frac{9b + 6k_L^2 b^3}{(r^2+b^2)^{\frac{5}{2}}} - \frac{9ik_L b + ik_L^3 b^3}{(r^2+b^2)^2} - \frac{3k_L^2 b}{(r^2+b^2)^{\frac{3}{2}}} \right] e^{ik_L\sqrt{r^2+b^2}} r dr \tag{A20}$$

$$+ \int_0^a \left[ -\frac{15b^3}{(r^2+b^2)^{\frac{7}{2}}} + \frac{15ik_T b^3}{(r^2+b^2)^3} + \frac{9b + 6k_T^2 b^3}{(r^2+b^2)^{\frac{5}{2}}} - \frac{9ik_T b + ik_T^3 b^3}{(r^2+b^2)^2} - \frac{3k^2 b}{(r^2+b^2)^{\frac{3}{2}}} \right] e^{ik_T\sqrt{r^2+b^2}} r dr + \frac{2\lambda + 7\mu}{15\mu(\lambda + 2\mu)}.$$

The first three terms on the righthand side of Eq. (A20) are regular functions of the cylinder radius $a$ and height $b$, which can be expressed collectively by a function $f(a,b)$. Thus the integral in Eq. (A20) can be expressed formally as:

$$P.S. \iiint_V G_{33,33}(\mathbf{r},\omega) dV = f(a,b) + S_{3333}, \tag{A21}$$

where

$$S_{3333} = \frac{2\lambda + 7\mu}{15\mu(\lambda + 2\mu)}. \tag{A22}$$

It is shown that the correct value of the field is exactly $f(a,b)$ [107]. In accordance with the conventional definition of Green's function, we can get the new definition which is valid both in and out of the source region:

$$\iiint_V G_{33,33}(\mathbf{r},\omega)dV := P.S.\iiint_V G_{33,33}(\mathbf{r},\omega)dV - S_{3333}, \tag{A23}$$

or equivalently:

$$G_{33,33}(x,x',\omega) := P.S.G_{33,33}(x,x',\omega) - S_{3333}\delta(x-x'). \tag{A24}$$

Following the same rational, we can get the correct definition of other components, and the results are summarized below:

$$G_{ii,ii}(x,x',\omega) := P.S.G_{ii,ii}(x,x',\omega) - S_{iiii}\delta(x-x'), \tag{A25}$$

$$G_{ii,jj}(x,x',\omega) := P.S.G_{ii,jj}(x,x',\omega) - S_{iijj}\delta(x-x'), \tag{A26}$$

$$G_{ij,ij}(x,x',\omega) := P.S.G_{ij,ij}(x,x',\omega) - S_{ijij}\delta(x-x'). \tag{A27}$$

Substituting from Eqs. (A25-A27) into the expression for the Green's tensor Eq. (37), we obtain correct definition of the principal value of the Green tensor, which is given by:

$$E_{i\alpha,j\beta}(\mathbf{r},\omega) := P.S.E_{i\alpha,j\beta}(\mathbf{r},\omega) - S_{i\alpha j\beta}\delta(\mathbf{r}), \tag{A28}$$

where

$$S_{\alpha ij\beta} = S_1 \delta_{\alpha i}\delta_{j\beta} + S_2(\delta_{\alpha j}\delta_{i\beta} + \delta_{\alpha\beta}\delta_{ij}), \tag{A29}$$

$$S_1 = -\frac{\lambda + \mu}{15(\lambda + 2\mu)\mu}, \quad S_2 = \frac{3\lambda + 8\mu}{30(\lambda + 2\mu)\mu}. \tag{A30}$$

It can also be expressed as:

$$S_{\alpha ij\beta} = \frac{-2(\lambda + \mu)\delta_{\alpha i}\delta_{j\beta} + (3\lambda + 8\mu)(\delta_{\alpha j}\delta_{i\beta} + \delta_{\alpha\beta}\delta_{ij})}{30(\lambda + 2\mu)\mu}. \tag{A31}$$

## Method 2: Calculation of the singularity in the wavenumber domain

From the calculations performed in the spatial domain, we know that the Green tensor takes the following form:

$$E^0_{\alpha ij\beta}(\mathbf{x},\mathbf{x}') := P.S.E^0_{\alpha ij\beta}(\mathbf{x},\mathbf{x}') - S_{\alpha ij\beta}\delta(\mathbf{x}-\mathbf{x}'). \tag{A32}$$

Taking the Fourier transform of Eq. (A32), we get:

$$\tilde{E}^0_{\alpha ij\beta}(\mathbf{k}) = P.S.\tilde{E}^0_{\alpha ij\beta}(\mathbf{k}) - S_{\alpha ij\beta}. \tag{A33}$$

In this section, we will show that the singularity can also be obtained by setting the zero-frequency limit of the convolution between $P.S.\tilde{E}^0_{\alpha ij\beta}(\mathbf{k})$ and the power spectrum function to be zero, i.e.:

$$\frac{1}{8\pi^3}\lim_{\mathbf{k}\to 0}\iiint_{V(\mathbf{s})} P.S.\tilde{E}^0_{\alpha ij\beta}(\mathbf{s},\omega)\tilde{P}(\mathbf{k}-\mathbf{s})d^3\mathbf{s} = 0. \tag{A34}$$

Substitution of (A33) into (A34) yields:

$$\frac{1}{8\pi^3}\lim_{\mathbf{k}\to 0}\iiint_{V(\mathbf{s})}[\tilde{E}^0_{\alpha ij\beta}(\mathbf{s}) + S_{\alpha ij\beta}]\tilde{P}(\mathbf{k}-\mathbf{s})d^3\mathbf{s} = 0, \tag{A35}$$

so we get:

$$S_{\alpha ij\beta} = -\frac{1}{8\pi^3}\lim_{\mathbf{k}\to 0}\iiint_{V(\mathbf{s})}\tilde{E}^0_{\alpha ij\beta}(\mathbf{s},\omega)\tilde{P}(\mathbf{k}-\mathbf{s})d^3\mathbf{s}. \tag{A36}$$

In the frequency-wavenumber domain, the second-order derivative of Green's function is:

$$\tilde{G}^0_{\alpha i,j\beta}(\mathbf{s},\omega) = -\frac{1}{\mu}\frac{\delta_{i\alpha}s_j s_\beta}{s^2 - k_T^2} - \frac{k_L^2 - k_T^2}{\rho\omega^2}\frac{s_i s_\alpha s_j s_\beta}{(s^2 - k_T^2)(s^2 - k_L^2)}, \tag{A37}$$

and the Fourier spectral of Green's tensor is:

$$\tilde{E}^0_{\alpha ij\beta}(\mathbf{s},\omega) = -\frac{k_T^2}{4\rho\omega^2}\frac{\delta_{i\alpha}s_js_\beta + \delta_{\alpha j}s_is_\beta + \delta_{\beta i}s_\alpha s_j + \delta_{\beta j}s_is_\alpha}{s^2 - k_T^2} - \frac{k_L^2 - k_T^2}{\rho\omega^2}\frac{s_is_\alpha s_js_\beta}{(s^2 - k_L^2)(s^2 - k_L^2)}.$$ (A38)

Meanwhile, we can obtain the explicit expression of $\tilde{P}(\mathbf{k}-\mathbf{s})$ from Eq. (71):

$$\tilde{P}(\mathbf{k}-\mathbf{s}) = \frac{8\pi a^3}{[1+a^2(\mathbf{k}-\mathbf{s})\cdot(\mathbf{k}-\mathbf{s})]^2}.$$ (A39)

Since $\lim_{k\to 0}\tilde{P}(\mathbf{k}-\mathbf{s}) = \tilde{P}(-\mathbf{s})$, we have

$$S_{\alpha ij\beta} = -\frac{1}{8\pi^3}\iiint_{V(\mathbf{s})}\tilde{E}^0_{\alpha ij\beta}(\mathbf{s},\omega)\tilde{P}(-\mathbf{s})d^3\mathbf{s}.$$ (A40)

These integrals are most conveniently calculated in the spherical coordinate system:

$$s_1 = s\sin\theta\cos\phi, \quad s_2 = s\sin\theta\sin\phi, \quad s_3 = s\cos\theta.$$ (A41)

Substituting from Eqs. (A38) and (A39) into Eq. (A40), we obtain:

$$S_{\alpha ij\beta} = \frac{1}{8\pi^3}\int_0^{+\infty}s^2 ds\int_0^\pi \sin\theta d\theta\int_0^{2\pi}d\phi\frac{8\pi a^3}{(1+a^2s^2)^2}\left[\frac{k_T^2}{4\rho\omega^2}\frac{\delta_{i\alpha}s_js_\beta + \delta_{\alpha j}s_is_\beta + \delta_{\beta i}s_\alpha s_j + \delta_{\beta j}s_is_\alpha}{s^2 - k_T^2} + \frac{k_L^2 - k_T^2}{\rho\omega^2}\frac{s_is_\alpha s_js_\beta}{(s^2 - k_L^2)(s^2 - k_L^2)}\right].$$ (A42)

We first calculate $S_{1111}$. Substitution of Eq. (A41) into (A42) yields:

$$S_{1111} = \frac{a^3}{\pi^2}\int_0^{+\infty}s^2 ds\int_0^\pi \sin\theta d\theta\int_0^{2\pi}d\phi\frac{1}{(1+a^2s^2)^2}\left[\frac{k_T^2}{\rho\omega^2}\frac{s^2\sin^2\theta\cos^2\phi}{s^2 - k_T^2} + \frac{k_L^2 - k_T^2}{\rho\omega^2}\frac{s^4\sin^4\theta\cos^4\phi}{(s^2 - k_L^2)(s^2 - k_L^2)}\right].$$ (A43)

After completing the integration over $\theta$ and $\phi$, we get:

$$S_{1111} = \frac{4}{3}\frac{a^3}{\pi}\frac{k_T^2}{\rho\omega^2}\int_0^{+\infty}\frac{s^4 ds}{(1+a^2s^2)^2(s^2 - k_T^2)} + \frac{4}{5}\frac{a^3}{\pi}\frac{k_L^2 - k_T^2}{\rho\omega^2}\int_0^{+\infty}\frac{s^6 ds}{(1+a^2s^2)^2(s^2 - k_L^2)(s^2 - k_T^2)}.$$ (A44)

After straightforward algebraic manipulations, Eq. (A44) can be rewritten as:

$$S_{1111} = \frac{2}{3}\frac{a^3}{\pi}\frac{1}{a^4}\frac{k_T^2}{\rho\omega^2}I_T + \frac{2}{5}\frac{a^3}{\pi}\frac{1}{a^4}\frac{k_L^2 - k_T^2}{\rho\omega^2}I_{LT},$$ (A45)

where:

$$I_T = \int_{-\infty}^{+\infty}\frac{s^4 ds}{(s+\frac{i}{a})^2(s-\frac{i}{a})^2(s-k_T)(s+k_T)}, \quad I_{LT} = \int_{-\infty}^{+\infty}\frac{s^6 ds}{(s+\frac{i}{a})^2(s-\frac{i}{a})^2(s-k_L)(s+k_L)(s-k_T)(s+k_T)}.$$ (A46)

These integrals can be calculated using the residual technique. Obviously, integral $I_{LT}$ has four first-order poles: $s = \pm k_L$ and $s = \pm k_T$, and two second-order poles: $s = \pm i/a$. We assume that the medium has a small damping, so the poles deviate from the real axis, as shown in Fig. A2.

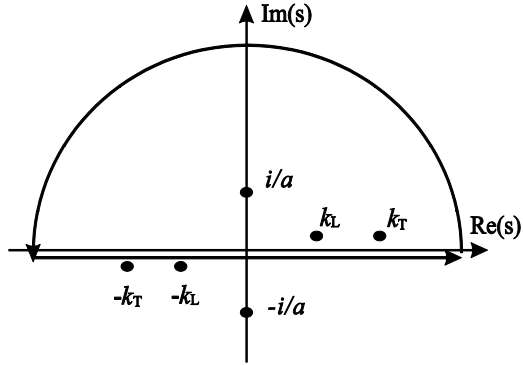

FIG. A2. Integral path in the complex $s$ plane.

Considering the condition that realistic waves must be out-going and with finite amplitude, we need to take the integral path formed by the upper semi-circle and the real axis to complete the integrals, so

$$I_T = 2\pi i\mathrm{Res}(I_T, s = k_T) + 2\pi i\mathrm{Res}(I_T, s = \frac{i}{a}),$$ (A47)

$$I_{LT} = 2\pi i \text{Res}(I_{LT}, s=k_L) + 2\pi i \text{Res}(I_{LT}, s=k_T) + 2\pi i \text{Res}(I_{LT}, s=\frac{i}{a}), \tag{A48}$$

where $\text{Res}(I, s=k)$ represents the residual of integral $I$ at its pole $s=k$.

From the residual theorem, we know:

$$\text{Res}(I_T, s=k_T) = \frac{k_T^3}{2\left(k_T^2 + \frac{1}{a^2}\right)^2}, \quad \text{Res}(I_T, s=\frac{i}{a}) = -\frac{ia}{4}, \tag{A49}$$

$$\text{Res}(I_{LT}, s=k_L) = \frac{k_L^5}{2(k_L^2 + \frac{1}{a^2})^2(k_L^2 - k_T^2)}, \quad \text{Res}(I_{LT}, s=k_T) = \frac{k_T^5}{2(k_T^2 + \frac{1}{a^2})^2(k_T^2 - k_L^2)}, \quad \text{Res}(I_{LT}, s=\frac{i}{a}) = -\frac{ia}{4}. \tag{A50}$$

Substituting from Eqs. (A49-A50) into Eqs. (A47-A48), we obtain the integrals $I_T$ and $I_{LT}$:

$$I_T = \frac{i\pi k_T^3}{\left(k_T^2 + \frac{1}{a^2}\right)^2} + \frac{\pi a}{2}, \quad I_{LT} = \frac{i\pi k_L^5}{(k_L^2 + \frac{1}{a^2})^2(k_L^2 - k_T^2)} + \frac{i\pi k_T^5}{(k_T^2 + \frac{1}{a^2})^2(k_T^2 - k_L^2)} + \frac{\pi a}{2}. \tag{A51}$$

Consequently, $S_{1111}$ is given by:

$$S_{1111} = \lim_{\omega \to 0}\left[\frac{2}{3}\frac{a^3}{\pi}\frac{1}{a^4}\frac{k_T^2}{\rho\omega^2}I_T + \frac{2}{5}\frac{a^3}{\pi}\frac{1}{a^4}\frac{k_L^2 - k_T^2}{\rho\omega^2}I_{LT}\right] = \frac{2\lambda + 7\mu}{15\mu(\lambda + 2\mu)}. \tag{A52}$$

On account of space limitations, we only give the detailed calculation for $S_{1111}$, other components can be calculated similarly and the results are exactly the same as that given by Method 1.

## Method 3: Calculation of the singularity using the static Green's tensor

It is seen from the calculations by Method 2 that the poles corresponding to the longitudinal and transverse wavenumbers, i.e., $s=k_L$ and $s=k_T$ do not contribute to the final results. This hints us that the singularity is only related to the static Green's tensor of the medium. Now, we prove this conjecture by direct calculation.

Let $\omega \to 0$, we have:

$$S_{\alpha ij\beta} = \frac{1}{8\pi^3}\int_0^{+\infty}s^2 ds \int_0^\pi \sin\theta d\theta \int_0^{2\pi}d\phi \frac{8\pi a^3}{(1+a^2 s^2)^2}\left[\frac{k_T^2}{4\rho\omega^2}\frac{1}{s^2}(\delta_{i\alpha}s_j s_\beta + \delta_{\alpha j}s_i s_\beta + \delta_{\beta i}s_\alpha s_j + \delta_{\beta j}s_i s_\alpha) + \frac{k_L^2 - k_T^2}{\rho\omega^2}\frac{1}{s^4}s_i s_\alpha s_j s_\beta\right]. \tag{A53}$$

Three nontrivial components are calculated as:

$$S_{1111} = \frac{a^3}{\pi^2}\int_0^{+\infty}s^2 ds \int_0^\pi \sin\theta d\theta \int_0^{2\pi}d\phi \frac{1}{(1+a^2s^2)^2}\left[\frac{k_T^2}{\rho\omega^2}\sin^2\theta\cos^2\phi + \frac{k_L^2 - k_T^2}{\rho\omega^2}\sin^4\theta\cos^4\phi\right] = \frac{2\lambda + 7\mu}{15\mu(\lambda + 2\mu)}, \tag{A54}$$

$$S_{1221} = \frac{a^3}{\pi^2}\frac{k_L^2 - k_T^2}{\rho\omega^2}\int_0^{+\infty}\frac{s^2 ds}{(1+a^2 s^2)^2}\int_0^\pi \sin^5\theta d\theta \int_0^{2\pi}\sin^2\phi\cos^2\phi d\phi = -\frac{\lambda + \mu}{15\mu(\lambda + 2\mu)}, \tag{A55}$$

$$S_{2233} = \frac{a^3}{\pi^2}\int_0^{+\infty}s^2 ds \int_0^\pi \sin\theta d\theta \int_0^{2\pi}d\phi \frac{1}{(1+a^2 s^2)^2}\left[\frac{k_T^2}{4\rho\omega^2}(\sin^2\theta\sin^2\phi + \cos^2\theta) + \frac{k_L^2 - k_T^2}{\rho\omega^2}\sin^2\theta\sin^2\phi\cos^2\theta\right] = \frac{3\lambda + 8\mu}{30\mu(\lambda + 2\mu)}. \tag{A56}$$

Comparing the three methods, we see Method 3 is the simplest way to obtain the singularity of the dynamic Green's tensor.